\documentclass[12pt]{article}
\pdfoutput=1 
\usepackage[utf8]{inputenc}
\usepackage{listings}
\usepackage{slashed}
\usepackage{cancel}
\usepackage{color, verbatim}
\usepackage{latexsym}
\usepackage{amssymb}
\usepackage{chngpage}
\usepackage{amsmath}
\usepackage{graphicx}
\usepackage{arydshln}
\usepackage[dvipsnames]{xcolor}
\usepackage{adjustbox}
\usepackage{easybmat}
\usepackage{bbm}
\usepackage{cite}
\usepackage{slashed}
\usepackage{bm}
\usepackage{adjustbox}
\usepackage{colortbl}
\usepackage{booktabs}
\usepackage{multirow} 
\usepackage{rotating}
\usepackage{mathtools}
\usepackage[inline]{enumitem}
\usepackage{lscape}
\usepackage[normalem]{ulem}
\usepackage{xspace}

\definecolor{codegreen}{rgb}{0,0.6,0}
\definecolor{codegray}{rgb}{0.5,0.5,0.5}
\definecolor{codegris}{rgb}{0.92,0.92,0.92}
\definecolor{codepurple}{rgb}{0.58,0,0.82}
\definecolor{backcolour}{rgb}{0.95,0.95,0.92}

\lstdefinestyle{mystyle}{
    backgroundcolor=\color{backcolour},   
    commentstyle=\color{codegreen},
    keywordstyle=\color{magenta},
    numberstyle=\tiny\color{codegray},
    stringstyle=\color{codepurple},
    basicstyle=\ttfamily\footnotesize,
    breakatwhitespace=false,         
    breaklines=true,                 
    captionpos=b,                    
    keepspaces=true,                 
    numbers=left,                    
    numbersep=5pt,                  
    showspaces=false,                
    showstringspaces=false,
    showtabs=false,                  
    tabsize=2
}

\newcommand{\lvec}[1]{\ensuremath{\overset{\leftarrow}{#1}}}

\newcommand{\lrvec}[1]{\ensuremath{\overset{\leftrightarrow}{#1}}}
\newcommand{\la}[1]{\tilde{\lambda}_{#1}}
\newcommand{\lab}[1]{\tilde{\lambda}^{\ast}_{#1}}
\newcommand{\mathematica}{\texttt{Mathematica}\xspace}
\newcommand{\form}{\texttt{FORM}\xspace}
\newcommand{\python}{\texttt{Python}\xspace}
\newcommand{\pip}{\texttt{pip}\xspace}
\newcommand{\conda}{\texttt{conda}\xspace}
\newcommand{\qgraf}{\texttt{QGRAF}\xspace}

\newcommand{\feynrules}{\texttt{FeynRules}\xspace}

\newcommand{\ftr}[1]{\lfloor {#1} \rfloor}

\newcommand{\OOp}[2]{\mathcal{O}_{#1}^{#2}}
\newcommand{\ROp}[2]{\mathcal{R}_{#1}^{#2}}
\newcommand{\EOp}[2]{\mathcal{E}_{#1}^{#2}}

\usepackage[colorlinks,citecolor=magenta,linkcolor=blue]{hyperref}
\newcommand{\ctoprule}{\toprule[0.5mm]}
\newcommand{\cbottomrule}{\bottomrule[0.5mm]}
\newcommand{\cmrule}{\cmidrule[0.25mm]}

\newcommand{\ii}{\ensuremath{\mathrm{i}}}

\newcommand{\mme}{\texttt{matchmakereft}\xspace}
\newcommand{\Mme}{\texttt{Matchmakereft}\xspace}

\newcommand{\RGEmaker}{\texttt{RGEmaker}\xspace}
\newcommand{\Matching}{\texttt{Matching}\xspace}

\newcommand{\Op}{\mathcal{O}}
\newcommand{\Lag}{\mathcal{L}}
\renewcommand{\theequation}{\thesection.\arabic{equation}}
\makeatletter
\@addtoreset{equation}{section}
\g@addto@macro\bfseries{\boldmath}
\newcommand\Label[1]{&\refstepcounter{equation}(\theequation)\ltx@label{#1}&}
\makeatother
\allowdisplaybreaks

\begin{document}
%
\thispagestyle{empty}
\begin{center}
    \includegraphics[width=.18\textwidth]{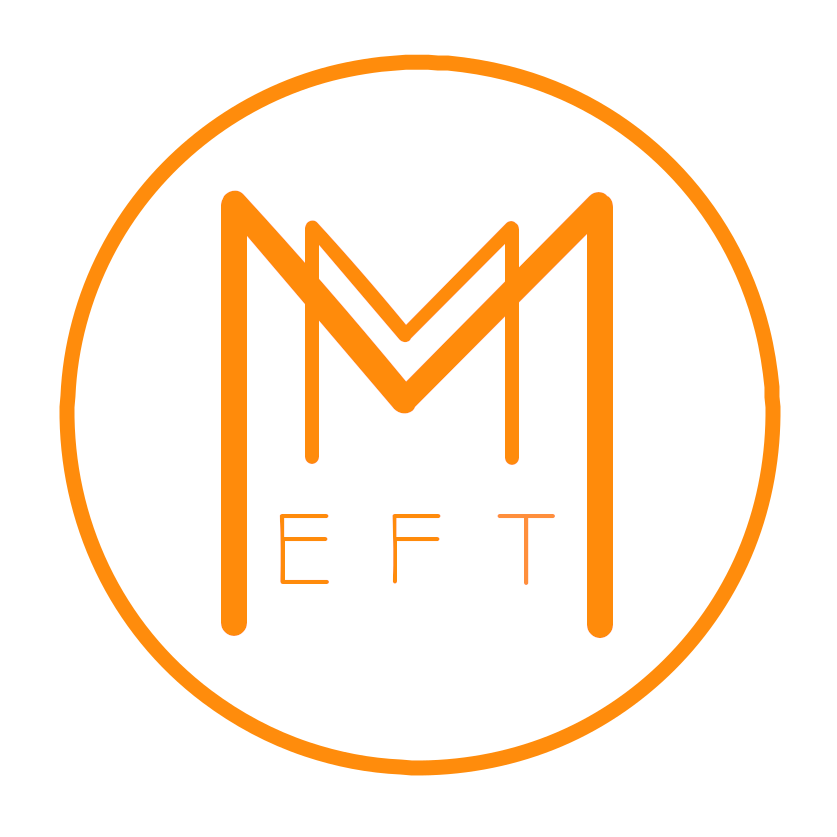}

{\Large \texttt{Matchmakereft}: automated tree-level and one-loop
  matching}
\vspace{0.8cm}

\textbf{
Adrián Carmona$^{\,a,b}$, Achilleas Lazopoulos$^{\,b}$, Pablo
Olgoso$^{\,a}$ and José Santiago$^{\,a}$}\\   
\vspace{1.cm}
{\em {$^a$ CAFPE and Departamento de F\'isica Te\'orica y del Cosmos,
Universidad de Granada, Campus de Fuentenueva, E--18071 Granada,
Spain}}\\[0.5cm] 
{\em {$^b$ Institute for Theoretical Physics, ETZ Zürich, 8093 Zürich,
    Switzerland}}    \\[0.2cm]

\vspace{0.5cm}
\end{center}
\begin{abstract}

We introduce \texttt{matchmakereft}, a fully automated tool to
compute the tree-level and  
one-loop matching of arbitrary models onto arbitrary effective
theories.
\Mme performs an off-shell matching, using diagrammatic methods and the
BFM when gauge theories are involved. 
The large redundancy inherent to the off-shell matching together with
explicit gauge invariance offers a significant number of non-trivial checks
of the results provided. 
These results are given in the physical basis but several intermediate
results, including the matching in the Green basis before and after
canonical normalization, are given for flexibility and the possibility
of further cross-checks. As a non-trivial example we provide the complete matching in the Warsaw basis up to one loop of an extension of the Standard Model with a charge $-1$ vector-like lepton singlet. \Mme has been built with generality,
flexibility and efficiency in mind. These ingredients allow \mme to
have many applications beyond the matching between models and
effective theories. Some of these applications include the
one-loop renormalization of arbitrary theories (including the calculation of
the one-loop renormalization group equations for arbitrary
theories); the translation between different Green bases for a fixed
effective theory or the check of (off-shell) linear independence of
the operators in an effective theory. All these applications are
performed in a fully automated way by \texttt{matchmakereft}.

\end{abstract}

\newpage

\tableofcontents

\section{Introduction}

Effective field theories (EFTs) are the most appropriate tool to perform
calculations in multi-scale problems when using mass-independent
renormalization schemes. The process of matching and 
running turns an often complicated, and sometimes not even
perturbatively convergent, multi-scale problem into a succession of
simpler single-scale calculations with the possibility of resummation of large
logarithms for a better perturbative
convergence~\cite{Georgi:1993mps}. The absence of direct experimental
indications of new physics beyond the Standard Model (SM) seems to
imply a hierarchy between the scale of new physics and the energies at
which experiment is performed. In these circumstances, EFTs are not
only applicable, but they become a powerful tool that allows us
to very efficiently solve the problem of comparing experimental data
with theoretical predictions. This is a highly non-trivial problem
that involves complicated, usually loop-level, calculations of many
experimental observables. Calculations that have to be repeated for
each new physics model and for each observable. The way EFTs simplify
this process is by splitting in two, mostly independent, steps. The
first one, the bottom-up approach to EFTs, provides a
model-independent parametrization of experimental observables that can
be systematically improved but has to be computed only once (for the
given precision) for each experimental observable. This efficient
parametrization can be provided in the form of global fits to
experimental data and a very important effort from the community has
been devoted to this task in the last few years (see~\cite{Hartland:2019bjb,Brivio:2019ius,Banerjee:2019twi,Aoude:2020dwv,Dawson:2020oco,Anisha:2020ggj,Falkowski:2020pma,Ellis:2020unq,Ethier:2021bye,Brivio:2021alv,Almeida:2021asy,Anisha:2021hgc} and references there in for some recent global fits). The second
step, the top-down approach to EFTs, sacrifices model-independence in
favour of model-discrimination. It consists of the process of
matching, in which the Wilson coefficients (WCs) of the EFT are computed in
terms of the parameters of the ultraviolet (UV) model. This process of
matching has to be repeated for each UV model but it can be
automated. 

When dealing with searches for new physics beyond the SM, the relevant
EFT seems to be the SMEFT (see~\cite{Brivio:2017vri} for a recent
review). The SMEFT WCs can be run down to the
electroweak scale thanks to the renormalization group equations (RGEs)
computed
in~\cite{Jenkins:2013zja,Jenkins:2013wua,Alonso:2013hga,Alonso:2014zka}
at dimension 
6 ignoring baryon and lepton number violating operators (see~\cite{Grojean:2013kd,Elias-Miro:2013mua} for calculations of a subset at dimension 6, \cite{Davidson:2018zuo,Chala:2021juk} for the inclusion of lepton-number violating operators, \cite{Alonso:2014zka} for baryon-number violating ones, \cite{Liao:2016hru} for dimension-7 operators and \cite{Chala:2021pll} for recent efforts towards the calculation of the
dimension 8 RGEs). The SMEFT has been matched to the low-energy EFT
(LEFT), the relevant EFT below the electroweak scale, both at tree
level~\cite{Jenkins:2017jig}
and at one-loop~\cite{Dekens:2019ept} and then run again using the
LEFT RGEs~\cite{Jenkins:2017dyc} to the relevant experimental energy
scale to compute the corresponding experimental observables.
This process of running in the SMEFT, matching to the LEFT and running
in the LEFT has been implemented in automated tools like
\texttt{DsixTools}~\cite{Celis:2017hod,Fuentes-Martin:2020zaz} or
\texttt{Wilson}~\cite{Aebischer:2018bkb}. This leaves the calculation
of the matching of arbitrary UV models onto the SMEFT as the only
missing step towards a fully automated calculation of the
phenomenological implications of new physics models.

We present in this article \texttt{matchmakereft}, \url{https://ftae.ugr.es/matchmakereft/}, a fully automated
tool to perform tree-level and one-loop matching of arbitrary UV
models onto arbitrary EFTs. The tree-level matching of the most
general extension of the SM onto the SMEFT at dimension 6 has been
recently computed in~\cite{deBlas:2017xtg}, building on previous
partial
results~\cite{delAguila:2000rc,delAguila:2008pw,
  delAguila:2010mx,deBlas:2014mba}. Going up to the one-loop order in
the matching is far more complex and some degree of automation is
needed. Functional methods, pioneered in~\cite{Henning:2014wua} extending
ideas from the 80s~\cite{Gaillard:1985uh,Cheyette:1987qz}, have seen
an impressive progress in the recent
years~\cite{Drozd:2015rsp,delAguila:2016zcb,Henning:2016lyp,
  Ellis:2016enq,Fuentes-Martin:2016uol,Zhang:2016pja,
  Ellis:2017jns,Kramer:2019fwz,Ellis:2020ivx,Angelescu:2020yzf,
  Cohen:2020fcu,Dittmaier:2021fls}
and they have 
resulted in computer tools that help with some of the most
computational-intensive steps of the
calculation~\cite{Cohen:2020qvb,Fuentes-Martin:2020udw} or that are
automated but apply only to specific sets of models~\cite{DasBakshi:2018vni}.
However, to the best of our knowledge, there is currently no code that
can provide the complete one-loop matching of arbitrary models onto
arbitrary EFTs in a fully automated
way.~\footnote{See~\cite{Uhlrich:2021ded} for alternative partial efforts
using a diagrammatic approach.} \Mme comes to fill this gap.

\Mme uses a diagrammatic approach to tree-level and one-loop matching,
performed in the background field method (BFM) when gauge theories are
involved. The matching is, as of version \texttt{1.0.0}, done off-shell which, together
with gauge invariance, provides a significant redundancy that results
in a number of non-trivial cross-checks of the calculation. It has been
designed with efficiency, generality and flexibility in mind, what
allows a number of applications beyond the direct matching of UV
models to EFTs. Current applications include the renormalization of
arbitrary (effective) theories, the calculation of the RGEs of
arbitrary (effective) theories, EFT basis translation and checks of (off-shell) linear independence of operators. All these
calculations are done in a fully automated way.

The rest of the article is organised as
follows. We describe the way \mme performs the tree-level and one-loop
calculation in Section \ref{underthehood}.
Model creation in \mme is explained in detail in Section~\ref{model_creation}. The different commands available in \mme are defined in Section \ref{manual} and common pitfalls when using \mme are described in Section~\ref{troubleshooting}. Some physical
applications are given in Section~\ref{applications} and we conclude
and provide some outlook in
Section~\ref{conclusions}. In the appendices we provide some more technical information, including comprehensive  installation instructions in Appendix~\ref{installation}, details about the handling of $\gamma_5$ in Appendix~\ref{sec:gamma5},  a minimal but complete example of the capabilities of \mme in Appendix~\ref{appendix:example}, as well as  the complete Green basis for the baryon-number preserving SMEFT, including our definition of evanescent operators, in \mme in Appendix~\ref{green_basis}.

\section{\texttt{Matchmakereft} in a nutshell\label{underthehood}}

\subsection{Model classification}

Models in \mme are classified according to two criteria. Depending on
their spectrum they can be \textbf{light models}, if only light (but not
necessarily massless) particles are present in the spectrum; or
\textbf{heavy models}, when there is at least one heavy particle in
the spectrum. 
Depending on their role in the process of matching we have
\textbf{UV models}, which can be light or heavy, that are to be
matched onto \textbf{EFTs}, 
which are necessarily light models. Models in \mme are created using
\feynrules~\cite{Alloul:2013bka}, as described in detail in Section~\ref{model_creation}.

\Mme performs an off-shell matching, in the BFM
when gauge theories are involved, of a UV model onto an EFT. This is
done by computing, in dimensional regularization in $D=4-2\epsilon$
space-time dimensions,   
 the hard-region contribution to the
one-light-particle-irreducible relevant 
amplitudes at tree and one-loop level in the
UV theory and equating it 
to the tree level contribution in the EFT for arbitrary kinematic
configurations of the external particles
(see~\cite{Manohar:2018aog,Cohen:2019wxr} for two recent
excellent reviews of the matching process). The amplitudes,
with only physical external light particles,
that have to be computed are fixed in an automated way by \mme but the
user has flexibility on changing this list as described in
Section~\ref{manual}. 
All the relevant diagrams in the UV model and the EFT are then
automatically computed by \qgraf~\cite{Nogueira:1991ex} and the
corresponding amplitudes are  
dressed by \mme using the Feynman rules computed during the creation
of the model.

\subsection{Amplitude calculation}

\Mme runs in two different modes, depending on whether the UV model is
light or heavy, called \RGEmaker and \Matching modes, respectively.
In \RGEmaker mode, which is used to compute the RGEs of an
arbitrary theory, the UV model contains no heavy 
particles and \mme computes the UV-divergent contribution
proportional to $1/\epsilon$ of the corresponding
one-particle-irreducible amplitudes. In \Matching mode, there are heavy
particles in the spectrum and 
both the finite and $1/\epsilon$ (both UV and IR) hard-region
contributions to the corresponding one-light-particle-irreducible
amplitudes are computed. In this case diagrams including only light 
particles are not included in the calculation, as they cancel in the matching.

The calculation of the hard region contribution to the amplitudes is
performed with \form~\cite{Ruijl:2017dtg} and
proceeds as follows:
\begin{itemize}
\item \textbf{Hard region expansion}. This corresponds to the following
  expansion of the integrand of the amplitude
  \begin{equation}
    k^2 \sim M^2 \gg p^2 \sim m^2,
  \end{equation}
  where $k$ represents the loop momentum, $M$ a heavy mass, $p$ any of
  the external momenta and $m$ a light mass. This is done by iterating
  the following identities
  \begin{align}
    \frac{1}{(k+p)^2-M^2} =& \frac{1}{k^2-M^2} \left [
      1-\frac{p^2+2k\cdot p}{(k+p)^2-M^2}\right],
    \nonumber \\
    \frac{1}{(k+p)^2-m^2} =& \frac{1}{k^2} \left [
      1-\frac{p^2+2k\cdot p-m^2}{(k+p)^2-m^2}\right].
  \end{align}
  These identities are imposed iteratively until the power of infrared
  (IR) scales (external
  momenta or light masses) is the correct one to match the maximum
  dimension of the operators appearing in the EFT as automatically
  computed in \mme.

\item \textbf{Tensor reduction}. Tensor reduction is performed by
  means of the following identities
\begin{eqnarray}
&& k^{\mu_1}k^{\mu_2}=g^{\mu_1\mu_2} \frac{k^2}{D}, \\
&& k^{\mu_1}k^{\mu_2}k^{\mu_3}k^{\mu_4}=g^{\mu_1\mu_2\mu_3\mu_4} \frac{k^4}{D^2+2D}, \\
&& k^{\mu_1}\ldots k^{\mu_6}=g^{\mu_1\ldots \mu_6} \frac{k^6}{D^3+6D^2+8D}, \\
&& k^{\mu_1}\ldots k^{\mu_8}=g^{\mu_1\ldots \mu_8}
  \frac{k^8}{D^4+12D^3+44D^2+48D}, \dots
\end{eqnarray}
where $g^{\mu_1 \ldots \mu_n}$ is the totally symmetric combination of
metric tensors.

\item \textbf{Dirac algebra}. Once the integrals have been reduced to
  scalar integrals we proceed to perform the corresponding Dirac
  algebra, in $D$ dimensions in \Matching mode (in 4
  dimensions in \RGEmaker mode). Version \texttt{1.0.0} of \mme uses an
  anticommuting $\gamma^5$ prescription as discussed in Section~\ref{gamma5:evanescent}. In the case of fermion number violating particles and/or
  interactions, we follow the rules proposed in
  Ref.~\cite{Denner:1992vza,Denner:1992me}. 
  
\item \textbf{Partial fractioning}. The following identity is used to
  separate propagators with different masses
\begin{equation}
\frac{1}{(k^2-m_1^2)(k^2-m_2^2)}=\frac{1}{m_1^2-m_2^2} \left[
  \frac{1}{k^2-m_1^2} - \frac{1}{k^2-m_2^2}\right],
\end{equation}
where masses can be light or heavy and one of them can be vanishing.
\item \textbf{Integration by parts}. After partial fractioning
  scaleless integrals are set to zero, except in \RGEmaker mode, 
  in which case we keep the UV poles using,
  \begin{equation}
   \int \frac{d^Dk}{(2\pi)^D} \frac{1}{k^4}=\frac{\ii}{(4\pi)^2}
     \frac{1}{\epsilon}+\ldots,
  \end{equation}
  before eliminating the remaining scaleless integrals.
  In \texttt{Matching} mode the following identity is
  used to reduce the massive integrals to tadpoles
\begin{equation}
\frac{1}{(k^2-m^2)^{n+1}}=\frac{D-2n}{2nm^2} \frac{1}{(k^2-m^2)^n}.
\end{equation}
At this point we are left with a tadpole integral
\begin{equation}
a_0(m)=\int\frac{d^Dk}{(2\pi)^D} \frac{1}{k^2-m^2} = \mathrm{i} \frac{m^2}{16\pi^2} \left[
\frac{1}{\bar{\epsilon}} + 1 - \log\left(\frac{m^2}{\mu^2}\right)\right] +\mathcal{O}(\epsilon),
\end{equation}
where
\begin{equation}
\frac{1}{\bar{\epsilon}} \equiv \frac{1}{\epsilon} + \gamma_E - \log (4\pi),
\end{equation}
with $\gamma_E\approx 0.5772$ the Euler-Mascheroni constant.
In \mme $1/\bar{\epsilon}$ is denoted by \texttt{invepsilonbar}.

\end{itemize}

\subsection{About $\gamma^5$ and evanescent operators\label{gamma5:evanescent}}

In order to ensure maximum generality \mme assumes no four-dimensional
properties when running in \Matching mode. In particular no Fierz
relations or reduction of products of three or more gamma matrices is
perform during the matching procedure. This means that all
evanescent structures, which are equivalent to operators in the Green
basis only in $D=4$ have to be explicitly defined as part of the Green
basis. As an example, we provide in
  Appendix~\ref{green_basis} the complete Green basis for the
  SMEFT at dimension 6 as needed for the matching with \mme of general
  theories with renormalizable couplings.~\footnote{Non-renormalizable
  theories can be matched also with \mme but an extension of the basis
  with a larger number of gamma matrices in four-fermion operators
  would be needed. Similarly, if bosonic evanescent operators appear in the process of matching a specific UV model they would have to be included in the EFT basis.} This extends the Green basis
  of~\cite{Gherardi:2020det} with the 
  general set of evanescent structures including fermionic operators. 

Regarding $\gamma^5$ it is well known that its implementation in
dimensional regularization schemes is problematic
(see~\cite{Gnendiger:2017pys} and references there in for an account
of the current status). The current version of \mme (\texttt{1.0.0}) implements an
anticommuting $\gamma^5$ together with an implementation of the
  hermiticity properties of the WCs that is enough
for the case of the SMEFT (and extensions with multiple scalar
doublets) as we discuss more carefully in Appendix~\ref{sec:gamma5}.

\subsection{Wilson coefficient matching}

Once the amplitudes have been computed, the output is written in two
files for each amplitude, one including the gauge structure and the
other including the kinematic and flavour structures. \Mme then uses
\mathematica match the amplitudes for all kinematic configurations
between the UV model and the EFT both at tree level and one loop. Once
the off-shell matching has been performed, that is the Wilson
coefficients of the Green basis have been computed in terms of the
couplings and masses of the UV model, \mme automatically performs a
canonical normalization of the results and then reduces the matching
to a physical basis as defined by the user (see Section~\ref{manual}
for details). Results at all three levels (Green basis with
non-canonical kinetic terms, canonically normalised Green basis and
Physical basis) are reported by \mme together with the corresponding
renormalisation of the gauge couplings as fixed by gauge boson
renormalisation in the BFM. If \RGEmaker mode
has been invoked then \mme can also automatically compute the beta
functions for all the WCs of the EFT used.

The off-shell matching used introduces a large degree of kinematic
redundancy. In gauge theories the BFM also introduces a large degree
of explicit gauge redundancy. These redundancies provide a very
powerful mechanism to cross-check the results obtained in \mme and
when any of them is not fulfilled \mme will issue a warning and provide some extra information that can be useful to debug the problem.

\section{Model creation\label{model_creation}}

Model creation is fully automated in \mme but it relies on the
explicit input from the user. Thus, it is the step that should be
performed with the greatest care, as it is the most likely culprit in
case of problems with the matching calculation. Model creation is greatly simplified by the use of
\feynrules but a few important points should be taken into account.

\subsection{Required files}

\Mme expects a number of files with all
  the relevant information to create the model. The detailed structure
  of each of these files will be defined below but we list them here
  first:~\footnote{The installation of \mme comes with a number of sample models that can be obtained with the command \texttt{copy\_models} (see below). We encourage the user to check these examples for details on how to implement new models.} 
  \begin{itemize} 
    \item Model files (compulsory): one or more files
      \verb+modfile1.fr, ..., modfilen.fr+ that define the
      model in \feynrules format. One of the files, the last one of
      the list during the creation of the model (see below) is
      special, as it will define the name of the \mme model, which
      is \verb+modelfilen_MM+, and the name that extra files with
      additional information need to have. \textbf{\Mme expects the Lagrangian of the to be defined as \texttt{Ltot}}. If a different name is used the model will not be created.
      
      \item Gauge information file (compulsory only if gauge groups
        are present): a file called \verb+modfilen.gauge+ that has the
        definition of all the gauge functions, including structure
        constants, group generators in different representations and
        Clebsch-Gordan coefficients, appearing in the
        model (see below for more information).
The user can choose any gauge basis of interest but they are
responsible for the consistency of the chosen basis. 
        \item Symmetry file (optional): a file called
          \verb+modfilen.symm+ that indicates possible symmetries in
          the parameters of the model. This is particularly important
          in the case of the EFT model and it is compulsory in this
          case if symmetries are present. The content of the file
          should be a \mathematica list in which the symmetries are
          given in the form of replacement rules. As an example we
          show the case of the symmmetries of the Wilson coefficient
          of the Weinberg operator (denoted by \verb+alphaWeinberg[i,j]=alphaWeinberg[j,i]+)
          and the four-lepton operator $\mathcal{O}_{\ell \ell}=\bar{\ell}_i \gamma^\mu \ell_j
          \bar{\ell}_k \gamma_\mu \ell_l$ (denoted by \verb+alphaOll[i,j,k,l]=alphaOll[k,l,i,j]+): 
\begin{lstlisting}[language=Mathematica]
listareplacesymmetry=
    {
    alphaWeinberg[i_, j_] -> alphaWeinberg[j, i],
    alphaWeinbergbar[i_, j_] -> alphaWeinbergbar[j, i],
    alphaOll[i_, j_, k_, l_] -> alphaOll[k, l, i, j]
   }
\end{lstlisting}
where the underscore denoting dummy indices on the left hand side of
the rules is compulsory and the \texttt{bar} at the end of a name
denotes complex conjugation.
          \item Redundancy file (compulsory): a file called
            \verb+modfilen.red+ that provides the redundancies that
            express the Wilson coefficient of a physical basis in
            terms of the ones in the Green basis. It is compulsory but
            it can be empty if no redundancies are needed (for example
            if the model is a UV model, if the physical and Green
            bases coincide or if we just want the results in the Green
            basis).
\item Hermiticity properties file: complex conjugation is a very
  important process in \mme, providing extra cross-checks of the
  correctness of the calculation. For that reason it is important to
  provide the information of which WCs have special
  (anti)hermiticity properties. A file called \verb+modfilen.herm+ can
  be provided by the user defining a \mathematica list called
  \verb+listahermiticity+ defining those WCs whose
  hermitian conjugate can be defined in terms of the original
  coefficient. As an example, for the case of the hermitian operator
  $(\mathcal{O}_{Hq}^{(1)})_{ij} = H^\dagger \ii  \lrvec{D}_\mu H
  \bar{\ell}_i \gamma^\mu \ell_j$, its Wilson coefficient, denoted
  \verb+alphaOHq1[i,j]+ satisfies
  \verb+alphaOHq1bar[i,j]=alphaOHq1[j,i]+. This is provided in
  the form of the hermiticity file as follows
\begin{lstlisting}
listahermiticity = {
                    alphaOHq1bar[i_,j_]->alphaOHq1[j,i]
                   }
\end{lstlisting}

  \end{itemize}
  
  \subsection{Gauge structure}
  \Mme is especially efficient when the
  matching is performed in the
  unbroken phase of gauge theories as it keeps gauge indices as dummy
  indices during the calculation of the amplitudes, replacing them
  with their explicit values only at the end of the calculation. When
  creating a model, all gauge functions, including structure
  constants, generators in different representations and Clebsh-Gordan
  coefficients need to have a specific name that does not correspond
  to any function already present in \mathematica or
  \feynrules. Structure constants and generators do not have to be
  defined as \feynrules parameters but Clebsch-Gordan coefficients
  do. The numerical values of these gauge functions are provided in a
  file wigh a \texttt{modfilen.gauge} extension with the definition of a
  mathematica list called \texttt{replacegaugedata} that consist of a
  list of substitutions in the form of \mathematica sparse arrays. As
  a simple example, the $SU(2)_L$ weak gauge group can be defined as
  follows in one of the \verb+.fr+ files:
\begin{lstlisting}
M$GaugeGroups = {
  SU2L == { 
    Abelian           -> False, 
    CouplingConstant  -> g2, 
    GaugeBoson        -> Wi, 
    StructureConstant -> fsu2, 
    Representations   -> {{Ta,SU2D}}
  } 
},
\end{lstlisting}
where we have defined the structure constant symbol and
one representation with generator symbol
\verb+Ta+ and index definition \verb+SU2D+.
Note that the
adjoint representation does not need to be explicitly defined as it is
defined by the structure constants and the definition of the
corresponding gauge bosons which, in this case reads (again, provided
in one of the \verb+.fr+ files),
\begin{lstlisting}
M$ClassesDescription = {
  V[2] == {
    ClassName        -> Wi,
    SelfConjugate    -> True,
    Indices          -> {Index[SU2W]},
    Mass             -> 0,
    FullName        -> "light"
  }
};
\end{lstlisting}
A few things are worth emphasizing from the above example. First, the
mass is set to zero because we are in the unbroken phase of the
SM. Second, we define physical fields in entire gauge multiplets,
rather than components.~\footnote{It is actually possible to define
also the field components separately as the physical fields. In fact,
this can be advantageous when creating complicated models that take
very long to generate in \feynrules.}  Finally, we assign the FullName
variable to 
``light'' (\verb+FullName->"light"+). This is compulsory in
\mme. Every particle has to be defined with \verb+FullName+ equal to
either \verb+"light"+ or \verb+"heavy"+ to define the
corresponding particle as light (and therefore to be kept in the EFT
model) or heavy (to be integrated out in the UV model).

The corresponding indices (not only gauge, also flavour indices if
present) have to be defined with a finite range within the \verb+.fr+
files. As an example the ones corresponding the the adjoint
(\verb+SU2W+) and fundamental (\verb+SU2D+) representations of
$SU(2)_L$, together with flavor indices for fermion generations (\verb+Generation+), can be defined as follows
\begin{lstlisting}
IndexRange[Index[SU2W]] = Range[3]; 
IndexRange[Index[SU2D]] = Range[2];
IndexRange[Index[Generation]] = Range[3];
IndexStyle[SU2W,n];
IndexStyle[SU2D,l];
IndexStyle[Generation, fl];
\end{lstlisting}
Only massless particles can have flavor indices in the current version of \mme (\texttt{1.0.0}).

In order to show how new particles with non-trivial quantum numbers
and Clebsch-Gordan coefficients are defined we show here the case of a
heavy scalar triplet under $SU(2)_L$ and the SM Higgs
\begin{lstlisting}
M$ClassesDescription = {
  S[105] == {
    ClassName        -> tphi,
    SelfConjugate    -> True,
    Indices          -> {Index[SU2W]},
    Mass             -> Mtphi,
    FullName        -> "heavy",
    QuantumNumbers -> {Y -> 0}
  },

  S[11] == {
    ClassName      -> Phi,
    Indices        -> {Index[SU2D]},
    SelfConjugate  -> False,
    Mass             -> muH,
    FullName        -> "light",
    QuantumNumbers -> {Y -> 1/2}
  }

};
\end{lstlisting}
where the new particle is defined as heavy and has a non-zero mass
while the SM Higgs boson is defined as light (but also has a non-vanishing
mass). We also see that $U(1)$ quantum numbers have to be defined
explicitly as is standard in \feynrules. A trilinear coupling between
the heavy scalar and two Higgs bosons, which has a non-trivial gauge
structure following the corresponding Clebsch-Gordan coefficients, has
to be defined explicitly like in the following example
\begin{lstlisting}
M$Parameters = {

  C223 == { 
    ParameterType     -> Internal,
    Indices -> {Index[SU2D],Index[SU2D],Index[SU2W]},
    ComplexParameter  -> True
  },

...
} 
\end{lstlisting}
and in the corresponding Lagrangian
\begin{lstlisting}
Ltot := Block[{ii,jj,nn},
    2 C223[ii,jj,nn] kappatphi tphi[nn] Phibar[ii] Phi[jj] + ...] 
\end{lstlisting}
As mentioned above the explicit values of the gauge
functions are given in a file called \verb+modfilen.gauge+ (assuming
the last \feynrules model file is called \verb+modfilen.fr+) which, in
the example we are showing would contain the following information
\begin{lstlisting}
replacegaugedata = {
   fsu2 -> SparseArray[Automatic, {3, 3, 3}, 0, 
           {1, {{0, 2, 4, 6}, {{2, 3}, {3, 2}, 
                {1, 3}, {3, 1}, {1, 2}, {2, 1}}}, 
                          {1, -1, -1, 1, 1, -1}}], 
       Ta -> SparseArray[Automatic, {3, 2, 2}, 0, 
           {1, {{0, 2, 4, 6}, {{1, 2}, {2, 1}, 
                {1, 2}, {2, 1}, {1, 1}, {2, 2}}}, 
                {1/2, 1/2, -I/2, I/2, 1/2, -1/2}}], 
       C223 -> SparseArray[Automatic, {2, 2, 3}, 0, 
           {1, {{0, 3, 6}, {{1, 3}, {2, 1}, {2, 2}, 
                {1, 1}, {1, 2}, {2, 3}}}, 
                {1/2, 1/2, -I/2, 1/2, I/2, -1/2}}]} 
\end{lstlisting}
where we have implemented the usual definitions,
$\mathtt{fsu2[i,j,k]}=\epsilon^{ijk}$,
$\mathtt{Ta[a,i,j]}=\sigma^a_{ij}/2=\mathtt{C223[i,j,a]}$.

\subsubsection{Background field method}

\Mme assumes that the BFM is used when gauge
theories are involved. Following the $SU(2)_L$ example, the associated quantum and ghost fields have to be defined on top of the definition of \verb+Wi+ given
above
\begin{lstlisting}
 V[102] == {
    ClassName        -> WiQuantum,
    SelfConjugate    -> True,
    Indices          -> {Index[SU2W]},
    Mass             -> 0,
    FullName        -> "light"
},
  U[1] == {
    ClassName       -> ghWi,
    SelfConjugate   -> False,
    Indices       -> {Index[SU2W]},
    Ghost           -> Wi,
    QuantumNumbers  -> {GhostNumber -> 1},
    Mass            -> 0,
    FullName        -> "light"
  },
...
\end{lstlisting}
and the Lagrangian involving the $SU(2)_L$ part is given by
\begin{lstlisting}
gotoBFM={Wi[a__]->Wi[a]+WiQuantum[a]};

Ltot := 
Block[{lag,mu,nu,ii,aa}, 
lag=-1/4 FS[Wi,mu,nu,ii] FS[Wi,mu,nu,ii]
    -ghWibar[aa].DC[(DC[ghWi[aa],mu]/.gotoBFM),mu]
    -DC[WiQuantum[mu,a],mu] DC[WiQuantum[nu,a],nu]/2;
lag/.gotoBFM
]
\end{lstlisting}
Note the compulsory use of \texttt{Ltot} for the name of the total Lagrangian of the model.

\subsection{Defining the EFT model}

As we have mentioned above, the EFT model has to be a \verb+light+
model (no heavy particles) and has to include all the relevant
operators in a Green basis. The \textbf{WCs in the EFT need to have a special format}. They have to be named \texttt{alphaXXX}, where \texttt{XXX} stands for an arbitrary number of alpha-numeric characters (see Section~\ref{protected_words}). 
The amplitudes that \mme computes are
defined by the operators in the EFT. In that sense, one does not need
to include all the operators of a Green basis but at least all the
operators in a certain class (same fields), including redundant and
evanescent operators.~\footnote{See section~\ref{gamma5:evanescent} for a discussion
about evanescent operators.} \Mme will then generate only the relevant
amplitudes to match these operators. Unless the user is absolutely sure that they do not need them, renormalizable operators, including kinetic and mass terms, have to be included in the EFT model. 

When performing the matching, \mme will check that all off-shell
kinematic configurations and all gauge directions are correctly
matched. If these checks are not satisfied, \mme will issue a warning,
and store the relevant information. This usually happens
because the user made a mistake when defining the models, either
because the model is not correctly defined or because there are
missing operators in the Green basis. See
Section~\ref{troubleshooting} for common problems and possible
solutions when running \mme.
When doing the external momentum expansion all operators of dimension
equal or smaller to the highest dimension of the operators appearing
in the EFT will be generated. Thus, one has to include all operators of smaller
dimensions within the same class. Failing to do so will make the
matching fail but the user can check that all problems appear in
sectors that are of no interest for them. Also sometimes some amplitudes are not correctly matched due to the use of an anticommuting $\gamma^5$. Our proposed solution should be enough to ensure the correct results in the SMEFT and similar EFTs but a warning will still be issued and the relevant information stored so that the user can check if the solution is correct or not. 

If the user is interested in the matching result in a physical basis,
they have to provide the corresponding redundancies to reduce the
WCs of the Green basis into the ones of the physical
basis. This is done in the file denoted \verb+modfilen.red+. As an
example, let's consider the following redundant operators in the SMEFT
at dimension 6
\begin{align}
  \Op_{HD}&=(H^\dagger D_\mu H)^\dagger (H^\dagger D_\mu H), \\
\ROp{BDH}{} &= (H^\dagger \lrvec{D}_\mu H) \partial_\nu B^{\mu\nu}
\to g_1 \Op_{HD} + \ldots,
\\
\ROp{2B}{} &=-\frac{1}{2} (\partial_\mu
B^{\mu\nu})(\partial^\rho B_{\rho\nu}) \to -\frac{g_1^2}{2} \Op_{HD} +
\ldots, 
\end{align}
where the $\to$ indicates an on-shell equivalence. These redudancies
imply the following on-shell condition for the corresponding Wilson
coefficients (with obvious notation)
\begin{equation}
  \verb+alphaOHD+ \to \verb+alphaOHD+ + 2 g_1 \verb+alphaRBDH+
  -\frac{g_1^2}{2} \verb+alphaR2B+,
\end{equation}
which we implement in the file
\verb+modfilen.red+ as follows
\begin{lstlisting}
finalruleordered={
  alphaOHD ->alphaOHD + 2*alphaRBDH*g1 - (alphaR2B*g1^2)/2,
  ...
  }
\end{lstlisting}

\subsection{Protected words \label{protected_words}}

There is quite a bit of flexibility in the process of model definition
in \mme but there are a few words that are protected and should be
used only for their specific purpose. In general, all variables in
\mme should be made of alphanumeric characters, without including any
special characters. The list of protected variables are the following
\begin{itemize}
  \item{\verb+alpha+}. All WCs should be defined as
    \verb+alphaXXX+ where \verb+XXX+ is an arbitrary string of
    alphanumeric characters. Similarly no other variable in the model
    should contain the substring \verb+alpha+. An exception to this
    rule is that when computing the RGEs of an EFT the Wilson
    coefficients in the UV model can be kept with their original
    \verb+alphaXXX+ name as they will be changed into \verb+WCXXX+
    automatically by \mme.
    \item{\verb+Ltot+}. Ltot should be used to define the complete
      Lagrangian of the model (and it should not be used for anything
      else).
\item \verb+Quantum+. Gauge bosons are split into a classical
  background and a quantum excitation. If the classical gauge boson is
  defined by \verb+ClassName->Vname+ then the quantum excitation has
  to be defined by \verb+ClassName->VnameQuantum+.
  \item \verb+invepsilonbar+ is used for the dimensional
    regularization variable $1/\bar{\epsilon}$ so it should not be use
    explicitly in the definition of a model. Similarly
    \verb+epsilonbar+ is used of $\bar{\epsilon}$.
      \item \verb+Eps[]+ denotes the \feynrules Levi-civita
        tensor. When used with four indices it is interpreted by \mme
        as the Minkowskian (with $+---$ metric signature) totally
        antisymmetric tensor and should therefore not be used for the
        Euclidean one (with a number of indices different from four it
        can be used as the Euclidean one). In case one needs to use
        the totally antisymmetric rank-4 tensor both with Minkowskian
        and Euclidean signatures, the latter should be explicitly
        defined as a gauge function and its numerical value defined in
        the corresponding  \verb+modfilen.gauge+ file.
    \item \verb+onelooporder+ is a dummy variable to identify the
      one-loop order contribution. 
    \item \verb+sSS+ is a dummy variable to identify the
      order in external momenta of a specific contribution.
      \item \verb+iCPV+$=\epsilon_{0123}$ is used to fix the sign
        convention for the Levi-Civita symbol.
\end{itemize}

\section{\Mme  usage\label{manual}}

An updated version of this manual can be found, once
\mme is installed, in the directory
\verb+matchmakereft-location/matchmaker/docs/+
where
\verb+matchmakereft-location+ is the directory listed under
\verb+Location+ when the command \verb+pip show matchmakereft+ is used or the analogous location in Anaconda (see Appendix~\ref{installation}).

\Mme can be run in two different ways. The same commands are
available in all different running modes although the syntax is
slightly different on each of them.
 
\subsection{\Mme commmand line interface}

The most straight-forward way to run \mme is via the command line
interface (CLI). This is obtained by just typing on the terminal (after \mme has been installed, see Appendix~\ref{installation} for details)
\begin{lstlisting}[numbers=none,backgroundcolor=\color{codegris}]
> matchmakereft
\end{lstlisting}
The user has then access to the CLI which looks as follows
\begin{lstlisting}[numbers=none,backgroundcolor=\color{codegris}]
Welcome to matchmakereft
Please refer to arXiv:2112.xxxxx when using this code 

matchmakereft> 
\end{lstlisting}
Inside the CLI tab-completion is available and all file paths can be absolute or relative.
The command \verb+help+ gives information on all
available commands.  The core commands in \mme CLI are:
\begin{lstlisting}[numbers=none,backgroundcolor=\color{codegris}]
matchmakereft> test_installation
\end{lstlisting} 
This command runs a number of
    minimal tests to check that \mme has been correctly installed. The
    process is verbose and provides information on what is being
    computed. It takes about 6 minutes to complete in a core-\texttt{i7@3.00
    GHz} laptop. 
\begin{lstlisting}[numbers=none,backgroundcolor=\color{codegris}]
matchmakereft> copy_models Location
\end{lstlisting} 
This command copies a number of
    sample models, including the complete baryon-number conserving
    SMEFT at dimension 6, in the directory \verb+Location+ (which can
    be \verb+.+ for the current directory).
\begin{lstlisting}[numbers=none,backgroundcolor=\color{codegris}]
matchmakereft> create_model modfile1.fr ... modfilen.fr
\end{lstlisting} 
This command
    creates a \mme model called \verb+modfilen_MM+ from the \feynrules
    model defined in one or more files with names
    \verb+modfile1.fr ... modfilen.fr+ as described in detail in
    Section~\ref{model_creation}. The \mme model is created in the
    same directory \verb+modfilen.fr+ is present.
    Both relative and absolute paths can be given as input.
\begin{lstlisting}[numbers=none,backgroundcolor=\color{codegris}]
matchmakereft> match_model_to_eft UVModelName EFTModelName
\end{lstlisting} 
This
    command performs the complete tree-level and one-loop matching of
    \mme UV model with name \verb+UVModelName+ onto a \mme EFT model
    with name \verb+EFTModelName+. The result of the matching is
    written in a file called \verb+MatchingResult.dat+ under file
    \verb+UVModelName+. Any possible problems with the matching are
    reported and stored in a file called \verb+MatchingProblems.dat+
    under the same directory. \Mme automatically checks if the
    matching is run in \RGEmaker or \texttt{Matching} mode.

    The result of the matching stored in \verb+MatchingResult.dat+ is
    a mathematica list called \verb+MatchingResult+ with four
    entries and the following structure
\begin{lstlisting}
MatchingResult={
{
{{GreenTree,GreenTreeProblems},{GreenLoop,GreenLoopProblems}},
{{NormGreenTree,NormGreenTreeProblems},{NormGreenLoop,NormGreenLoopProblems}},
{PhysTreeLoop},
{GaugeCouplingMatching}
}
\end{lstlisting}
where \verb+GreenTree+ and \verb+GreenLoop+ stands for the tree level
and one-loop matching in the Green
basis, respectively and \verb+GreenTreeProblems,GreenLoopProblems+ are
filled if problems were found in the process of impossing hermiticity
as discussed in Section~\ref{gamma5:evanescent}. The second level,
with the \verb+Norm+ prefix stand for the matching in the Green basis
(again separately for tree level and one-loop) after canonical
normalization. The third level, denoted \verb+PhysTreeLoop+ stands for
the matching in the physical basis in which the tree level and
one-loop contributions have been merged into a single expression (with
the dummy variable \verb+onelooporder+ denoting the one-loop
contribution). If no physical basis is defined by providing an empty
\verb+modfilen.red+ file then the Green basis is used as a physical
basis. Finally, the fourth level, denoted \verb+GaugeCouplingMatching+
provides the redefinition of the gauge couplings after matching as
fixed by the corresponding gauge boson canonical normalization in the
background gauge.
\begin{lstlisting}[numbers=none,backgroundcolor=\color{codegris}]
matchmakereft> match_model_to_eft_onlytree UVModelName EFTModelName
\end{lstlisting} 
Identical
    to \verb+match_model_to_eft+ but only the tree level matching is
    computed. With this feature, \mme can be used as an automated
    basis translator, as one can simply use the corresponding EFT in a
    different basis as UV model and the matching will provide the
    complete translation between the two bases (see Section~\ref{applications} for an explicit example).
\begin{lstlisting}[numbers=none,backgroundcolor=\color{codegris}]
matchmakereft> compute_rge_model_to_eft UVModelName EFTModelName
\end{lstlisting} 
This
    command runs \verb+match_model_to_eft UVModelName EFTModelName+ in
    \RGEmaker mode (the UV model has to be a light model) and then
    computes the beta functions for the WCs of the EFT
    given the UV model. They are stored in a file called
    \verb+RGEResult.dat+ under directory \verb+UVModelName+. We define the beta function of a WC $C$ as
    \begin{equation}
        \beta(C)=\mu \frac{\mathrm{d}C}{\mathrm{d} \mu}.
    \end{equation}
\begin{lstlisting}[numbers=none,backgroundcolor=\color{codegris}]
matchmakereft> clean_model ModelName
\end{lstlisting} 
\Mme is designed for the
    maximal efficiency so that if a specific process has been already
    computed it is not computed again. If for any reason the user
    wants to recreate the calculation of all the amplitudes this
    command should be invoked to clean the previous calculations.
\begin{lstlisting}[numbers=none,backgroundcolor=\color{codegris}]
matchmakereft> check_linear_dependence EFTModelName
\end{lstlisting} 
Given a set of operators, defined as an EFT model in directory \verb+EFTModelName+,
    this command checks if they are off-shell linearly independent or
    not. This command is useful when finding a Green basis as
    sometimes the off-shell relations between different operators are
    difficult to obtain analytically. If the set is not linearly independent \mme will provide the relations between the different WCs (see Section~\ref{applications} for an explicit example). 
\begin{lstlisting}[numbers=none,backgroundcolor=\color{codegris}]
matchmakereft> exit
\end{lstlisting} 
This command exits the CLI.    

For the sake of flexibility the following commands are also available
to perform independently some of the steps of the calculations:
\begin{lstlisting}[numbers=none,backgroundcolor=\color{codegris}]
matchmakereft> match_model_to_eft_amplitudes UVModelName EFTModelName
\end{lstlisting} 
This
  command is used to compute all the relevant amplitudes in the UV
  model and the EFT but no calculation of the WCs is
  attempted. 
\begin{lstlisting}[numbers=none,backgroundcolor=\color{codegris}]
matchmakereft> match_model_to_eft_amplitudes_onlytree UVModelName EFTModelName
\end{lstlisting} 
This
  command is identical to \verb+match_model_to_eft_amplitudes+ but
  performs only the tree level calculation.
\begin{lstlisting}[numbers=none,backgroundcolor=\color{codegris}]
matchmakereft> compute_wilson_coefficients UVModelName EFTModelName
\end{lstlisting} 
This
  command should be run after the call to
  \verb+match_model_to_eft_amplitudes+ or
  \verb+match_model_to_eft_amplitudes_onlytree+ and it computes the
  WCs to complete the matching.

\subsection{\Mme as a \python module}

The python CLI described in the previous section provides an interactive experience to user. However, \mme can be also run by importing \mme into a python script, a iPython shell or a Jupyter notebook as a module and then running the same commands as in the CLI adding the parameters of the corresponding function as a string. As an example, the commands to create a model stored in model \verb+UVmodel.fr+ and to match it to the EFT stored in directory \verb+EFT_MM+ (which we assume has been already created) are given e.g. by
\begin{lstlisting}
from matchmakereft.libs.mm_offline import *
create_model("UVmodel.fr")
match_model_to_eft("UVmodel_MM EFT_MM")
\end{lstlisting}

All other commands in the CLI are also available to use as functions in a script that imports \verb+matchmakereft+.

\section{Troubleshooting in \mme \label{troubleshooting}}

\Mme provides a significant number of cross-checks that usually catch
problems with the installation or with the definition of the models.
When a problem is encountered, \mme tries to provide a useful warning
message that can be used to figure out the origin of the problem.
If the user encounters a problem that cannot be solved from the information provided by \mme we encourage them to check the troubleshooting section in the latest \mme manual and the \texttt{Gitlab} \mme issue tracker (\url{https://gitlab.com/m4103/matchmaker-eft/-/issues}) to see if the problem has been encountered by other users and a solution is available. If no solution can be found, the issue tracker should be use to pose questions to the \mme developers or to file possible bugs.

Most of the times an unsuccessful matching is due to a badly defined
model. Some common pitfalls are:
\begin{itemize}
\item The complete Lagrangian of the model has to be named \verb+Ltot+. Using a different name results in \mme not creating the model properly.
\item Operators badly defined in \feynrules (a common example is
  indices not properly contracted). This results in wrong Feynman
  rules that lead to incorrect matching.
  \item Model generation takes too long. This can happen with
    complicated models, in particular with effective operators of high
    mass dimension. As an example, the generation of the SMEFT model
    can easily take more than 30 minutes in a core-\texttt{i7@3.00
    GHz} laptop. In this
    case it is useful to compute the Feynman rules directly with
    \feynrules to check that the model does not have any obvious
    problems. Also sometimes expanding in the gauge components can
    significantly speed up model creation (at the expense of a reduced
    gauge degeneracy and therefore a smaller set of cross-checks). 
  \item \qgraf not running correctly. This could happen when vertices
    with a larger number of particles than the limit set in \qgraf are
    present. The solution is to modify correspondingly the limit in
    the \qgraf source and compile it again.
    \item \form not running correctly. This is normally due to
      variables not being correctly defined (again due to an incorrect
      implementation of the model). Running directly with \form the
      offending file can give hints on what is happening.
      \item \form taking too long to run. Amplitudes with many
        external legs usually involve a very large number of diagrams
        that can take a long time to compute. The simplest solution is
        to not include the corresponding operators in the EFT model if
        the user is not interested in their matching (in the SMEFT
        case at dimension 6 the operator $\Op_H=(H^\dagger H)^3$ is
        usually the one that takes the longest to be matched). Other
        options could imply splitting the diagrams into smaller sets
        and combining the result after the calculation (this is how
        the beta function of the $(H^\dagger H)^4$ operator was
        computed in~\cite{Chala:2021pll}) but this process is not
        straight-forward. We expect to automate this procedure in
        future versions of \mme.

\item All amplitudes are computed but the matching is
  unsuccessful. This can be due to a number of reasons, the most
  common ones being: the WCs of the EFT model or the
  couplings in the UV model have
  some symmetry properties that have not been implemented in the
  corresponding \verb+modfilen.symm+ file; the hermiticity properties
  of the couplings in the EFT or UV models have not been properly
  defined, either in the definition of the model itself or in the
  corresponding \verb+modfilen.herm+ file; there are some missing
  operators in the Green basis of the EFT model.

\end{itemize}

\section{Physics applications \label{applications}}

A preliminary version of \mme has been already used in a number of
physical
applications~\cite{Chala:2020wvs,Chala:2021pll,Chala:2021wpj} but we
list here some of the new applications that \mme has. 

\subsection{Cross-checks}

As we have emphasized, the large redundancy inherent in the off-shell
matching in the BFM gives us confidence on the
correctness of the results computed with \mme. Nevertheless we have
tested \mme against some of the few available complete one-loop
matching results in the literature. We have found complete agreement
except when explicitly described. The list of models we have compared
to include:
\begin{itemize}
\item \RGEmaker mode:
  \begin{itemize}
  \item Complete RGEs for the ALP-SMEFT up to mass dimension-5 as
  computed in~\cite{Chala:2020wvs}. Exact agreement was found up to a typo in the original reference.
  \item RGEs for the purely bosonic and two-fermion operators in the
    Warsaw basis~\cite{Grzadkowski:2010es} as computed in~\cite{Jenkins:2013zja,Jenkins:2013wua,Alonso:2013hga} and implemented in \texttt{DSixTools}~\cite{Celis:2017hod,Fuentes-Martin:2020zaz}. Complete agreement was found. 
  \end{itemize}
\item \Matching mode:
  \begin{itemize}
    \item Scalar singlet. The complete matching up to one-loop order
      of an extension of the SM with a scalar singlet was recently
      completed in~\cite{Haisch:2020ahr}, after several partial
      attempts~\cite{Henning:2014wua,Jiang:2018pbd}. We have found complete agreement with the results in~\cite{Haisch:2020ahr}.
      \item Type-I see-saw model, as computed in~\cite{Zhang:2021jdf}. Complete agreement was found.
      
        \item Scalar leptoquarks, as computed
          in~\cite{Gherardi:2020det}. We have found some minor differences that we are discussing with the authors.
          \item Charged scalar electroweak singlet, as computed
            in~\cite{Bilenky:1993bt}. We agree with the result except for a sign in Eqs. (4.14), the terms with Pauli matrices in (4.15), (B.4) and (B.5) (the latter is the culprit of the opposite sign in terms with Pauli matrices) and a factor of 2 in Eq. (4.17) and of 4 in (B.7). We have contacted the authors about these differences.
  \end{itemize}
  
\end{itemize}

\subsection{Complete one-loop matching of a new charged vector-like
  lepton singlet}

In this section we provide the complete tree-level and one-loop
matching of an extension of the SM with a new hypercharge $-1$,
electroweak singlet vector-like lepton $E$. The Lagrangian is given by
\begin{equation}
  \Lag =\Lag_{\rm SM} + \bar{E}(\ii \slashed{D} -M_E) E
  -\big[\la{i} \bar{\ell}_i \phi E_R + \mathrm{h.c.}\big],
\end{equation}
where $\ell_i$ and $\phi$ stand for the SM lepton doublets and the Higgs
boson, respectively and $i$ is a SM flavor index. See~\cite{Guedes:2021oqx} for direct experimental limits on such an extension of the SM.

This model is included in the distribution of \mme and can be obtained via the \verb+copy_models+ command. Once the model is downloaded, and inside the corresponding directory, the following commands will generate the complete one-loop matching, including the complete matching in the Greeen basis. We use the CLI as an example and replace the output given by \mme with ... ,
\begin{lstlisting}[numbers=none,backgroundcolor=\color{codegris}]
matchmakereft> create_model UnbrokenSM_BFM.fr VLL_Singlet_Y_m1_BFM.fr
...
matchmakereft> match_model_to_eft VLL_Singlet_Y_m1_BFM_MM SMEFT_Green_Bpreserving_MM
...
\end{lstlisting}

The non vanishing WCs in the Warsaw basis, including one-loop accuracy are given in the next sections. In order to reduce clutter we only write explicitly flavor indices when necessary. Also we use the following notation
\begin{equation}
\ftr{\la{}\lab{}}\equiv \la{i}\lab{i},
\qquad
\ftr{\la{}\mathcal{M}\lab{}}\equiv \la{i}\mathcal{M}_{ij}\lab{j},
\end{equation}
with $\mathcal{M}_{ij}$ and arbitrary matrix with flavor indices. We also define
\begin{equation}
    L_E\equiv \log(\mu^2/M_E^2)
\end{equation}.

The tree level result agrees with the calculation in~\cite{delAguila:2008pw} (when taking into account the different notation in the Yukawa coupling).

\subsubsection{SM couplings}
The SM couplings receive the following (one-loop) corrections:
\begin{align}
\mu_H^2=&\mu_H^{(0)\,2}+ \frac{\ftr{\lab{} \la{}}}{16\pi^2}\left[ 2 M_E^2
  -\frac{1}{2} \mu_H^{(0)\,2} -\frac{1}{3}
  \frac{\mu_H^{(0)\,4}}{M_E^2} - (\mu_H^{(0)\,2}-2M_E^2) L_E\right],\label{VLL:muH2}
\\
\lambda=&\lambda^{(0)}
+\frac{1}{16\pi^2} \left[
  \frac{\ftr{\lab{}\la{}}
    (5g_2^{(0)\,2} \mu_H^{(0)\,2} -6
    \lambda^{(0)}(4\mu_H^{(0)\,2}+3M_E^2) +18 \mu_H^{(0)\,2}
    \ftr{\lab{}\la{}})}
       {18 M_E^2}\right.
       \nonumber \\       &\qquad \qquad\quad\left.
	+\frac{(2M_E^2-\mu_H^{(0)\,2})\ftr{\lab{} Y_{e}^{(0)} Y_{e}^{(0)\,\dagger}\la{}}
  }{M_E^2}
  \right]
\\ \nonumber
&\phantom{\lambda^{(0)}}-\frac{1}{16\pi^2} \left[
  \frac{\ftr{\lab{}\la{}}
    [-g_2^{(0)\,2} \mu_H^{(0)\,2} +6
    \lambda^{(0)}M_E^2 -3 M_E^2
    \ftr{\lab{}\la{}}]
    }{3 M_E^2}
    \right. \nonumber \\ &  \qquad\qquad\quad \left.+\frac{
		2 (-M_E^2+\mu_H^{(0)\,2})\ftr{\lab{} Y_{e}^{(0)} Y_{e}^{(0)\,\dagger}\la{}}
  }{ M_E^2}  
  \right] L_E,
\\
Y_u=&Y_u^{(0)} -\frac{1}{16\pi^2}
\left[
  \left( \frac{1}{4} + \frac{\mu_H^{(0)\,2}}{3M_E^2}\right)
  Y_u^{(0)} \ftr{\lab{}\la{}}
  +\frac{1}{2}Y_u^{(0)} \ftr{\lab{}\la{}}L_E
  \right],
\\
Y_d=&Y_d^{(0)} -\frac{1}{16\pi^2}
\left[
  \left( \frac{1}{4} + \frac{\mu_H^{(0)\,2}}{3M_E^2}\right)
  Y_d^{(0)} \ftr{\lab{}\la{}}
  +\frac{1}{2}Y_d^{(0)} \ftr{\lab{}\la{}}L_E
  \right],
\\
	(Y_{e})_{ij}=&(Y_{e}^{(0)})_{ij}
-\frac{1}{16\pi^2}
\left[
  \left( \frac{1}{4} + \frac{\mu_H^{(0)\,2}}{3M_E^2}\right)
	(Y_{e}^{(0)})_{ij} \ftr{\lab{}\la{}}
  +
  \left( \frac{3}{8} + \frac{3\mu_H^{(0)\,2}}{4M_E^2}\right)
	\la{i} \lab{k} (Y_{e}^{(0)})_{kj}
  \right]
\nonumber \\
	&\phantom{(Y_{e}^{(0)})_{ij}}
-\frac{1}{16\pi^2}
\left[
	\frac{1}{2}(Y_{e}^{(0)})_{ij} \ftr{\lab{}\la{}}
  +\left(\frac{1}{4}+\frac{\mu_H^{(0)\,2}}{2M_E^2}\right)
	\la{i} \lab{k} (Y_{e}^{(0)})_{kj}
  \right]L_E, \label{VLL:Yl}
\end{align}
where the $(0)$ superscript denotes the original parameters in the SM
Lagrangian. All other SM couplings receive no corrections. In the
following we express our results in terms of the physical SM couplings,
the ones on the left hand side of Eqs. (\ref{VLL:muH2}-\ref{VLL:Yl}).

\subsubsection{Bosonic operators}

Turning now to dimension 6 bosonic operators, we obtain the following non-vanishing WCs.
\begin{align}
  \alpha_{HW}=&\frac{1}{16\pi^2} \frac{g_2^2 \ftr{\la{}\lab{}}}{24M_E^2},
  \\
  \alpha_{HB}=&\frac{1}{16\pi^2} \frac{g_1^2 \ftr{\la{}\lab{}}}{8M_E^2},
  \\
  \alpha_{HWB}=&-\frac{1}{16\pi^2} \frac{g_1 g_2 \ftr{\la{}\lab{}}}{6M_E^2}, 
\\
\alpha_{H\square}=&
\frac{1}{16\pi^2 } \frac{1}{M_E^2}
\left[
  -\frac{g_1^4}{30} +\left(\frac{13 g_1^2}{72}-\frac{5g_2^2}{24}
  -\frac{\ftr{\la{}\lab{}}}{3}\right) \ftr{\la{}\lab{}}
	+\frac{3}{2} \ftr{\lab{}Y_{e} Y_{e}^\dagger \la{}}
  \right]
\nonumber \\
+&\frac{1}{16\pi^2 } \frac{1}{M_E^2}
\left[
  \left(\frac{g_1^2}{12}-\frac{g_2^2}{4}
  \right) \ftr{\la{}\lab{}}
	+ \ftr{\lab{}Y_{e} Y_{e}^\dagger \la{}}
  \right] L_E,
\\
\alpha_{HD}=&
\frac{1}{16\pi^2 } \frac{1}{M_E^2}
\left[
  -\frac{2g_1^4}{15} +\left(\frac{13 g_1^2}{18}
  -\frac{\ftr{\la{}\lab{}}}{2}\right) \ftr{\la{}\lab{}}
	+\frac{1}{2} \ftr{\lab{}Y_{e} Y_{e}^\dagger \la{}}
  \right]
\nonumber \\
+&\frac{1}{16\pi^2 } \frac{1}{M_E^2}
\left[
  \frac{g_1^2}{3} \ftr{\la{}\lab{}}
	+ \ftr{\lab{}Y_{e} Y_{e}^\dagger \la{}}
  \right]L_E,
\\
\alpha_{H}=&
\frac{1}{16\pi^2 } \frac{1}{M_E^2}
\bigg[
  \bigg(
  \frac{4\lambda^2}{3}-\frac{5\lambda g_2^2}{9}-2 \lambda
  \ftr{\la{}\lab{}}
  +\frac{\ftr{\la{}\lab{}}\ftr{\la{}\lab{}}}{3}
	+2\ftr{\lab{}Y_{e} Y_{e}^\dagger \la{}}
  \bigg) \ftr{\la{}\lab{}}
  \nonumber\\&
	\phantom{\frac{1}{16\pi^2 } \frac{1}{M_E^2}}  +2 \lambda \ftr{\lab{}Y_{e} Y_{e}^\dagger \la{}}
	-2 \ftr{\lab{}Y_{e} Y_{e}^\dagger Y_{e} Y_{e}^\dagger \la{}}
  \bigg]
\nonumber \\
+&
\frac{1}{16\pi^2 } \frac{1}{M_E^2}
\left[
  -\frac{2\lambda g_2^2}{3}
  \ftr{\la{}\lab{}}
	+4 \lambda \ftr{\lab{}Y_{e} Y_{e}^\dagger \la{}}
	-2 \ftr{\lab{}Y_{e} Y_{e}^\dagger Y_{e} Y_{e}^\dagger \la{}}  
  \right]L_E.
\end{align}
All other bosonic operators do not receive any corrections up to one loop.

\subsubsection{Bi-fermion operators}

Regarding operators in the Warsaw basis with two fermion fields, the non-vanishing contributions in our model are the following.
\begin{align}
  (\alpha_{eW})_{ij}=&-\frac{1}{16\pi^2} \frac{g_2}
  {24 M_E^2}
	\la{i} \lab{k}  (Y_{e})_{kj},
  \\
  (\alpha_{eB})_{ij}=&-\frac{1}{16\pi^2} \frac{g_1}
  {12 M_E^2}
	\la{i} \lab{k}  (Y_{e})_{kj},
  \\
  (\alpha_{Hq}^{(1)})_{ij}=&
  \frac{ g_1^2}{16\pi^2} \frac{1}{M_E^2}
  \left[
    -\frac{g_1^2}{45} +\frac{13}{216} \ftr{\la{}\lab{}}
+
    \frac{1}{36} \ftr{\la{}\lab{}}
     L_E
    \right]\delta_{ij},
  \\
  (\alpha_{Hq}^{(3)})_{ij}=&
  \frac{ g_2^2}{16\pi^2} \frac{1}{M_E^2}
  \left[
    -\frac{5 }{72} \ftr{\la{}\lab{}}
- \frac{1}{12} \ftr{\la{}\lab{}}
     L_E
    \right]\delta_{ij},
  \\
  (\alpha_{Hu})_{ij}=&
  \frac{ g_1^2}{16\pi^2} \frac{1}{M_E^2}
  \left[
    -\frac{4g_1^2}{45} +\frac{13}{54} \ftr{\la{}\lab{}}
+ \frac{1}{9} \ftr{\la{}\lab{}} L_E
    \right]\delta_{ij},
  \\
  (\alpha_{Hd})_{ij}=&
  \frac{ g_1^2}{16\pi^2} \frac{1}{M_E^2}
  \left[
    \frac{2g_1^2}{45} -\frac{13}{108} \ftr{\la{}\lab{}}
- \frac{1}{18} \ftr{\la{}\lab{}} L_E
    \right]\delta_{ij},
  \\
  (\alpha_{H\ell}^{(1)})_{ij}=&
  -\frac{\la{i} \lab{j}}{4M_E^2}
  +\frac{1}{16\pi^2} \frac{1}{M_E^2}
  \left[
    \left(\frac{g_1^4 }{15}-
    \frac{13 g_1^2 \ftr{\la{}\lab{}}}{72} \right)\delta_{ij}
    +\left(
    \frac{31 g_1^2}{288} -\frac{33 g_2^2}{32}+\frac{13}{16} \ftr{\la{}\lab{}}
    \right) \la{i}\lab{j}
    \right.\nonumber\\&\left.
\phantom{  -\frac{\la{i} \lab{j}}{4M_E^2}
  +\frac{1}{16\pi^2} \frac{1}{M_E^2}
	}    -\frac{1}{2} \la{i} \lab{k} (Y_{e})_{kl} (Y_{e}^\dagger)_{lj}
	-\frac{1}{2} (Y_{e})_{ik} (Y_{e}^\dagger)_{kl}\la{l} \lab{j}
    \right]
  \nonumber  \\
  &
\phantom{ -\frac{\la{i} \lab{j}}{4M_E^2}}  
+\frac{1}{16\pi^2} \frac{1}{M_E^2}
  \left[
    -\frac{g_1^2\ftr{\la{}\lab{}}}{12}\delta_{ij}
    +
    \left(
    \frac{25 g_1^2}{48}-\frac{9g_2^2}{16}+\frac{3\ftr{\la{}\lab{}}}{8}
    \right)\la{i}\lab{j}
    \right.\nonumber\\&\left.    
\phantom{  -\frac{\la{i} \lab{j}}{4M_E^2} +\frac{1}{16\pi^2} \frac{1}{M_E^2}
}    -\frac{1}{2} \la{i} \lab{k} (Y_{e})_{kl} (Y_{e}^\dagger)_{lj}
	-\frac{1}{2} (Y_{e})_{ik} (Y_{e}^\dagger)_{kl}\la{l} \lab{j}    
\right]L_E,
  \\
  (\alpha_{H\ell}^{(3)})_{ij}=&
  -\frac{\la{i} \lab{j}}{4M_E^2}
  +\frac{1}{16\pi^2} \frac{1}{M_E^2}
  \left[
   - 
    \frac{5 g_2^2 \ftr{\la{}\lab{}}}{72}\delta_{ij} 
    +\left(
    \frac{9 g_1^2}{32} +\frac{77 g_2^2}{288}+\frac{5}{16} \ftr{\la{}\lab{}}
    \right) \la{i}\lab{j}
    \right]
  \nonumber  \\
  &
\phantom{ -\frac{\la{i} \lab{j}}{4M_E^2}}  
  +\frac{1}{16\pi^2} \frac{1}{M_E^2}
  \left[
    -\frac{g_2^2\ftr{\la{}\lab{}}}{12}\delta_{ij}
    +
    \left(
    \frac{9g_1^2}{16}
    +\frac{7g_2^2}{48}+\frac{3\ftr{\la{}\lab{}}}{8}
    \right)\la{i}\lab{j}
\right]L_E,
  \\
  (\alpha_{He})_{ij}=&
  \frac{1}{16\pi^2} \frac{1}{M_E^2}
  \left[
     g_1^2 \left(
    \frac{2g_1^2}{15}-\frac{13 \ftr{\la{}\lab{}}}{36}
    \right)\delta_{ij}
	+\frac{1}{24} (Y_{e}^\dagger)_{ik} \la{k} \lab{l} (Y_{e})_{lj}
    \right.\nonumber\\&\left.    
\phantom{  \frac{1}{16\pi^2} \frac{1}{M_E^2}
}    +\left(
    -\frac{g_1^2}{6}  \ftr{\la{}\lab{}} \delta_{ij}
	+ \frac{1}{4} (Y_{e}^\dagger)_{ik} \la{k} \lab{l} (Y_{e})_{lj}
\right)    L_E
    \right],
  \\
  (\alpha_{uH})_{ij}=&
  \frac{1}{16\pi^2} \frac{1}{M_E^2}
  \left[
    \left(
    \frac{2}{3} \lambda
    -\frac{5g_2^2}{36}-\frac{1}{2}\ftr{\la{}\lab{}}
    \right)\ftr{\la{}\lab{}}
	+\frac{1}{2} \ftr{\lab{}Y_{e} Y_{e}^\dagger \la{}}
\right.\nonumber\\ 
&\left.
\phantom{  \frac{1}{16\pi^2} \frac{1}{M_E^2}
}+\left(
    -\frac{g_2^2}{6} \ftr{\la{}\lab{}}
	+\ftr{\lab{}Y_{e} Y_{e}^\dagger \la{}}
    \right)L_E
    \right](Y_u)_{ij},
  \\
  (\alpha_{dH})_{ij}=&
  \frac{1}{16\pi^2} \frac{1}{M_E^2}
  \left[
    \left(
    \frac{2}{3} \lambda
    -\frac{5g_2^2}{36}-\frac{1}{2}\ftr{\la{}\lab{}}
    \right)\ftr{\la{}\lab{}}
	+\frac{1}{2} \ftr{\lab{}Y_{e} Y_{e}^\dagger \la{}}
\right.\nonumber \\ 
&\left.
\phantom{\frac{1}{16\pi^2} \frac{1}{M_E^2}}+\left(
    -\frac{g_2^2}{6} \ftr{\la{}\lab{}}
	+\ftr{\lab{}Y_{e} Y_{e}^\dagger \la{}}
    \right)L_E
    \right](Y_d)_{ij},
  \\
  (\alpha_{eH})_{ij}=&
  \frac{\la{i} \lab{k} (Y_{e})_{kj}}{2M_E^2}
  +
  \frac{1}{16\pi^2} \frac{1}{M_E^2}
  \left[
   \left(
    \frac{2}{3} \lambda
    -\frac{5g_2^2}{36}-\frac{1}{2}\ftr{\la{}\lab{}}
	\right)\ftr{\la{}\lab{}} (Y_{e})_{ij}
	+\frac{1}{2} \ftr{\lab{}Y_{e} Y_{e}^\dagger \la{}} (Y_{e})_{ij}
    \right.\nonumber\\&\left.    
    + \left(
    5 \lambda -\frac{14}{16} \ftr{\la{}\lab{}}
	\right)\la{i} \lab{k} (Y_{e})_{kj}
	+\frac{37}{24} (Y_{e})_{ik} (Y_{e}^\dagger)_{kl}\la{l} \lab{m}
	(Y_{e})_{mj}
	-\frac{1}{4} \la{i} \lab{k} (Y_{e})_{kl} (Y_{e}^\dagger)_{lm}(Y_{e})_{mj}
    \right]\nonumber\\&
    + \frac{1}{16\pi^2} \frac{1}{M_E^2}
  \left[
	  -\frac{g_2^2}{6}\ftr{\la{}\lab{}} (Y_{e})_{ij}    
	+ \ftr{\lab{}Y_{e} Y_{e}^\dagger \la{}} (Y_{e})_{ij}
    + \left(
    4 \lambda -\frac{3}{4} \ftr{\la{}\lab{}}
	\right)\la{i} \lab{k} (Y_{e})_{kj}
    \right.\nonumber\\&\left.    
\phantom{\frac{1}{16\pi^2} \frac{1}{M_E^2}}   
+\frac{3}{4} (Y_{e})_{ik} (Y_{e}^\dagger)_{kl}\la{l} \lab{m}
	(Y_{e})_{mj}
    \right] L_E.
\end{align}

\subsubsection{Four-fermion operators}

Finally, the following four-fermion operators receive non-vanishing WCs. 
\begin{align}
  (\alpha_{qq}^{(1)})_{ijkl}=&
 - \frac{1}{16\pi^2} \frac{g_1^4}{270 M_E^2} \delta_{ij} \delta_{kl},
  \\
  (\alpha_{uu})_{ijkl}=&
  -\frac{1}{16\pi^2} \frac{8g_1^4}{135 M_E^2} \delta_{ij} \delta_{kl},
  \\
  (\alpha_{dd})_{ijkl}=&
  -\frac{1}{16\pi^2} \frac{2g_1^4}{135 M_E^2} \delta_{ij} \delta_{kl},
  \\
  (\alpha_{ud}^{(1)})_{ijkl}=&
  -\frac{1}{16\pi^2} \frac{8g_1^4}{135 M_E^2} \delta_{ij} \delta_{kl},
  \\
  (\alpha_{qu}^{(1)})_{ijkl}=&
  -\frac{1}{16\pi^2} \frac{1}{M_E^2}\left[
    \frac{4g_1^4}{135} \delta_{ij} \delta_{kl}
    +\frac{1}{18} \ftr{\la{}\lab{}}(Y_u)_{il} (Y_u^\dagger)_{kj} 
    \right],
  \\
  (\alpha_{qu}^{(8)})_{ijkl}=&
  -\frac{1}{16\pi^2}\frac{1}{3M_E^2} 
      \ftr{\la{}\lab{}}(Y_u)_{il} (Y_u^\dagger)_{kj},
  \\
  (\alpha_{qd}^{(1)})_{ijkl}=&
  \frac{1}{16\pi^2} \frac{1}{M_E^2}\left[
    \frac{2g_1^4}{135} \delta_{ij} \delta_{kl}
    -\frac{1}{18} \ftr{\la{}\lab{}}(Y_d)_{il} (Y_d^\dagger)_{kj} 
    \right],
  \\
  (\alpha_{qd}^{(8)})_{ijkl}=&
  -\frac{1}{16\pi^2}\frac{1}{3M_E^2} 
      \ftr{\la{}\lab{}}(Y_d)_{il} (Y_d^\dagger)_{kj},
  \\
  (\alpha_{quqd}^{(1)})_{ijkl}=&
  \frac{1}{16\pi^2}\frac{1}{3M_E^2} 
      \ftr{\la{}\lab{}}(Y_u)_{ij} (Y_d)_{kl},
  \\
  (\alpha_{\ell \ell})_{ijkl}=&
  \frac{1}{16\pi^2}\frac{1}{M_E^2}
  \left[
    -\frac{g_1^4}{30}\delta_{ij}\delta_{kl}
    +\frac{25g_1^2+11g_2^2}{288} (\delta_{ij} \la{k} \lab{l}+
    \la{i}\lab{j} \delta_{kl})
    \right.\nonumber\\&\left.    
    -\frac{11 g_2^2}{144}(\delta_{il}\la{k} \lab{j} + \delta_{jk}
    \la{i} \lab{l})
    -\frac{1}{8} \la{i} \lab{j} \la{k} \lab{l}
    +\frac{3}{16}(\la{i} \lab{l} (Y_{e})_{km}(Y_{e}^\dagger)_{mj}
    +\la{k} \lab{j} (Y_{e})_{im}(Y_{e}^\dagger)_{ml})
    \right]
  \nonumber \\
  &+
  \frac{1}{16\pi^2}\frac{1}{M_E^2}
  \left[
    \frac{g_1^2+g_2^2}{48} (\delta_{ij} \la{k} \lab{l}+
    \la{i}\lab{j} \delta_{kl})
    -\frac{g_2^2}{24}(\delta_{il}\la{k} \lab{j} + \delta_{jk}
    \la{i} \lab{l})
    \right.\nonumber\\&\left.    
    \phantom{\frac{1}{16\pi^2}\frac{1}{M_E^2}}
    +\frac{1}{8}(\la{i} \lab{l} (Y_{e})_{km}(Y_{e}^\dagger)_{mj}
    +\la{k} \lab{j} (Y_{e})_{im}(Y_{e}^\dagger)_{ml})
    \right] L_E,
  \\
  (\alpha_{ee})_{ijkl}=&
  -\frac{1}{16\pi^2}\frac{2g_1^4}{15M_E^2} 
     \delta_{ij} \delta_{kl},
  \\
  (\alpha_{\ell e})_{ijkl}=&
  \frac{1}{16\pi^2}\frac{1}{M_E^2}
  \left[
    -\frac{2g_1^4}{15}\delta_{ij} \delta_{kl}
    +\frac{25g_1^2}{72} \la{i} \lab{j} \delta_{kl}
    -\frac{1}{6} (Y_{e})_{il} (Y_{e}^\dagger)_{kj}
    -\frac{3}{8} \la{i}\lab{j} (Y_{e}^\dagger)_{km} (Y_{e})_{ml}
    \right]
  \nonumber \\
  +&
  \frac{1}{16\pi^2}\frac{1}{M_E^2}
  \left[
    +\frac{g_1^2}{12} \la{i} \lab{j} \delta_{kl}
    -\frac{1}{4} \la{i}\lab{j} (Y_{e}^\dagger)_{km} (Y_{e})_{ml}
    \right] L_E,
  \\
  (\alpha_{\ell q}^{(1)})_{ijkl}=&
  \frac{1}{16\pi^2}\frac{1}{M_E^2}
  \left[
    \frac{g_1^4}{45}\delta_{ij} \delta_{kl}
    +\frac{25g_1^2}{432} \la{i} \lab{j} \delta_{kl}
    +\frac{3}{16} \la{i}\lab{j}
    \Big((Y_d)_{km}(Y_d^\dagger)_{ml}-(Y_u)_{km}(Y_u^\dagger)_{ml}\Big)
    \right]
  \nonumber \\
  +&
  \frac{1}{16\pi^2}\frac{1}{M_E^2}
  \left[
    -\frac{g_1^2}{72} \la{i} \lab{j} \delta_{kl}
    +\frac{1}{8} \la{i}\lab{j}
    \Big((Y_d)_{km}(Y_d^\dagger)_{ml}-(Y_u)_{km}(Y_u^\dagger)_{ml}\Big)
    \right] L_E,
  \\
  (\alpha_{\ell q}^{(3)})_{ijkl}=&
  \frac{1}{16\pi^2}\frac{1}{M_E^2}
  \left[
    -\frac{11g_2^2}{144} \la{i} \lab{j} \delta_{kl}
    +\frac{3}{16} \la{i}\lab{j}
    \Big((Y_d)_{km}(Y_d^\dagger)_{ml}+(Y_u)_{km}(Y_u^\dagger)_{ml}\Big)
    \right]
  \nonumber \\
  +&
  \frac{1}{16\pi^2}\frac{1}{M_E^2}
  \left[
    -\frac{g_2^2}{24} \la{i} \lab{j} \delta_{kl}
    +\frac{1}{8} \la{i}\lab{j}
    \Big((Y_d)_{km}(Y_d^\dagger)_{ml}+(Y_u)_{km}(Y_u^\dagger)_{ml}\Big)
    \right] L_E,
  \\
  (\alpha_{eu})_{ijkl}=&
  \frac{1}{16\pi^2}\frac{8g_1^4}{45M_E^2} 
     \delta_{ij} \delta_{kl},
  \\
  (\alpha_{ed})_{ijkl}=&
  -\frac{1}{16\pi^2}\frac{4g_1^4}{45M_E^2} 
     \delta_{ij} \delta_{kl},
  \\
  (\alpha_{qe})_{ijkl}=&
  \frac{1}{16\pi^2}\frac{2g_1^4}{45M_E^2} 
     \delta_{ij} \delta_{kl},
  \\
  (\alpha_{\ell u})_{ijkl}=&
  \frac{1}{16\pi^2}\frac{1}{M_E^2}
  \left[
    \frac{4g_1^4}{45} \delta_{ij}\delta_{kl}
    -\frac{25g_1^2}{108} \la{i} \lab{j} \delta_{kl}
    +\frac{3}{8} \la{i}\lab{j}
    (Y_u^\dagger)_{km}(Y_u)_{ml}
    \right]
  \nonumber \\
  +&
  \frac{1}{16\pi^2}\frac{1}{M_E^2}
  \left[
    -\frac{g_1^2}{18} \la{i} \lab{j} \delta_{kl}
    +\frac{1}{4} \la{i}\lab{j}
    (Y_u^\dagger)_{km}(Y_u)_{ml}
    \right] L_E,
  \\
  (\alpha_{\ell d})_{ijkl}=&
  \frac{1}{16\pi^2}\frac{1}{M_E^2}
  \left[
    -\frac{2g_1^4}{45} \delta_{ij}\delta_{kl}
    +\frac{25g_1^2}{216} \la{i} \lab{j} \delta_{kl}
    -\frac{3}{8} \la{i}\lab{j}
    (Y_d^\dagger)_{km}(Y_d)_{ml}
    \right]
  \nonumber \\
  +&
  \frac{1}{16\pi^2}\frac{1}{M_E^2}
  \left[
    \frac{g_1^2}{36} \la{i} \lab{j} \delta_{kl}
    -\frac{1}{4} \la{i}\lab{j}
    (Y_d^\dagger)_{km}(Y_d)_{ml}
    \right] L_E,
  \\
  (\alpha_{\ell edq})_{ijkl}=&
  \frac{1}{16\pi^2}\frac{1}{3M_E^2}\ftr{\la{}\lab{}} 
     (Y_{e})_{ij} (Y_d^\dagger)_{kl}, 
  \\
  (\alpha_{\ell equ}^{(1)})_{ijkl}=&
  -\frac{1}{16\pi^2}\frac{1}{3M_E^2}\ftr{\la{}\lab{}} 
     (Y_{e})_{ij} (Y_u)_{kl}.
\end{align}

\subsection{Basis translation}

Given a specific EFT, defined by its field content and symmetries, there are different options to choose a basis of operators (either Green or physical). Different bases are useful for different purposes and it is very useful to have a systematic way to translate the results from one basis to another. The \verb+Rosetta+ code~\cite{Falkowski:2015wza} can be used to translate between popular physical bases for the dimension 6 SMEFT. However a more general approach, applicable to different EFTs and any two bases (not necessarily physical) would be very welcome. \Mme can do this in a straight-forward way by performing the tree-level matching of the new basis (as a UV model) onto the old one (as an EFT). 

As a trivial example we consider the following two operators that appear in the one-loop integration of a scalar singlet as shown in~\cite{Jiang:2018pbd}
\begin{align}
O_R=& (H^\dagger H) (D_\mu H^\dagger D^\mu H)
\nonumber \\
\to&2\lambda \Op_H+\frac{1}{2}\Op_{H\square}
+\frac{1}{2}\Big(
(Y_u)_{ij} (\Op_{uH})_{ij}
+(Y_d)_{ij} (\Op_{dH})_{ij}
+(Y_e)_{ij} (\Op_{eH})_{ij}
+\mathrm{h.c.}\Big)
, \\
O_T=& \frac{1}{2} (H^\dagger \lrvec{D}_\mu H)^2
\to-2\Op_{HD} -\frac{1}{2}\Op_{H\square},
\end{align}
where in the second equality we have written these operators in terms of the corresponding ones in the Warsaw basis (see Appendix~\ref{green_basis} for the definition of the operators).

When using \mme to match at tree level a UV model consisting of the SM plus the two operators in the new basis, $O_R$ and $O_T$, onto the SMEFT in the basis described in Appendix~\ref{green_basis} we obtain the following tree-level matching in the physical basis (we use $\beta$ for the WCs of the operators in the new basis) 

\begin{align}
\alpha_{H}=& 2\lambda \beta_R, & \alpha_{HD}=& -2 \beta_T, & \alpha_{H\square}=&\frac{1}{2}( \beta_R-\beta_T), \\
\alpha_{uH}=& \frac{1}{2} \beta_R (Y_u)_{ij}, &
\alpha_{dH}=& \frac{1}{2} \beta_R (Y_d)_{ij}, &
\alpha_{eH}=& \frac{1}{2} \beta_R (Y_e)_{ij},
\end{align}
which exactly reproduce the above equations. This is of course just a minimal example to show the application of \mme to basis translation but complete bases (both Green and physical) can be translated in an automated way using this procedure. 

\subsection{Off-shell operator independence}

Constructing a Green basis, that can match arbitrary off-shell amplitudes, is in principle straight-forward. One has to simply write all possible operators and then eliminate those that are related to others by integration by parts (momentum conservation at the amplitude level) or gauge (Fierz) identities. However, this procedure is sometimes quite cumbersome in practice, in particular when there are many (space-time) indices involved. \Mme can be used in this case to check if the operators defined in the EFT are linearly independent for arbitrary off-shell kinematics or not. This is done by checking the rank of the system of equations obtained by matching the EFT to all vanishing amplitudes. If several operators are linearly dependent, the rank will be smaller than the number of operators and \mme will solve the system of equations to provide the relationship of the list of the dependent operators in terms of a particular set of independent ones. This is achieved by defining an EFT with all the relevant operators (including possibly linearly dependent ones) and running the command 
\verb+check_linear_dependece EFTModel+ (see Section~\ref{manual} for details).
As a word of caution it should be emphasized that \mme makes no assumption about the space-time dimension. Thus, operators that are linearly dependent in $D=4$ but not in arbitrary $D$ are listed as linearly independent in \mme. 

\section{Conclusions and outlook\label{conclusions}}

EFTs have become a standard tool in quantum field theory to address multi-scale problems. 
Using a mass-independent renormalization scheme, the process of computing observables in models with disparate scales consists of integrating out particles at the highest mass threshold, running the WCs by means of the RGEs down to the next threshold, integrating out the particles at that threshold and so on until we reach the energies at which experimental measurements are performed. We then compute the corresponding experimental observable using the EFT relevant at that scale.

The current situation in particle physics seems to indicate that indeed, physics beyond the SM, if present, is likely to be at a scale much larger than the energies at which we are performing our experiments. 
In searches beyond the SM, the relevant EFT seems to be the SMEFT at energies above the electroweak scale and the LEFT below that scale. The process of running in the SMEFT, matching to the LEFT and running in the LEFT down to experimental energies has been recently completed, up to mass dimension 6 and one-loop order, and it is now available in computer tools. The matching of arbitrary models to the (dimension 6) SMEFT has been also recently solved at tree level. Thus, the only missing piece to perform a complete one-loop analysis of the implications of experimental data on new physics models is the one-loop matching of arbitrary models onto the SMEFT.

In this article we have introduced \mme, a computer tool that performs tree-level and one-loop matching of arbitrary models onto arbitrary EFTs. \Mme is robust, efficient, flexible and fully automated and can be trivially used to perform the step to one-loop matching of new physics models onto the SMEFT to complete the program outlined above. Due to its flexibility \mme has many more applications than just the matching of new physics models onto the SMEFT. First it can match any model onto any EFT. Thus, it can be used to match new physics models to operators of dimension higher than 6 in the SMEFT (see~\cite{Chala:green8} for the definition of the bosonic sector of a complete Green basis for the SMEFT at dimension 8), or to other EFTs beyond the SMEFT that can be also phenomenologically interesting, including for instance new light (or heavy  particles) like an axion-like particle or a right-handed neutrino~\cite{Chala:2020vqp}. \Mme is also able to match divergences in EFTs to compute the RGEs of arbitrary theories. Again this includes other EFTs beyond the dimension-6 SMEFT (or LEFT) but also renormalizable (or not) arbitrary theories. Other applications of \mme include an automated basis translation between two (Green or physical) bases of an EFT or the reduction of an off-shell linearly dependent set of operators to a minimal Green basis. All these applications are performed in an automated way, with minimal interaction from the user, that only needs to provide the information of the UV and the EFT models. 

Version \texttt{1.0.0} of \mme performs an off-shell matching in the background-field gauge, thus providing a significant (kinematic and gauge) redundancy that is used to perform numerous cross-checks of the computed WCs. There are nevertheless some limitations in version \texttt{1.0.0} that we plan to overcome in future versions of \mme. Among these some of the most significant are the following:
\begin{itemize}
    \item Flavor indices are currently not supported for massive particles. There is an experimental version of this in \mme and we expect to implement it in the near future.
    \item Amplitudes with many external particles (as required to match operators with many fields) can have a very large number of Feynman diagrams at one loop. When the number is very large the calculation can become very slow. We plan to introduce new procedures to deal with this issue, including the splitting of groups of Feynman diagrams that contribute to a single amplitude and the parallelization of the calculation to allow an efficient computation in multi-core or multi-node computer systems. 
\end{itemize}

\Mme has already been used in several physical applications and we have performed an extensive number of non-trivial cross-checks to ensure the robustness of the calculations produced by \mme. Its flexibility and efficiency will allow the particle physics community to analyze in a fully systematic and automated way the one-loop phenomenology or arbitrary new physics models.

\section*{Acknowledgments}
This has been a long project that has benefited from discussions with
many people. We would like to thank F. del Águila, J. de Blas, C. Bobeth,
M. Ciuchini, A. Freytas, J. Fuentes-Martín, G. Guedes, M. Gorbahn, U. Haisch,
Z. Kunstz, M. Neubert, 
U. Nierste, M. Pérez-Victoria, A. Pich, E. Salvioni, A. Signer and
M. Trott for useful discussions. 
We are especially grateful to C. Anastasiou for collaboration in the initial
stages of the project, to M. Chala, for continuos discussions,
suggestions and requests that allowed us to ``milk \mme for all
its worth'', to T. Hahn, for patiently explaining many \form tricks,
and to G.P. Passarino, for daring us to ``finish
  \mme already!''. We would like to thank Nicolasa Navarrete for the design of the logo and her creative support. 
This work has been partially supported by 
the Ministry of Science and Innovation and SRA (10.13039/501100011033)
under grant PID2019-106087GB-C22, 
by the Junta de Andaluc\'ia grants FQM 101, A-FQM-211-UGR18 and
P18-FR-4314 (FEDER). AC acknowledges funding from the European Union's Horizon
2020 research and innovation programme under the Marie
Sklodowska-Curie grant agreement No 754446 and UGR Research and
Knowledge Transfer Found - Athenea3i. PO is funded by an FPU grant from the Spanish government.

\appendix

\section{\Mme installation \label{installation}}

\subsection{Prerequisites}
In order to be able to run \mme, some prerequisites need to be
met. First of all, the following programs need to be installed: 
	\begin{itemize}
		\item \mathematica, version 10 or higher.  	
		\item \form\,:\,  Installation can be checked by
                  typing \verb+form -v+ in a terminal. Binaries or
                  the source code can be 
                  downloaded from 
                  \url{http://www.nikhef.nl/~form/} 

		\item \qgraf\,:\, Installation can be checked by
                  typing \verb+qgraf+ in a terminal. Binaries or the
                  source code can be 
                  downloaded from
                  \url{http://cfif.ist.utl.pt/~paulo/qgraf.html}. 
Please note that versions of \qgraf
earlier than \texttt{3.5} had a maximun multiplicity of vertices equal
to 6 particles. If vertices with more than 6 particles are needed
(this can happen if dimension-8 operators are considered for instance) the
source code has to be modified accordingly and compiled again. The
maximum multiplicity of \qgraf \texttt{v3.5} is 8.
	\end{itemize}
The binaries of both \form\, and \qgraf\, need to be located in some
path that is included in the binary path of the system, in such a  way
that they can be executed from any possible location. It is
also necessary to have installed 
\begin{itemize}
\item \python\, (\verb+3.5+ or higher)\,:\, Installation can be checked by
typing \verb+python --version+ in a terminal. In
some systems \verb+python3+ might have to be
explicitly invoked.
\item \feynrules\, the mathematica package, is needed for the creation of \mme models. It can be downloaded from   \url{https://feynrules.irmp.ucl.ac.be/}.

\end{itemize}

\Mme is available both in the Python Package Index (PyPI) \url{https://pypi.org/project/matchmakereft/} as well as in the Anaconda Python distribution \url{https://anaconda.org/matchmakers/matchmakereft}. Moreover,
\begin{itemize}

		      \item \pip. 
		      In order to access packages from PyPI, \texttt{pip} needs to be installed on the system. It is available in most Linux distributions and can be easily installed on MacOS and Windows. 
                        Installation can be checked by typing
                        \verb+pip -V+ in a terminal. In some systems
                        \texttt{pip3} has to be explicitly invoked.  General information about installing packages fcan be found on \url{https://packaging.python.org/en/latest/tutorials/installing-packages/}.
\item \conda. Instructions about Anaconda installation on different OS can be found on \url{https://docs.anaconda.com/anaconda/install/index.html}. Explicit information about installing \texttt{conda} packages can be found on \url{https://docs.anaconda.com/anacondaorg/user-guide/howto/}.
\end{itemize}

\subsection{Installing {\rm \mme}}
\subsubsection{PyPI}

Once \pip\, is installed in the system, \mme can be installed
by just typing in a terminal (we denote the terminal prompt as \verb+>+)
\begin{lstlisting}[numbers=none,backgroundcolor=\color{codegris}]
> python3 -m pip install matchmakereft --user
\end{lstlisting}
or by typing 
\begin{lstlisting}[numbers=none,backgroundcolor=\color{codegris}]
> pip install matchmakereft --user 
\end{lstlisting}
Alternatively, if the installation has been downloaded from the
project web page, it can be installed via
\begin{lstlisting}[numbers=none,backgroundcolor=\color{codegris}]
> python3 -m pip install matchmakereft-x.x.x.tar.gz  --user
\end{lstlisting}
where the \verb+x.x.x+ correspond to the version being installed.

We can get information about \mme by writing 
\begin{lstlisting}[numbers=none,backgroundcolor=\color{codegris}]
> pip3 show matchmakereft
Name: matchmakereft
Version: 1.0.0
Summary: One loop matching
Home-page: https://ftae.ugr.es/matchmakereft/
Author: Adrian Carmona, Achilleas Lazopoulos, Pablo Olgoso, Jose Santiago
Author-email: adrian@ugr.es, lazopoulos@itp.phys.ethz.ch, pablolgoso@ugr.es, jsantiago@ugr.es
License: Creative Commons Attribution-Noncommercial-Share Alike license
Location: /home/user/.local/lib/python3.9/site-packages
Requires: requests, yolk3k, setuptools
Required-by: 
\end{lstlisting}
If \mme is already installed in the system, it is possible to check for possible updates by writing
\begin{lstlisting}[numbers=none,backgroundcolor=\color{codegris}]
> pip install --upgrade matchmakereft --user
\end{lstlisting}
whereas one can remove it by typing
\begin{lstlisting}[numbers=none,backgroundcolor=\color{codegris}]
> pip uninstall matchmakereft
\end{lstlisting}
\subsubsection{Anaconda}
For users employing Anaconda python distribution, \mme can be installed by just typing
\begin{lstlisting}[numbers=none,backgroundcolor=\color{codegris}]
conda install -c matchmakers matchmakereft 
\end{lstlisting}

\subsection{Updating the system path}

Once \mme is installed, we have to ensure that the corresponding
executable is included in the user path. This can be easily achieved
by copying the following two lines in the \verb+~/.bashrc+ file (or
the equivalent one if a different shell is used)
\begin{lstlisting}
export PY_USER_BIN=$(python -c 'import site; print(site.USER_BASE + "/bin")')
export PATH=$PY_USER_BIN:$PATH
\end{lstlisting}
If the user prefers to install without the \verb+--user+ option (s)he
should ensure that the directory where the executables are installed
is included in the path.

\section{Dealing with $\gamma_5$}
\label{sec:gamma5}

When computing the matching equations with \mme, we might encounter loop integrals that involve fermionic traces with chiral projectors, $P_{L,R}$. Among such traces, those that contain an odd number of $\gamma_5$ and six or more $\gamma$-matrices are known to be ambiguous, due to the impossibility of finding a regulator that is Lorenz invariant and preserves the chiral structure of the theory. In \mme we regulate integrals within Dimensional Regularization, so we have to face the question of how to compute these traces consistently. 

Limiting ourselves, at first, to matching renormalizable UV models to the SMEFT, the cases we have to deal with, in practice, are few. In order to have a trace in the first place, there can be no external fermions in the related diagram. Moreover, the ambiguity is proportional to the fully antisymmetric $\epsilon_{\mu\nu\rho\sigma}$ tensor, and can therefore contribute to the Wilson Coefficient of one of the few CP violating bosonic operators of the SMEFT, with two or three field strength tensors. We therefore need  two or three external gauge bosons. In order to have a trace with at least 6 $\gamma$-matrices we, therefore need four or three internal fermionic propagators respectively. In summary, we only need to worry about boxes contributing to $H^\dagger H X_{\mu\nu}\tilde{X}^{\mu\nu}$ type operators or triangles contributing to $X^{\mu}_{\nu}X^{\nu}_{\rho}\tilde{X}_{\rho}^\mu$. 
Triangle diagrams contributing to $X^{\mu}_{\nu}X^{\nu}_{\rho}\tilde{X}_{\rho}^\mu$, with fermions in the internal lines, however, are non-ambiguous because there are no $\gamma_5$'s involved in the vertices: all heavy fermions have to be vector-like\footnote{Triangle diagrams with only light particles are ambiguous, but they do not concern us during matching. The corresponding ambiguities are fixed by the anomaly cancellation mechanism that we assume any EFT has built in.}.

Next, let's look at the box diagrams contributing to $H^\dagger H X_{\mu\nu}\tilde{X}^{\mu\nu}$. There are four internal fermionic propagators, of which one or more can correspond to heavy particles in the UV model. As mentioned, the corresponding traces have ambiguous and non-ambiguous parts. The unambiguous part can be computed in any $\gamma_5$-scheme. The ambiguous part is proportional to $D-4$ and is only of interest if it multiplies one of the singularities of the integral. Moreover, since the UV theory is renormalizable, the maximum number of $\gamma$-matrices is 6, which means that terms in the numerator of such integrals with mass insertions have a lower number of $\gamma$'s and are therefore unambiguous. There are two types of singularities: UV singularities corresponding to the UV structure of the full theory, and IR singularities that appear after the hard region expansion is performed. The latter correspond to the UV singularities of the SMEFT.

The generic structure of the potentially ambiguous part of such box integrals is, up to couplings and group theory factors:
\begin{equation}
    I^{\mu\nu} \sim \int_k T_{\mu_1\ldots\mu_4 \mu \nu} \prod_{i=1}^4 \frac{(k+q_i)^{\mu_i}}{(k+q_i)^2-m_i^2},
\end{equation}
where $\int_k \equiv \int \frac{d^Dk}{(2\pi)^D}$, and  $T_{\mu_1\ldots\mu_4 \mu \nu}$ denotes a trace with an odd number of $\gamma_5$ insertions and six $\gamma$'s, e.g. 
\begin{equation}
    \mathrm{Tr}[\gamma_{\mu_1}\gamma_5 \gamma_{\mu_2}\gamma_\mu \gamma_{\mu_3}\gamma_5 \gamma_\nu \gamma_{\mu_4}\gamma_5].
\end{equation} 

Performing the hard region expansion and the  tensor reduction we get 
\begin{align}
    I^{\mu\nu} &\sim \int_k T_{\mu_1\ldots\mu_4 \mu \nu}  \frac{1}{\prod_{i=1}^4 (k^2-m_i^2)} \nonumber
    \\ 
    & \left(k^4 \frac{g^{\mu_1\mu_2\mu_3\mu_4}}{D(D+2)} (1 + \sum_i\frac{q_i^2}{k^2-m_i^2}) + k^6 \frac{g^{\mu_1\mu_2\mu_3\mu_4\rho\sigma}}{D(D+2)(D+4)} \sum_{i\leq j}\frac{4 q_i^\rho q_j^\sigma}{(k^2-m_i^2)(k^2-m_j^2)} \right. \nonumber\\
    &  \left.-2k^4 \sum_{i,m}q_i^{\mu_i}q_m^\rho \frac{g^{{\{\hat{\mu}_i\}\rho}}}{D(D+2)}\frac{1}{k^2-m_m^2}
    + \sum_{i < j} k^2 \frac{g^{\{\hat{\mu}_i,\hat{\mu}_j\}}}{D}q^{\mu_i}q^{\mu_j}
     \right)\nonumber\\
     &\sim
     \int_k  T_{\mu_1\ldots\mu_4 \mu \nu} \frac{g^{\mu_1\mu_2\mu_3\mu_4}}{D(D+2)} \frac{k^4}{\prod_{i=1}^4(k^2-m_i^2)} \nonumber
    +  \int_k  T_{\mu_1\ldots\mu_4 \mu \nu} \frac{k^2}{\prod_{i=1}^4 (k^2-m_i^2)} \frac{1}{D(D+2)}
    \nonumber\\
    & \left( g^{\mu_1\mu_2\mu_3\mu_4} 
 \sum_i q_i^2 \left(1+\frac{m_i^2}{k^2-m_i^2}\right) +  \frac{g^{\mu_1\mu_2\mu_3\mu_4\rho\sigma}}{(D+4)} \sum_{i\leq j}
    4 q_i^\rho q_j^\sigma\prod_{l=i,j}\left(1+\frac{m_l^2}{(k^2-m_l^2)}\right) \right. 
    \nonumber\\
    &  \left. -2\sum_{i,m}q_i^{\mu_i}q_m^\rho g^{\{\hat{\mu}_i\}\rho}(1+\frac{m_m^2}{k^2-m_m^2})
    + \sum_{i < j}  (D+2)g^{\{\hat{\mu}_i,\hat{\mu}_j\}}q^{\mu_i}q^{\mu_j}
     \right),
\end{align}
where $\{\hat{\mu}_i\} = \mu_j\mu_k\mu_r$ with $j,k,r \neq i$, and $\{\hat{\mu}_i \hat{\mu}_j \} = \mu_k\mu_r$ with $k,r\neq i,j$.

The first term, of $\mathcal{O}{(p^0)}$, is UV singular. It does not lead to ambiguities, though, since $g^{\mu_1\mu_2\mu_3\mu_4}$ makes the $\gamma_5$-dependent part of the trace vanish. The rest of the terms, of order $\mathcal{O}(p^2)$, are UV finite, but can be IR singular, depending on the number of heavy propagators: for two, three or four heavy propagators, there is no singularity. For one heavy and three light propagators, however, we do have an IR singularity. 

We, therefore, conclude that ambiguous contributions from $\gamma_5$-odd traces in box integrals are present only in diagrams with one heavy and three light propagators. They appear as a product of the ambiguous $(D-4)$ coefficient of the trace multiplying a $1/ \epsilon$ pole of IR type resulting from the hard region expansion. 
We would like to remain within the naive anti-commuting $\gamma_5$ scheme when computing traces in \mme. To this end, we wish to fix the ambiguous contributions of the $\gamma_5$-odd traces in box integrals with one heavy massive fermion, a posteriori. We note that the Wilson coefficient we are trying to match must be real: all operators of the type $H^\dagger H X_{\mu\nu}\tilde{X}^{\mu\nu}$ are hermitian. Therefore any contribution from $\gamma_5$-odd traces, being imaginary, must be multiplied by a purely imaginary product of couplings of the full theory. But, in the case that ambiguities are present, i.e. when the UV theory has Yukawa terms with the Higgs boson, one heavy and one light fermion, leading to box diagrams with a single fermionic propagators, the corresponding Yukawa couplings are complex conjugates of each other, by virtue of the hermiticity of the UV Lagrangian. As a result, the product of all couplings is real and, therefore,  the $\gamma_5$-odd contributions are purely imaginary. They can be set to zero by hand, at the end of the computation. Note that this does not imply that the ambiguous contributions are zero: it is the sum of ambiguous and non-ambiguous traces that are set to zero. 



Note that this procedure also works if the effective theory is not the SMEFT. In such cases, more than one scalar fields might be present, allowing for non-hermitian operators of the type $\phi_1^\dagger \phi_2 X_{\mu\nu} \tilde{X}^{\mu\nu}$.
The constraint that the corresponding wilson coefficient is real does not apply any more. It is, however still true that the wilson coefficient of this operator should equal the complex conjugate of the wilson coefficient of the hermitian conjugate operator $\phi_2^\dagger \phi_1 X_{\mu\nu} \tilde{X}^{\mu\nu}$. This is, again, enough to remove the ambiguities.

\section{A minimal complete example\label{appendix:example}}

In this section we demonstrate many of the features of {\tt matchmakereft} with a concrete example involving two scalar fields, a light, but not massless, field $\phi$ and a heavy field $\Phi$.
Our model is described by the Lagrangian: 
\begin{equation}
    \mathcal{L} = \frac{1}{2}(\partial_\mu \phi)^2 -\frac{1}{2}m_L^2 \phi^2 + \frac{1}{2} (\partial_\mu \Phi)^2 - \frac{1}{2} M_H^2 \Phi^2 - \frac{\lambda_0}{4!}\phi^4 - \frac{\lambda_2}{4}\phi^2 \Phi^2 - \frac{\kappa}{2}\phi^2 \Phi,
    \label{lagrangian:uv}
\end{equation}
which we want to match to the EFT Lagrangian without the heavy scalar, 
\begin{equation}
    \mathcal{L}_{\rm EFT} = \frac{\alpha_{4k}}{2}(\partial_\mu \phi)^2 
    -\frac{\alpha_2}{2} \phi^2
    - \frac{\alpha_4}{4!}\phi^4 
    - \frac{\alpha_6}{6!}\phi^6 
    - \frac{\tilde{\alpha}_6}{4!}\phi^3 \partial^2\phi
    -\frac{\hat{\alpha}_6}{2}\left(\partial^2 \phi \right)^2.
\end{equation}
We will use this Lagrangian during off-shell matching. Subsequently, the kinetic term can be canonically normalized, and the redundant operators can be eliminated. Two of the three operators of dimension 6 are redundant. We choose $\phi^6$ as the independent operator. Using equations of motion we can readily find that:
\begin{align}
    \phi^3 \partial^2\phi &\to -\alpha_2\phi^4 - \frac{1}{3!}\alpha_4\phi^6, \\
    \left(\partial^2 \phi \right)^2 &\to  \alpha_2^2 \phi^2  + \frac{\alpha_2\alpha_4}{3}\phi^4 + \frac{\alpha_4}{36}\phi^6.
\end{align}
Eliminating these operators from the Lagrangian would induce the shifts
\begin{eqnarray}
\alpha_2 & \to & \alpha_2 + \alpha_2^2 \hat{\alpha}_6\\
\alpha_4 & \to & \alpha_4 - \tilde{\alpha}_6 \alpha_2 +4 \alpha_2\alpha_4\hat{\alpha}_6 \\
\alpha_6 & \to & \alpha_6 - 5\tilde{\alpha}_6\alpha_4 +10 \hat{\alpha}_6\alpha_4^2
\label{reductiontogreeneqs}
\end{eqnarray}
The coupling $\kappa$ of this model is a dimensionful coupling, and is expected to be parametrically of the order of the heavy mass scale $M_H$. Thus,  $\frac{\kappa}{M_H}$ is of $\mathcal{O}(1)$ and is kept throughout the matching procedure consistently. 

The Feynrules file for the UV model, saved at {\tt two\_scalars.fr}, is shown below.
\begin{lstlisting}[language=Mathematica]
(* --- Contents of Feynrules file two_scalars.fr --- *)
M$ModelName = "two_scalars";
(* **** Particle classes **** *)
M$ClassesDescription = {
S[1] == {ClassName -> phiH, SelfConjugate -> True, Mass -> MH,
         FullName -> "heavy"},
S[2] == {ClassName -> phi, SelfConjugate -> True, Mass -> mL,
         FullName -> "light"}
};
(* *****   Parameters   ***** *)
M$Parameters = {
MH == {ParameterType -> Internal, ComplexParameter -> False},
mL == {ParameterType -> Internal, ComplexParameter -> False},
V == {ParameterType -> Internal, ComplexParameter -> False},
lambda0 == {ParameterType -> Internal, ComplexParameter -> False},
kappa == {ParameterType -> Internal, ComplexParameter -> False},
lambda2 == {ParameterType -> Internal, ComplexParameter -> False}
};
(* *****   Lagrangian   ***** *)
Ltot := Block[{mu}, 
  + 1/2 * del[phi,mu] *  del[phi,mu]  + 1/2 *del[phiH,mu] * del[phiH,mu]
  - 1/2 * MH^2 * phiH^2 - 1/2 * mL^2 * phi^2 
  - lambda0 / 24 * phi^4 - kappa / 2  * phi^2 * phiH 
  - lambda2 / 4 * phi^2 * phiH^2
  ];
\end{lstlisting}

 Note that we use the keyword {\tt FullName} to characterize each field as {\tt "heavy"} or {\tt "light"}. This is mandatory: {\tt matchmaker} uses this keyword to distinguish between fields that are integrated out and those that are light and are also present in the EFT. Also note that all the parameters that are used in the Lagrangian, masses as well as couplings, must be declared. In this example all parameters are real. 

The Feynrules file for the EFT model, saved at {\tt one\_scalar.fr}, is: 
\begin{lstlisting}[language=Mathematica]
(* --- Contents of Feynrules file for the EFT model  one_scalar.fr --- *)
M$ModelName = "one_scalar";
(* **** Particle classes **** *)
M$ClassesDescription = {
S[2] == {ClassName -> phi, SelfConjugate -> True, Mass -> 0,
         FullName -> "light"}
};
(* *****   Parameters   ***** *)
M$Parameters = {
alpha4kin == {ParameterType -> Internal, ComplexParameter -> False},
alpha2mass == {ParameterType -> Internal, ComplexParameter -> False},
alpha4 == {ParameterType -> Internal, ComplexParameter -> False},
alpha6 == {ParameterType -> Internal, ComplexParameter -> False},
alpha6Rtilde == {ParameterType -> Internal, ComplexParameter -> False},
alpha6Rhat == {ParameterType -> Internal, ComplexParameter -> False}
};
(* *****   Lagrangian   ***** *)
Ltot := Block[{mu,mu2},
  1/2 * alpha4kin * del[phi,mu] * del[phi,mu] 
  - 1/2 * alpha2mass * phi^2
  - alpha4/24 * phi^4 
  - alpha6 * phi^6/720 
  - alpha6Rtilde/24 * phi^3 * del[del[phi,mu],mu] 
  - alpha6Rhat/2 * del[del[phi,mu],mu]  *  del[del[phi,mu2],mu2] 
  ];
\end{lstlisting}

Note that we have included WCs (denoted by {\tt alpha}) also for the kinetic and mass terms (squared), as well as for all operators that are redundant solely due to the equations of motion. 

In order for {\tt matchmakereft} to perform the reduction to the physical basis, we need to provide a set of relations that express the redundant WCs in terms of the irreducible ones, see Eq.~(\ref{reductiontogreeneqs}).  This is done at {\tt one\_scalar.red}: 
\begin{lstlisting}[language=Mathematica]
(* --- Contents of one_scalar.red --- *)
finalruleordered = {
	alpha6 -> - alpha6Rtilde * alpha4 *5 + alpha6Rhat * alpha4^2 * 10 + alpha6 ,
	alpha4 -> alpha4 - alpha6Rtilde * alpha2mass + 4 * alpha6Rhat * alpha2mass * alpha4 ,
	alpha4kin -> alpha4kin ,
	alpha2mass -> alpha2mass +  alpha6Rhat * alpha2mass^2
	}
\end{lstlisting}

Note that {\bf only} the  WCs corresponding to physical operators, among those appearing in the EFT Lagrangian and defined in the file \verb+one_scalar.fr+, must be present on the left hand side of the replacement rules in this file. The WCs corresponding to redundant operators appear only on the right hand side. When these rules are used, both redundant and non-redundant WCs have been matched and are known as functions of the parameters of the UV theory. The rules are therefore instructions on how to update the non-redundant WCs, to include the effect of the redundant ones.

With these files prepared we are ready to proceed with matching. In the matching directory, where {\tt two\_scalars.fr},{\tt one\_scalar.fr},{\tt one\_scalar.red} are present, we can run {\tt matchmakereft}: 
\begin{lstlisting}[numbers=none,backgroundcolor=\color{codegris}]
>matchmakereft
\end{lstlisting}
upon which we enter the python interface 
\begin{lstlisting}[numbers=none,backgroundcolor=\color{codegris}]
Welcome to matchmakereft
Please refer to arXiv:2112.xxxxx when using this code 

matchmakereft> 
\end{lstlisting}
We first need to create the matchmaker models, i.e. the directories with all the necessary information for the UV and the EFT models. We do this by 
\begin{lstlisting}[numbers=none,backgroundcolor=\color{codegris}]
matchmakereft> create_model two_scalars.fr
\end{lstlisting}
which has the response
\begin{lstlisting}[numbers=none,backgroundcolor=\color{codegris}]
Creating model two_scalars_MM. This might take some time depending on the complexity of the model
Model two_scalars_MM created
It took 7 seconds to create it.
\end{lstlisting}
We can now observe that the directory {\tt two\_scalars\_MM} is created. We proceed with creating the EFT model
\begin{lstlisting}[numbers=none,backgroundcolor=\color{codegris}]
matchmakereft> create_model one_scalar.fr
Creating model one_scalar_MM. This might take some time depending on the complexity of the model
Model one_scalar_MM created
It took 7 seconds to create it.
\end{lstlisting}
The {\tt one\_scalar\_MM} directory is now created as well, and we are ready for the matching calculation. This is performed by the {\tt match\_model\_to\_eft} command:
\begin{lstlisting}[numbers=none,backgroundcolor=\color{codegris}]
matchmakereft> match_model_to_eft two_scalars_MM one_scalar_MM
\end{lstlisting}
Upon completion, the results of the matching are stored in the UV model directory, {\tt two\_scalars\_MM} in our case. The file {\tt two\_scalars\_MM/MatchingProblems.dat} contains troubleshooting information in case the matching procedure failed. In our case it is an empty list, indicating no problems:
\begin{lstlisting}
problist = {}
\end{lstlisting}

The result of the matching procedure is in {\tt two\_scalars\_MM/MatchingResults.dat}, a Mathematica file with a list of lists of replacement rules. 
The off-shell matching gives the following results for the WCs of the Green basis. At tree level the non-vanishing contributions are,
\begin{eqnarray}
\alpha_2^{(0)} &=& m_L^2, \\
\alpha_4^{(0)} &=&\lambda _0 -\frac{4 \kappa^2 m_L^2}{M_H^4}-\frac{3 \kappa^2}{M_H^2},\\
\alpha_6^{(0)} &=& \frac{1}{M_H^2} \left(
   \frac{60 \kappa^4}{M_H^4}-\frac{20 \lambda _0 \kappa^2}{M_H^2}+\frac{45 \lambda _2 \kappa^2}{M_H^2}
   \right), \\
\tilde{\alpha}_6^{(0)} &=& \frac{4 \kappa^2}{M_H^4}.
\end{eqnarray}

At one loop we get (we define $L_M \equiv \log(\frac{\mu^2}{M_H^2})$)
\begin{align}
\alpha_2^{(1)} &=
-\frac{1}{16\pi^2}
   \left(1+ L_M \right) \left(\frac{\kappa^2 m_L^4}{M_H^4}+\frac{\kappa^2
   m_L^2}{M_H^2}+\frac{1}{2} \lambda _2 M_H^2+\kappa^2\right),\\
\alpha_{4k}^{(1)} &= 1 + \frac{1}{16\pi^2} \left(\frac{5 \kappa^2 m_L^2}{2
   M_H^4}+\frac{\kappa^2}{2
   M_H^2}\right)+\frac{1}{16\pi^2}\frac{ \kappa^2 m_L^2
     }{M_H^4}L_M,\\
    \alpha_4^{(1)} &= 
    \frac{1}{16\pi^2}
    \left(
    \frac{48 \kappa^4 m_L^2}{ M_H^6}
    +\frac{24 \lambda _2 \kappa^2 m_L^2}{ M_H^4}
    -\frac{12 \lambda _0 \kappa^2 m_L^2}{ M_H^4}
    +\frac{18 \kappa^4}{ M_H^4}
    +\frac{18 \lambda _2 \kappa^2}{ M_H^2}
    -\frac{6 \lambda _0 \kappa^2}{ M_H^2}
    \right)
    \\
    & 
    +
    \frac{1}{16\pi^2} \left(\frac{36 \kappa^4 m_L^2}{ M_H^6}+\frac{18 \lambda _2 \kappa^2 m_L^2}{ M_H^4}
    -\frac{12 \lambda _0 \kappa^2 m_L^2}{ M_H^4}
    +\frac{12 \kappa^4}{ M_H^4}
    +\frac{12 \lambda _2 \kappa^2}{ M_H^2}
    -\frac{6 \lambda _0 \kappa^2}{ M_H^2}
    -\frac{3 \lambda _2^2}{2}\right)L_M,
\\
\alpha_6^{(1)} &= 
\frac{1}{16\pi^2M_H^2}\left(
-\frac{1290 \kappa^6}{M_H^6}+\frac{720 \lambda _0 \kappa^4}{M_H^4}-\frac{1665 \lambda _2 \kappa^4}{M_H^4}+\frac{360 \lambda _0 \lambda _2 \kappa^2}{M_H^2}-\frac{90 \lambda _0^2 \kappa^2}{M_H^2}-\frac{495 \lambda _2^2 \kappa^2}{M_H^2}+\frac{15 \lambda _2^3}{2}
\right)
\\
&
 + \frac{1}{16\pi^2 M_H^2} \left(-\frac{810 \kappa^6}{M_H^6}+\frac{540 \lambda _0 \kappa^4}{M_H^4}-\frac{945 \lambda _2 \kappa^4}{M_H^4}+\frac{270 \lambda _0 \lambda _2 \kappa^2}{M_H^2}-\frac{90 \lambda _0^2 \kappa^2}{M_H^2}-\frac{270 \lambda _2^2 \kappa^2}{M_H^2}\right)L_M,
\\
\tilde{\alpha}^{(1)}_6 &=
\frac{1}{16\pi^2 M_H^2}\left(
-\frac{107 \kappa^4}{3 M_H^4}+\frac{9 \lambda _0 \kappa^2}{M_H^2}-\frac{77 \lambda _2 \kappa^2}{3 M_H^2}+\frac{\lambda _2^2}{3}
\right)
\nonumber \\
&+\frac{1}{16\pi^2M_H^2} 
\left(
-\frac{14 \kappa^4}{M_H^4}+\frac{2 \lambda _0 \kappa^2}{M_H^2}-\frac{16 \lambda _2 \kappa^2}{M_H^2}\right)L_M,
\\    
\hat{\alpha}_6^{(1)} &= -\frac{\kappa^2}{96 \pi ^2 M_H^4}.
\end{align}

As can be seen in the equations above, the kinetic operator receives a correction and therefore $\phi$ is no longer canonically normalized. A field redefinition is needed to obtain a canonically normalized theory on which we can apply the corresponding redundancies to go to the physical basis. \Mme does these two processes (canonical normalization and going to the physical basis) automatically. The resulting WCs in the physical basis read, up to one loop order
\begin{align}
\alpha_2 &= m_L^2 -\frac{1}{16\pi^2} \left(\frac{11 \kappa^2 m_L^4}{3
   M_H^4}+\frac{3 \kappa^2 m_L^2}{2 M_H^2}+\frac{1}{2}
   \lambda _2 M_H^2+\kappa^2\right)
   \nonumber \\
   &\phantom{=m_L^2}-\frac{1}{16\pi^2}
    \left(\frac{2 \kappa^2
   m_L^4}{M_H^4}+\frac{\kappa^2 m_L^2}{M_H^2}+\frac{1}{2}
   \lambda _2 M_H^2+\kappa^2\right)L_M,
   \label{matching:a2}\\
\alpha_4 &= \lambda _0 -\frac{4 \kappa^2 m_L^2}{M_H^4}-\frac{3 \kappa^2}{M_H^2} 
   \\
   & + 
\frac{1}{16\pi^2} \left(\frac{332 \kappa^4 m_L^2}{3
   M_H^6}-\frac{80 \lambda _0 \kappa^2 m_L^2}{3
   M_H^4}+\frac{149 \lambda _2 \kappa^2 m_L^2}{3
   M_H^4}-\frac{\lambda _2^2 m_L^2}{3 M_H^2}+\frac{25 \kappa^4}{M_H^4}-\frac{7 \lambda _0 \kappa^2}{M_H^2}+\frac{20
   \lambda _2 \kappa^2}{M_H^2}\right)
    \nonumber\\
   & + 
   \frac{1}{16\pi^2}
   \left(\frac{60 \kappa^4 m_L^2}{M_H^6}-\frac{16
   \lambda _0 \kappa^2 m_L^2}{M_H^4}+\frac{34 \lambda _2
   \kappa^2 m_L^2}{M_H^4}+\frac{16 \kappa^4}{M_H^4}-\frac{6 \lambda _0 \kappa^2}{M_H^2}+\frac{14
   \lambda _2 \kappa^2}{M_H^2}-\frac{3 \lambda
   _2^2}{2}\right)L_M, \nonumber
   \label{matching:alpha4}
   \\
\alpha_6 &=  \frac{1}{M_H^2}\left( \frac{60 \kappa^4}{M_H^4}-\frac{20
   \lambda _0 \kappa^2}{M_H^2}+\frac{45 \lambda _2 \kappa^2}{M_H^2}\right)
   \nonumber\\
   & + \frac{1}{16\pi^2 M_H^2} \left(-\frac{1560 \kappa^6
   m_L^2}{M_H^{8}}+\frac{440 \lambda _0 \kappa^4
   m_L^2}{M_H^6}-\frac{1635 \lambda _2 \kappa^4 m_L^2}{2
   M_H^6}-\frac{2320 \kappa^6}{M_H^6}+\frac{3610 \lambda _0
   \kappa^4}{3 M_H^4}
   \right.
   \nonumber\\
    &
   \left. \phantom{\frac{1}{16\pi^2 M_H^2}}
   -\frac{4955 \lambda _2 \kappa^4}{2
   M_H^4}-\frac{410 \lambda _0^2 \kappa^2}{3
   M_H^2}-\frac{490 \lambda _2^2 \kappa^2}{M_H^2}+\frac{1465
   \lambda _0 \lambda _2 \kappa^2}{3 M_H^2}+\frac{15 \lambda
   _2^3}{2 }-\frac{5 \lambda _0 \lambda _2^2}{3
   }\right)
   \nonumber\\
    &
   +\frac{1}{16\pi^2 M_H^2}  \left(-\frac{960
   \kappa^6 m_L^2}{M_H^{8}}+\frac{320 \lambda _0 \kappa^4 m_L^2}{M_H^6}-\frac{495 \lambda _2 \kappa^4
   m_L^2}{M_H^6}-\frac{1260 \kappa^6}{M_H^6}+\frac{760
   \lambda _0 \kappa^4}{M_H^4}
   \right.
   \nonumber\\
    &
   \left.\phantom{\frac{1}{16\pi^2 M_H^2}}
   -\frac{1425 \lambda _2 \kappa^4}{M_H^4}-\frac{100 \lambda _0^2 \kappa^2}{M_H^2}-\frac{240 \lambda _2^2 \kappa^2}{M_H^2}+\frac{350 \lambda _0 \lambda _2 \kappa^2}{M_H^2}\right)L_M.
\end{align}

Next, we would like to compute the RGE equations for both models. For each model we need a new pair of Feynrule files as a starting point. We start for the RGEs for the UV model. The corresponding UV model, located in the file {\tt rge\_two\_scalars\_uv.fr} is the same as {\tt two\_scalars.fr} with the very crucial difference that the heavy field H is also declared as "light" here.  
\begin{lstlisting}[language=Mathematica]
S[1] == {ClassName-> H, SelfConjugate->True, Mass->MH,
    FullName->"light"}
\end{lstlisting}
 The target model is now at {\tt rge\_two\_scalars\_eft.fr}. It consists of an EFT Lagrangian with the field $H$ considered as a light field: 
 \begin{lstlisting}[language=Mathematica]
 (* ---  Contents of rge_two_scalars_eft.fr  --- *)
M$ModelName = "rge_two_scalars_eft";
(* **** Particle classes **** *)
M$ClassesDescription = {
S[1] == {ClassName ->  phiH, SelfConjugate -> True, Mass -> 0,
         FullName -> "light"},
S[2] == {ClassName -> phi,SelfConjugate -> True,Mass -> 0,
         FullName -> "light"}
};
(* *****   Parameters   ***** *)
M$Parameters = {
alpha4kinphi == {ParameterType -> Internal, ComplexParameter -> False},
alpha4kinH == {ParameterType -> Internal, ComplexParameter -> False},
alpha2MH == {ParameterType -> Internal, ComplexParameter -> False},
alpha2ML == {ParameterType -> Internal, ComplexParameter -> False},
alpha4 == {ParameterType -> Internal, ComplexParameter -> False},
alphaV == {ParameterType -> Internal, ComplexParameter -> False}, 
alpha1 == {ParameterType -> Internal, ComplexParameter -> False},
alpha4H == {ParameterType -> Internal, ComplexParameter -> False},
alpha2 == {ParameterType -> Internal, ComplexParameter -> False},
alpha3 == {ParameterType -> Internal, ComplexParameter -> False}
};
(* *****   Lagrangian   ***** *)
Ltot := Block[{mu},
  alphaV *  phiH
  +1/2 * alpha4kinphi* del[phi,mu] *  del[phi,mu]  
  + 1/2 * alpha4kinH * del[phiH,mu] * del[phiH,mu] 
  - 1/2 * alpha2MH * phiH^2 
  -1/2 * alpha2ML * phi^2 
  - alpha4 / 24 * phi^4 
  -  alpha1 / 2  * phi^2 * phiH 
  - alpha2 / 4 * phi^2 * phiH^2
  -alpha3 * phiH^3
  -alpha4H * phiH^4
];
\end{lstlisting}
Note the presence of the new interaction terms $\Phi^3, \Phi^4$: all possible operators of dimension up to 4, compatible with the symmetries should appear here. Also note that we now have a WC also for the kinetic term of the $\Phi$ field. 

We don't need a {\tt rge\_two\_scalars\_eft.red} file at all, since all operators of dimension four present are physical. We can now create the two models with 
\begin{lstlisting}[numbers=none,backgroundcolor=\color{codegris}]
matchmakereft> create_model rge_two_scalars_uv.fr
\end{lstlisting}
and
\begin{lstlisting}[numbers=none,backgroundcolor=\color{codegris}]
matchmakereft> create_model rge_two_scalars_eft.fr
\end{lstlisting}
and then we can proceed with the RGE computation via 
\begin{lstlisting}[numbers=none,backgroundcolor=\color{codegris}]
matchmakereft> compute_rge_model_to_eft  rge_two_scalars_uv_MM rge_two_scalars_eft_MM
\end{lstlisting}
We then get, as before, an empty {\tt MatchingProblems.dat} file, as well as a {\tt MatchingResults.dat} file in the {\tt rge\_two\_scalars\_uv\_MM} directory. Moreover, there is another file produced, {\tt RGEResult.dat} which contains the explicit form of the beta functions for our UV model. 
\begin{lstlisting}[language=Mathematica]
RGEResult = {
\[Beta][alphaV] -> -1/32*(kappa*mL^2)/Pi^2
\[Beta][alpha4kinphi] -> 0, 
\[Beta][alpha4kinH] -> 0, 
\[Beta][alpha2MH] -> kappa^2/(16*Pi^2)+(lambda2*mL^2)/(16*Pi^2), 
\[Beta][alpha2ML] -> kappa^2/(8*Pi^2)+(lambda2*MH^2)/(16*Pi^2) + 
       (lambda0*mL^2)/(16*Pi^2), 
\[Beta][alpha4] -> (3*lambda0^2)/(16*Pi^2) + 
       (3*lambda2^2)/(16*Pi^2), 
\[Beta][alpha1] ->(lambda0*kappa)/(16*Pi^2) + (kappa*lambda2)/(4*Pi^2), 
\[Beta][alpha2] -> (lambda0*lambda2)/(16*Pi^2) + lambda2^2/(4*Pi^2), 
\[Beta][alpha3] -> (kappa*lambda2)/(32*Pi^2), 
\[Beta][alpha4H] -> lambda2^2/(128*Pi^2)}
\end{lstlisting}

Let us now touch upon the topic of tadpole contributions, which will also explain the reason for the $\alpha_V \Phi$ operator defined in line 25 of {\tt rge\_two\_scalars\_eft.fr}. In our model there are tadpole contributions to many 1LPI diagrams, due to the operator $\kappa \phi^2 \Phi$. They would vanish, up to one loop, if the mass of the light field was set to zero, but are non-vanishing otherwise. The resulting finite contributions are taken into account by our matching procedure. However, the corresponding poles, which contribute to the beta functions of $m_L$ and $\kappa$ are disregarded, during the RGE computation. However, there is an easy way to account for them. By adding the $\alpha_V \Phi$ term in the Lagrangian of {\tt rge\_two\_scalars\_eft.fr}, we induce the computation of the one-point function beta-function: as we see from line 2 of {\tt RGEResult.dat}, we have 
\begin{equation}
\beta(\alpha_V) =  -\frac{1}{32\pi^2} \kappa m_L^2.
\end{equation}  
If, instead of working with tadpole contributions, we would have shifted the field $\Phi \to \Phi + V$ we would have induced an explicit  linear term $ V M_H^2 \Phi$ and, due to the operators $\phi^2 \Phi$ and $\phi^2 \Phi^2$, we would modify the mass term for the light scalar, $m_L$ and the $\kappa$ coupling: 
\begin{equation}
\tilde{m}_L^2 = m_L^2 - \kappa V + \frac{1}{2}\kappa V^2\;\;\;\; , \;\;\;\; \tilde{\kappa} = \kappa-\lambda_2 V.
\end{equation}
By setting the tree-level contribution of $V$ such that it cancels loop corrections order by order, we could eliminate tadpoles from the theory completely. Instead, here, we include their finite part in the calculation, and we absorb the pole in the renormalization of $m_L^2$ and $\kappa$. We should therefore modify the beta functions we read from {\tt RGEResult.dat}, to account for the tadpole pole, by 
\begin{eqnarray}
\delta \beta(M_L^2) &=& \kappa \beta(V) = \frac{\kappa}{M_H^2} \beta(\alpha_V) = - \frac{\kappa^2}{16\pi^2}\frac{m_L^2}{M_H^2},
\\
\delta \beta(\kappa) &=& \lambda_2 \beta(V) = \frac{\lambda_2}{M_H^2} \beta(\alpha_V) = - \frac{\kappa \lambda_2}{16\pi^2}\frac{m_L^2}{M_H^2}.
\end{eqnarray}
Reading the results of {\tt RGEResult.dat} and adding these contributions we get
\begin{align}
\beta(m_L^2) &= \frac{\lambda _2 M_H^2}{16 \pi ^2}+\frac{\kappa^2}{8 \pi
   ^2}+\frac{\lambda _0 m_L^2}{16 \pi ^2} - \frac{\kappa^2}{16\pi^2}\frac{m_L^2}{M_H^2},
   \label{rge:mL}
\\
\beta(m_H^2) &= \frac{\kappa^2}{16 \pi ^2}+\frac{\lambda _2 m_L^2}{16 \pi
   ^2}, \\
\beta(\lambda_0) &= \frac{3 \lambda _0^2}{16 \pi ^2}+\frac{3 \lambda _2^2}{16 \pi
   ^2},
\\
\beta(\kappa) &= \frac{\lambda _0 \kappa}{16 \pi ^2}+\frac{\lambda _2 \kappa}{4 \pi ^2} - \frac{\kappa \lambda_2}{16\pi^2}\frac{m_L^2}{M_H^2},
\\
\beta(\lambda_2) &= \frac{\lambda _2^2}{4 \pi ^2}+\frac{\lambda _0 \lambda _2}{16
   \pi ^2}.
\end{align}

We can perform the same procedure, but now with the EFT model. We create the {\tt rge\_one\_scalar\_uv.fr }
\begin{lstlisting}[language=Mathematica]
M$ModelName = "rge_one_scalar_uv";
(* **** Particle classes **** *)
M$ClassesDescription = {
S[2] == {ClassName -> phi,SelfConjugate -> True,Mass -> mL,
         FullName -> "light"}
};
(* *****   Parameters   ***** *)
M$Parameters = {
mL == {ParameterType -> Internal, ComplexParameter -> False},
a4 == {ParameterType -> Internal, ComplexParameter -> False},
a6 == {ParameterType -> Internal, ComplexParameter -> False}
};
(* *****   Lagrangian   ***** *)
Ltot := Block[{mu,mu2},
  1/2  * del[phi,mu] * del[phi,mu] - 1/2 * mL^2 * phi^2
  -a4/24 phi^4 - a6 * phi^6/720 
  ];
  \end{lstlisting}
which contains the physical operators of the EFT model. Note that we have changed the names of the couplings to something other than {\tt alphaXXX}\footnote{If {\tt alphaXXX} names are used then \mme automatically changes them to {\tt WCXXX}.}. We also create {\tt rge\_one\_scalar\_eft.fr} and {\tt rge\_one\_scalar\_eft.red} that are identical with {\tt one\_scalar.fr} and {\tt one\_scalar.red} defined above.  
Once again, we can create the two models and run {\tt compute\_rge\_model\_to\_eft}. The result is 
\begin{align}
\beta(\alpha_2) &= \frac{\alpha _2 \alpha _4}{16 \pi ^2},
\label{rge:alpha2}
\\
\beta(\alpha_4) &= \frac{3 \alpha _4^2}{16 \pi ^2}+\frac{\alpha _2 \alpha _6}{16
   \pi ^2}, \\
\beta(\alpha_6) &= \frac{15 \alpha _4 \alpha _6}{16 \pi ^2}.
\end{align}

We can now check whether the matching conditions and the RGE equations are consistent with each other: if we match at $\mu=M_H$ and we evolve all couplings using the RGEs of the UV and the EFT model, to a lower scale $Q$, we should find the same expressions as when we match directly at $\mu=Q$. Let's see how this works in the case of the mass coefficient, $\alpha_2$. The matching condition for $\alpha_2$, at scale $\mu = M_H$, gives, see Eq.~(\ref{matching:a2}),
\begin{equation}
\alpha_2(M_H) = m_L^2(M_H) -\frac{1}{16\pi^2} \left(\frac{11 \kappa^2 m_L^4}{3
   M_H^4}+\frac{3 \kappa^2 m_L^2}{2 M_H^2}+\frac{1}{2}
   \lambda _2 M_H^2+\kappa^2\right).
  \end{equation}
We can use Eq.~(\ref{rge:alpha2}) to evolve $\alpha_2$ from $M_H$ to a lower scale $Q$. In the leading-log approximation it reads, 
\begin{equation}
\alpha_2(Q) = \alpha_2(M_H) + \frac{1}{2} L_Q \beta(\alpha_2) = \alpha_2(M_H) + \frac{1}{2} L_Q  \frac{\alpha _2 \alpha _4}{16 \pi ^2},
\end{equation}
where $L_Q \equiv \log(\frac{Q^2}{M_H^2})$. 
Replacing the tree-level matched values of $\alpha_2$ and $\alpha_4$ we get 
\begin{eqnarray}
\alpha_2(Q) &=& m_L^2(M_H)
-\frac{1}{16\pi^2} \left(\frac{11 \kappa^2 m_L^4}{3
   M_H^4}+\frac{3 \kappa^2 m_L^2}{2 M_H^2}+\frac{1}{2}
   \lambda _2 M_H^2+\kappa^2\right)
   \\
   & &
-\frac{L_Q}{16\pi^2} \left(
\frac{2\kappa^2 m_L^4}{ M_H^4}
+\frac{3 \kappa^2 m_L^2}{2 M_H^2}
-\frac{\lambda _0 m_L^2}{2}
\right).
\end{eqnarray}
If the matching condition for $\alpha_2$ had no tree-level contribution, this would be the full expression for $\alpha_2(Q)$. In our case, however, there is a tree-level contribution, $m_L^2(M_H)$. We need to use the RGE equation of the UV model, for $m_L$, Eq.~(\ref{rge:mL}) to evolve it to the scale $Q$. We then get 
\begin{eqnarray}
\alpha_2(Q) &=& m_L^2(Q)-\frac{1}{16\pi^2} \left(\frac{11 \kappa^2 m_L^4}{3
   M_H^4}+\frac{3 \kappa^2 m_L^2}{2 M_H^2}+\frac{1}{2}
   \lambda _2 M_H^2+\kappa^2\right) 
   \\
   & &
   -\frac{L_Q}{16\pi^2} \left(
   \frac{2\kappa^2 m_L^4}{ M_H^4}
   +\frac{3 \kappa^2 m_L^2}{2 M_H^2}
   +\frac{\lambda _2 M_H^2}{2}
   +2\kappa^2
\right).
\label{eq:a2evolved}
\end{eqnarray}
Had we match directly at the scale $\mu=Q$ using Eq.~(\ref{matching:a2}), we would have found exactly the same result.  

Similarly for $\alpha_4$, matching at $\mu=M_H$ and evolving to $\mu=Q$, with the help of the RGEs for both the UV and the EFT couplings, gives
\begin{eqnarray}
\alpha_4(Q) &=&\lambda _0 -\frac{4 \kappa^2 m_L^2}{M_H^4}-\frac{3 \kappa^2}{M_H^2} +
\frac{1}{16\pi^2}\left( \frac{332 \kappa^4 m_L^2}{3 M_H^6}+\frac{149 \lambda _2 \kappa^2 m_L^2}{3 M_H^4}-\frac{80 \lambda _0 \kappa^2 m_L^2}{3 M_H^4}
\right. \nonumber
\\
& &
\left.
-\frac{\lambda _2^2 m_L^2}{3 M_H^2}+\frac{25 \kappa^4}{M_H^4}+\frac{20 \lambda _2 \kappa^2}{M_H^2}-\frac{7 \lambda _0 \kappa^2}{M_H^2}
\right)
  +\frac{1}{16\pi^2} L_Q  \left( 
\frac{60 \kappa^4 m_L^2}{M_H^6}
+\frac{34 \lambda _2 \kappa^2 m_L^2}{M_H^4}
\right. \nonumber
\\
& &
\left.
-\frac{16 \lambda _0 \kappa^2 m_L^2}{M_H^4}+\frac{16 \kappa^4}{M_H^4}+\frac{14 \lambda _2 \kappa^2}{M_H^2}-\frac{6 \lambda _0 \kappa^2}{M_H^2}-\frac{3 \lambda _2^2}{2}
 \right),
\end{eqnarray}
in agreement with what we would get by matching directly at $\mu=Q$, see Eq.~(\ref{matching:alpha4}).


\section{SMEFT Green basis \label{green_basis}}
Here we present a Green basis of the dimension 6 SMEFT including two and four-fermion evanescent operators. As far as possible, we will follow the notation and the conventions used in the model file included with \texttt{matchmakereft}. We present first the renormalizable SM Lagrangian, which reads
\begin{eqnarray}
	\mathcal{L}_{\rm SM}&=& 
-\frac{1}{4} G^A_{\mu\nu} G^{A\,\mu\nu}
-\frac{1}{4} W^I_{\mu\nu} W^{I\,\mu\nu}
-\frac{1}{4} B_{\mu\nu} B^{\mu\nu}
+(D_\mu H)^\dagger D^\mu H-m^2 H^\dagger H
-\lambda (H^\dagger H)^2 \nonumber \\
&+& \mathrm{i} [ 
\bar{\ell}\cancel{D} \ell+\bar{e}\cancel{D}e
+\bar{q}\cancel{D} q+\bar{u}\cancel{D}u
+\bar{d}\cancel{D} d]
-[
\bar{\ell}Y_e e H
+\bar{q}Y_u u \tilde{H}
+\bar{q}Y_d d H
+\mathrm{h.c.}]~.
\end{eqnarray}
Hereinafter, we omit flavor and gauge indices whenever possible. Otherwise, we use $i,j,k,l,\ldots$ as flavor indices and $A,B,C,\ldots$ and $I,J,K,\ldots$ for the adjoint representation of $SU(3)$ and $SU(2)$, respectively. We will use on the other hand $a,b,c,\ldots$ and $r,s,t,\ldots$ for the fundamental representation of the color and the electroweak group, respectively. In the model file coming along with \texttt{matchmakereft} we use a slightly different notation for the couplings in the renormalizable Lagrangian. For instance,
\begin{equation}
Y_{u}\to \texttt{alphaOlambdau},\ \lambda\to \texttt{alphaOlambda}, \ m^2\to \texttt{alphaOmuH2},\ \ldots\,.
\end{equation}
We refer the reader to the model file \texttt{SMEFT\_Green\_Bpreserving.fr} for more details.

The doublet $\tilde{H}$ is defined by $\tilde{H}=\ii \sigma^2 H^{\ast}$ as usual and we assume the following definition for the covariant derivative
\begin{equation}
D_\mu q=(\partial_\mu 
- \mathrm{i} g_3 T^A G^A_\mu
- \mathrm{i} g_2 \frac{\sigma^I}{2} W^I_\mu
- \mathrm{i} g_1 Y B_\mu)q,
\end{equation}
where $T^A=\lambda^A/2$ with $\lambda^A$ and $\sigma^I$ the Gell-Mann and
Pauli matrices, respectively. Correspondingly, the field strength tensors are
\begin{eqnarray}
G^A_{\mu\nu}&=& \partial_\mu G^A_\nu-\partial_\nu G^A_\mu
+g_3 f^{ABC} G^B_\mu G^C_\nu, \\
W^I_{\mu\nu}&=& \partial_\mu W^I_\nu-\partial_\nu W^I_\mu
+g_2 \epsilon^{IJK} W^J_\mu W^K_\nu, \\
B_{\mu\nu}&=& \partial_\mu B_\nu-\partial_\nu B_\mu,
\end{eqnarray}
and their covariant derivatives
\begin{eqnarray}
(D_\rho G_{\mu\nu})^A&=& \partial_\rho G^A_{\mu\nu} + g_3 f^{ABC}
G^B_\rho G^C_{\mu\nu}, \\
(D_\rho W_{\mu\nu})^I&=& \partial_\rho W^I_{\mu\nu} + g_2 \epsilon^{IJK}
W^J_\rho W^K_{\mu\nu}, \\
(D_\rho B_{\mu\nu})&=& \partial_\rho B_{\mu\nu},
\end{eqnarray}
with $f^{ABC}$ and $\epsilon^{IJK}$ the $SU(3)$ and $SU(2)$ structure constants, respectively.  

The $B$-number preserving dimension 6 Green basis is presented in Tables~\ref{tab:orbosonic}-\ref{tab:e4fermC}. We follow the convention of denoting operators in the Warsaw basis~\cite{Grzadkowski:2010es} by $\mathcal{O}$, whereas redundant and evanescent operators are denoted by $\mathcal{R}$ and $\mathcal{E}$, respectively. We should stress that in our case, redundant and evanescent operators do not vanish when using equations of motion or going to $D=4$ dimensions, being reduced  to the operators in the physical basis in this case. We follow the notation of~\cite{Gherardi:2020det} regarding redundant operators and include evanescent operators with two and four fermions as well as possibly featuring charge conjugation. We use the shorthand notation $\gamma^{\mu_1 \ldots \mu_n}  \equiv \gamma^{\mu_1} \ldots \gamma^{\mu_n}$ with no (anti)symmetrization. In the case of four fermion operators with charge conjugation we do not include for the moment structures with three gammas, which we plan to include in the near future. 

We denote by $\verb+iCPV+=\epsilon_{0123}\in\{1,-1\}$ with $\epsilon^{\alpha\beta\mu\nu}$ the Levi-Civita tensor, in such a way that for $D=4$ e.g.
\begin{equation}
	\sigma^{\mu\nu}\epsilon_{\mu\nu\rho\sigma}=2 \texttt{iCPV}\, \ii \sigma_{\rho\sigma}\gamma_5,
\end{equation}
with $\sigma^{\mu\nu}=\frac{\ii}{2}[\gamma^{\mu},\gamma^{\nu}]$ and
\begin{equation}
    \gamma_5=\ii \gamma^0\gamma^1\gamma^2\gamma^3=\texttt{iCPV}\frac{i}{4!}\epsilon_{\mu\nu\alpha\beta}\gamma^{\mu}\gamma^{\nu}\gamma^{\alpha}\gamma^{\beta}.
    \end{equation}
At any rate, dual tensors are defined by 
\begin{equation}
    \tilde{X}_{\mu\nu}=\frac{1}{2}\epsilon_{\mu\nu\alpha\beta}X^{\mu\nu},\quad \mathrm{with}\quad X=G,W,B.
\end{equation}

We follow~\cite{Dekens:2019ept} for our definitions of evanescent operators, with the difference that we do not define them to be zero. Considering at most three gamma matrices, they read
\begin{align}
	E_{LR}^{(2)}&=	P_L\gamma^{\mu\nu}P_L\otimes P_R\gamma_{\mu\nu}P_R=4(1+a_{\rm ev}\epsilon)P_L\otimes P_R,\label{eq:evrl2},\\
	E_{RL}^{(2)}&= P_R\gamma^{\mu\nu}P_R\otimes P_L\gamma_{\mu \nu}P_L=4(1+a_{\rm ev}^{\prime}\epsilon)P_R\otimes P_L,\label{eq:evlr2}\\
	E_{LL}^{(3)}&=	P_R\gamma^{\mu \nu \lambda} P_L\otimes P_R\gamma_{\mu\nu \lambda} 	P_L=4(4-b_{\rm ev}\epsilon)P_R\gamma^{\mu}P_L\otimes P_R\gamma_{\mu}P_L,\\
	E_{RR}^{(3)}&=	P_L\gamma^{\mu \nu \lambda} P_R\otimes P_L\gamma_{\mu \nu \lambda} P_R=4(4-b_{\rm ev}^{\prime}\epsilon)P_L\gamma^{\mu}P_R\otimes P_L\gamma_{\mu}P_R,\\
 E_{LR}^{(3)}&=	P_R\gamma^{\mu\nu\lambda} P_L\otimes P_L\gamma_{\mu \nu \lambda} P_R=4(1+c_{\rm ev}\epsilon)P_R\gamma^{\mu}P_L\otimes P_L\gamma_{\mu}P_R,\\
	E_{RL}^{(3)}&=	P_L\gamma^{\mu\nu \lambda} P_R\otimes P_R\gamma_{\mu\nu\lambda} P_L=4(1+c_{\rm ev}^{\prime}\epsilon)P_L\gamma^{\mu}P_R\otimes P_R\gamma_{\mu}P_L,
\end{align}
while
\begin{align}
	P_L\gamma^{\mu\nu}P_L\otimes P_L\gamma_{\mu\nu}P_L&=(4-2\epsilon)P_L\otimes P_L-P_L\sigma^{\mu\nu}P_L\otimes P_L\sigma_{\mu\nu}P_L,\label{eq:notev}\\
	P_R\gamma^{\mu\nu}P_R\otimes P_R\gamma_{\mu\nu}P_R&=(4-2\epsilon)P_R\otimes P_R-P_R\sigma^{\mu\nu}P_R\otimes P_R\sigma_{\mu\nu}P_R,
	\label{eq:notev2}
\end{align}
assuming the following basis of the space of Lorentz singlets and pseudo-singlets in $D=4$
\begin{align}
	\Big\{\Gamma_1^i\otimes \Gamma_2^i\,|\, i=1,\ldots,10 \Big\}&=\Big\{P_L\otimes P_L,P_R\otimes P_R,P_L\otimes P_R, P_R\otimes P_L,P_R\gamma^{\mu}P_L\otimes P_R\gamma_{\mu} P_L,\nonumber\\
	&P_L\gamma^{\mu}P_R\otimes P_L\gamma_{\mu} P_R,P_R\gamma^{\mu}P_L\otimes P_L\gamma_{\mu} P_R,P_L\gamma^{\mu}P_R\otimes P_R\gamma_{\mu} P_L,\phantom{\Big\}}\nonumber\\
	&P_L\sigma^{\mu\nu}P_L\otimes P_L\sigma_{\mu\nu}P_L,P_R\sigma^{\mu\nu}P_R\otimes P_R\sigma_{\mu\nu}P_R\Big\}.
\end{align}
In order to be consistent with flavor, we further assume that  $a_{\rm ev}=a_{\rm ev}^{\prime}, c_{\rm ev}=c_{\rm ev}^{\prime}$. Therefore, at the end of the day the one-loop matching will be dependant on four free parameters: $\texttt{iCPV},\texttt{aEV},\texttt{bEV}$ and $\texttt{cEV}$. However, physical observables computed with the obtained dimension-6 EFT can not depend on such parameters.

One final important remark should be made. In the model file \texttt{.fr} of the SMEFT used by \texttt{matchmakereft} to do the matching, the operator of the Warsaw basis $\OOp{\ell equ}{(3)}$ is replaced by $\EOp{\ell equ}{[2]}$ since \texttt{matchmakereft} and FORM use always $\gamma^{\mu\nu}$ instead of its antisymmetric version $\sigma^{\mu\nu}$. After doing the matching, such operator is reduced onto the one present in the Warsaw basis by the relation (we use $\alpha$, $\beta$ and $\gamma$ for the WCs of physical, redundant and evanescent operators, respectively)
\begin{align}
     \left(\alpha_{\ell equ}^{(3)} \right)_{ijkl}&=
     \left(\gamma_{\ell equ}^{[2]}\right)_{ijkl} -\left(1-\frac{\epsilon}{4}\right)\left(\gamma_{\ell equ}^{[2]}\right)_{ilkj} -\frac{1}{8}\left(\gamma_{\ell uqe}\right)_{ilkj} \nonumber\\
     &-\frac{1}{8}\left(\gamma_{ue\ell q}^{c}\right)_{ljik}-\left(1-\frac{\epsilon}{4}\right)\left(\gamma_{ue\ell q}^{c\,[2]}\right)_{ljik}.
     \end{align}

\begin{table}[hb]
\begin{centering}
\begin{tabular}{cccccc}
\ctoprule
	\multicolumn{2}{c}{\cellcolor[gray]{0.90}$\boldsymbol{X^{3}}$} & \multicolumn{2}{c}{\cellcolor[gray]{0.90}$\boldsymbol{X^{2}H^{2}}$} & \multicolumn{2}{c}{\cellcolor[gray]{0.90}$\boldsymbol{H^{2}D^{4}}$}\tabularnewline
	\cmrule{1-6} 
	$\OOp{3G}{}$  &$f^{ABC}G_{\mu}^{A\nu}G_{\nu}^{B\rho}G_{\rho}^{C\mu}$  & $\OOp{HG}{}$  & $G_{\mu\nu}^{A}G^{A\mu\nu}(H^{\dagger}H)$  & $\ROp{DH}{}$  & $(D_{\mu}D^{\mu}H)^{\dagger}(D_{\nu}D^{\nu}H)$\tabularnewline
	\cmrule{5-6}  
$\OOp{\widetilde{3G}}{}$  & $f^{ABC}\widetilde{G}_{\mu}^{A\nu}G_{\nu}^{B\rho}G_{\rho}^{C\mu}$  & $\OOp{H\widetilde{G}}{}$  & $\widetilde{G}_{\mu\nu}^{A}G^{A\mu\nu}(H^{\dagger}H)$  & \multicolumn{2}{c}{\cellcolor[gray]{0.90}$\boldsymbol{H^{4}D^{2}}$}\tabularnewline
\cmrule{5-6}  
	$\OOp{3W}{}$  & $\epsilon^{IJK}W_{\mu}^{I\nu}W_{\nu}^{J\rho}W_{\rho}^{K\mu}$  & $\OOp{HW}{}$  & $W_{\mu\nu}^{I}W^{I\mu\nu}(H^{\dagger}H)$  &$\OOp{H\square}{}$  & $(H^{\dagger}H)\square(H^{\dagger}H)$\tabularnewline
	$\OOp{\widetilde{3W}}{}$  &$\epsilon^{IJK}\widetilde{W}_{\mu}^{I\nu}W_{\nu}^{J\rho}W_{\rho}^{K\mu}$  & $\OOp{H\widetilde{W}}{}$  & $\widetilde{W}_{\mu\nu}^{I}W^{I\mu\nu}(H^{\dagger}H)$  &$\OOp{HD}{}$  & $(H^{\dagger}D^{\mu}H)^{\dagger}(H^{\dagger}D_{\mu}H)$\tabularnewline
\cmrule{1-2} 
	\multicolumn{2}{c}{\cellcolor[gray]{0.90}$\boldsymbol{X^{2}D^{2}}$} & $\OOp{HB}{}$  & $B_{\mu\nu}B^{\mu\nu}(H^{\dagger}H)$  & $\ROp{HD}{\prime}$  & $(H^{\dagger}H)(D_{\mu}H)^{\dagger}(D^{\mu}H)$\tabularnewline
\cmrule{1-2}  
	$\ROp{2G}{}$  & $-\frac{1}{2}(D_{\mu}G^{A\mu\nu})(D^{\rho}G_{\rho\nu}^{A})$  &$\OOp{H\widetilde{B}}{}$  &$\widetilde{B}_{\mu\nu}B^{\mu\nu}(H^{\dagger}H)$  & $\ROp{HD}{\prime\prime}$  & $(H^{\dagger}H)D_{\mu}(H^{\dagger}\ii\overleftrightarrow{D}^{\mu}H)$\tabularnewline
\cmrule{5-6}  
	$\ROp{2W}{}$  & $-\frac{1}{2}(D_{\mu}W^{I\mu\nu})(D^{\rho}W_{\rho\nu}^{I})$  &$\OOp{HWB}{}$  & $W_{\mu\nu}^{I}B^{\mu\nu}(H^{\dagger}\sigma^{I}H)$  & \multicolumn{2}{c}{\cellcolor[gray]{0.90}$\boldsymbol{H^{6}}$}\tabularnewline
\cmrule{5-6} 
	$\ROp{2B}{}$  & $-\frac{1}{2}(\partial_{\mu}B^{\mu\nu})(\partial^{\rho}B_{\rho\nu})$  &$\OOp{H\widetilde{W}B}{}$  & $\widetilde{W}_{\mu\nu}^{I}B^{\mu\nu}(H^{\dagger}\sigma^{I}H)$  & $\OOp{H}{}$  & $(H^{\dagger}H)^{3}$\tabularnewline
\cmrule{3-4} 
 &  & \multicolumn{2}{c}{\cellcolor[gray]{0.90}$\boldsymbol{H^{2}XD^{2}}$} &  & \tabularnewline
\cmrule{3-4}  
	&  & $\ROp{WDH}{}$  & $D_{\nu}W^{I\mu\nu}(H^{\dagger}\ii\overleftrightarrow{D}_{\mu}^{I}H)$  &  & \tabularnewline
	&  & $\ROp{BDH}{}$  & $\partial_{\nu}B^{\mu\nu}(H^{\dagger}\ii\overleftrightarrow{D}_{\mu}H)$  &  & \tabularnewline
\cbottomrule
\end{tabular}
\par\end{centering}
\caption{\label{tab:orbosonic} Physical and redundant bosonic operators.}
\end{table}

\begin{table}[p]
\begin{centering}
\begin{tabular}{cccccc}
\ctoprule
	\multicolumn{2}{c}{\cellcolor[gray]{0.90}$\boldsymbol{\psi^{2}D^{3}}$} & \multicolumn{2}{c}{\cellcolor[gray]{0.90}$\boldsymbol{\psi^{2}XD}$} & \multicolumn{2}{c}{\cellcolor[gray]{0.90}$\boldsymbol{\psi^{2}DH^{2}}$}\tabularnewline
	\cmrule{1-6} 
	$\ROp{qD}{}$  & $\frac{\ii}{2}\overline{q}\left\{ D_{\mu}D^{\mu},\slashed D\right\} q$  & $\ROp{Gq}{}$  & $(\overline{q}T^{A}\gamma^{\mu}q)D^{\nu}G_{\mu\nu}^{A}$  & $\OOp{Hq}{(1)}$  & $(\overline{q}\gamma^{\mu}q)(H^{\dagger}\ii\overleftrightarrow{D}_{\mu}H)$\tabularnewline
	$\ROp{uD}{}$  & $\frac{\ii}{2}\overline{u}\left\{ D_{\mu}D^{\mu},\slashed D\right\} u$  & $\ROp{Gq}{\prime}$  & $\frac{1}{2}(\overline{q}T^{A}\gamma^{\mu}\ii\overleftrightarrow{D}^{\nu}q)G_{\mu\nu}^{A}$  & $\ROp{Hq}{\prime(1)}$  & $(\overline{q}\,\ii\overleftrightarrow{\slashed D}q)(H^{\dagger}H)$\tabularnewline
	$\ROp{dD}{}$  & $\frac{\ii}{2}\overline{d}\left\{ D_{\mu}D^{\mu},\slashed D\right\} d$  & $\ROp{\widetilde{G}q}{\prime}$  & $\frac{1}{2}(\overline{q}T^{A}\gamma^{\mu}\ii\overleftrightarrow{D}^{\nu}q)\widetilde{G}_{\mu\nu}^{A}$  & $\ROp{Hq}{\prime\prime(1)}$  & $(\overline{q}\gamma^{\mu}q)\partial_{\mu}(H^{\dagger}H)$\tabularnewline
	$\ROp{\ell D}{}$  & $\frac{\ii}{2}\overline{\ell}\left\{ D_{\mu}D^{\mu},\slashed D\right\} \ell$  & $\ROp{Wq}{}$  & $(\overline{q}\sigma^{I}\gamma^{\mu}q)D^{\nu}W_{\mu\nu}^{I}$  &$\OOp{Hq}{(3)}$  &$(\overline{q}\sigma^{I}\gamma^{\mu}q)(H^{\dagger}\ii\overleftrightarrow{D}_{\mu}^{I}H)$\tabularnewline
	$\ROp{eD}{}$  & $\frac{\ii}{2}\overline{e}\left\{ D_{\mu}D^{\mu},\slashed D\right\} e$  & $\ROp{Wq}{\prime}$  & $\frac{1}{2}(\overline{q}\sigma^{I}\gamma^{\mu}\ii\overleftrightarrow{D}^{\nu}q)W_{\mu\nu}^{I}$  & $\ROp{Hq}{\prime(3)}$  & $(\overline{q}\,\ii\overleftrightarrow{\slashed D}^{I}q)(H^{\dagger}\sigma^{I}H)$\tabularnewline
\cmrule{1-2}  
	\multicolumn{2}{c}{\cellcolor[gray]{0.90}$\boldsymbol{\psi^{2}HD^{2}}+\textbf{h.c.}$} & $\ROp{\widetilde{W}q}{\prime}$  & $\frac{1}{2}(\overline{q}\sigma^{I}\gamma^{\mu}\ii\overleftrightarrow{D}^{\nu}q)\widetilde{W}_{\mu\nu}^{I}$  & $\ROp{Hq}{\prime\prime(3)}$  & $(\overline{q}\sigma^{I}\gamma^{\mu}q)D_{\mu}(H^{\dagger}\sigma^{I}H)$\tabularnewline
\cmrule{1-2}  
	$\ROp{uHD1}{}$  & $(\overline{q}u)D_{\mu}D^{\mu}\widetilde{H}$  & $\ROp{Bq}{}$  & $(\overline{q}\gamma^{\mu}q)\partial^{\nu}B_{\mu\nu}$  & $\OOp{Hu}{}$  & $(\overline{u}\gamma^{\mu}u)(H^{\dagger}\ii\overleftrightarrow{D}_{\mu}H)$\tabularnewline
	$\ROp{uHD2}{}$  & $(\overline{q}\,\ii\sigma_{\mu\nu}D^{\mu}u)D^{\nu}\widetilde{H}$  & $\ROp{Bq}{\prime}$  & $\frac{1}{2}(\overline{q}\gamma^{\mu}\ii\overleftrightarrow{D}^{\nu}q)B_{\mu\nu}$  & $\ROp{Hu}{\prime}$  & $(\overline{u}\,\ii\overleftrightarrow{\slashed D}u)(H^{\dagger}H)$\tabularnewline
	$\ROp{uHD3}{}$  & $(\overline{q}D_{\mu}D^{\mu}u)\widetilde{H}$  & $\ROp{\widetilde{B}q}{\prime}$  & $\frac{1}{2}(\overline{q}\gamma^{\mu}i\overleftrightarrow{D}^{\nu}q)\widetilde{B}_{\mu\nu}$  & $\ROp{Hu}{\prime\prime}$  & $(\overline{u}\gamma^{\mu}u)\partial_{\mu}(H^{\dagger}H)$\tabularnewline
	$\ROp{uHD4}{}$  & $(\overline{q}D_{\mu}u)D^{\mu}\widetilde{H}$  & $\ROp{Gu}{}$  & $(\overline{u}T^{A}\gamma^{\mu}u)D^{\nu}G_{\mu\nu}^{A}$  & $\OOp{Hd}{}$  & $(\overline{d}\gamma^{\mu}d)(H^{\dagger}\ii\overleftrightarrow{D}_{\mu}H)$\tabularnewline
	$\ROp{dHD1}{}$  & $(\overline{q}d)D_{\mu}D^{\mu}H$  & $\ROp{Gu}{\prime}$  & $\frac{1}{2}(\overline{u}T^{A}\gamma^{\mu}\ii\overleftrightarrow{D}^{\nu}u)G_{\mu\nu}^{A}$  & $\ROp{Hd}{\prime}$  & $(\overline{d}\,\ii\overleftrightarrow{\slashed D}d)(H^{\dagger}H)$\tabularnewline
	$\ROp{dHD2}{}$  & $(\overline{q}\,\ii\sigma_{\mu\nu}D^{\mu}d)D^{\nu}H$  & $\ROp{\widetilde{G}u}{\prime}$  & $\frac{1}{2}(\overline{u}T^{A}\gamma^{\mu}\ii\overleftrightarrow{D}^{\nu}u)\widetilde{G}_{\mu\nu}^{A}$  & $\ROp{Hd}{\prime\prime}$  & $(\overline{d}\gamma^{\mu}d)\partial_{\mu}(H^{\dagger}H)$\tabularnewline
	$\ROp{dHD3}{}$  & $(\overline{q}D_{\mu}D^{\mu}d)H$  & $\ROp{Bu}{}$  & $(\overline{u}\gamma^{\mu}u)\partial^{\nu}B_{\mu\nu}$  & $\OOp{Hud}{}$  & $(\overline{u}\gamma^{\mu}d)(\widetilde{H}^{\dagger}\ii D_{\mu}H)$\tabularnewline
	$\ROp{dHD4}{}$  & $(\overline{q}D_{\mu}d)D^{\mu}H$  & $\ROp{Bu}{\prime}$  & $\frac{1}{2}(\overline{u}\gamma^{\mu}\ii\overleftrightarrow{D}^{\nu}u)B_{\mu\nu}$  & $\OOp{H\ell}{(1)}$  & $(\overline{\ell}\gamma^{\mu}\ell)(H^{\dagger}\ii\overleftrightarrow{D}_{\mu}H)$\tabularnewline
	$\ROp{eHD1}{}$  & $(\overline{\ell}e)D_{\mu}D^{\mu}H$  & $\ROp{\widetilde{B}u}{\prime}$  & $\frac{1}{2}(\overline{u}\gamma^{\mu}\ii\overleftrightarrow{D}^{\nu}u)\widetilde{B}_{\mu\nu}$  & $\ROp{H\ell}{\prime(1)}$  & $(\overline{\ell}i\overleftrightarrow{\slashed D}\ell)(H^{\dagger}H)$\tabularnewline
	$\ROp{eHD2}{}$  & $(\overline{\ell}\,\ii\sigma_{\mu\nu}D^{\mu}e)D^{\nu}H$  & $\ROp{Gd}{}$  & $(\overline{d}T^{A}\gamma^{\mu}d)D^{\nu}G_{\mu\nu}^{A}$  & $\ROp{H\ell}{\prime\prime(1)}$  & $(\overline{\ell}\gamma^{\mu}\ell)\partial_{\mu}(H^{\dagger}H)$\tabularnewline
	$\ROp{eHD3}{}$  & $(\overline{\ell}D_{\mu}D^{\mu}e)H$  & $\ROp{Gd}{\prime}$  & $\frac{1}{2}(\overline{d}T^{A}\gamma^{\mu}\ii\overleftrightarrow{D}^{\nu}d)G_{\mu\nu}^{A}$  & $\OOp{H\ell}{(3)}$  & $(\overline{\ell}\sigma^{I}\gamma^{\mu}\ell)(H^{\dagger}\ii\overleftrightarrow{D}_{\mu}^{I}H)$\tabularnewline
	$\ROp{eHD4}{}$  & $(\overline{\ell}D_{\mu}e)D^{\mu}H$  & $\ROp{\widetilde{G}d}{\prime}$  & $\frac{1}{2}(\overline{d}T^{A}\gamma^{\mu}\ii\overleftrightarrow{D}^{\nu}d)\widetilde{G}_{\mu\nu}^{A}$  & $\ROp{H\ell}{\prime(3)}$  & $(\overline{\ell}\ii\overleftrightarrow{\slashed D}^{I}\ell)(H^{\dagger}\sigma^{I}H)$\tabularnewline
\cmrule{1-2} 
	\multicolumn{2}{c}{\cellcolor[gray]{0.90}$\boldsymbol{\psi^{2}XH}+\textbf{h.c.}$} & $\ROp{Bd}{}$  & $(\overline{d}\gamma^{\mu}d)\partial^{\nu}B_{\mu\nu}$  & $\ROp{H\ell}{\prime\prime(3)}$  & $(\overline{\ell}\sigma^{I}\gamma^{\mu}\ell)D_{\mu}(H^{\dagger}\sigma^{I}H)$\tabularnewline
\cmrule{1-2} 
	$\OOp{uG}{}$  & $(\overline{q}T^{A}\sigma^{\mu\nu}u)\widetilde{H}G_{\mu\nu}^{A}$  & $\ROp{Bd}{\prime}$  & $\frac{1}{2}(\overline{d}\gamma^{\mu}i\overleftrightarrow{D}^{\nu}d)B_{\mu\nu}$  & $\OOp{He}{}$  &$(\overline{e}\gamma^{\mu}e)(H^{\dagger}\ii\overleftrightarrow{D}_{\mu}H)$\tabularnewline
	$\OOp{uW}{}$  & $(\overline{q}\sigma^{\mu\nu}u)\sigma^{I}\widetilde{H}W_{\mu\nu}^{I}$  & $\ROp{\widetilde{B}d}{\prime}$  & $\frac{1}{2}(\overline{d}\gamma^{\mu}\ii\overleftrightarrow{D}^{\nu}d)\widetilde{B}_{\mu\nu}$  & $\ROp{He}{\prime}$  & $(\overline{e}\,\ii\overleftrightarrow{\slashed D}e)(H^{\dagger}H)$\tabularnewline
	$\OOp{uB}{}$  & $(\overline{q}\sigma^{\mu\nu}u)\widetilde{H}B_{\mu\nu}$  & $\ROp{W\ell}{}$  & $(\overline{\ell}\sigma^{I}\gamma^{\mu}\ell)D^{\nu}W_{\mu\nu}^{I}$  & $\ROp{He}{\prime\prime}$  & $(\overline{e}\gamma^{\mu}e)\partial_{\mu}(H^{\dagger}H)$\tabularnewline
\cmrule{5-6}
	$\OOp{dG}{}$  & $(\overline{q}T^{A}\sigma^{\mu\nu}d)HG_{\mu\nu}^{A}$  & $\ROp{W\ell}{\prime}$  & $\frac{1}{2}(\overline{\ell}\sigma^{I}\gamma^{\mu}\ii\overleftrightarrow{D}^{\nu}\ell)W_{\mu\nu}^{I}$  & \multicolumn{2}{c}{\cellcolor[gray]{0.90}$\boldsymbol{\psi^{2}H^{3}}+\textbf{h.c.}$}\tabularnewline
\cmrule{5-6}  
	$\OOp{dW}{}$  & $(\overline{q}\sigma^{\mu\nu}d)\sigma^{I}HW_{\mu\nu}^{I}$  & $\ROp{\widetilde{W}\ell}{\prime}$  & $\frac{1}{2}(\overline{\ell}\sigma^{I}\gamma^{\mu}\ii\overleftrightarrow{D}^{\nu}\ell)\widetilde{W}_{\mu\nu}^{I}$  & $\OOp{uH}{}$  &$(H^{\dagger}H)\overline{q}\widetilde{H}u$\tabularnewline
	$\OOp{dB}{}$  & $(\overline{q}\sigma^{\mu\nu}d)HB_{\mu\nu}$  & $\ROp{B\ell}{}$  & $(\overline{\ell}\gamma^{\mu}\ell)\partial^{\nu}B_{\mu\nu}$  & $\OOp{dH}{}$  & $(H^{\dagger}H)\overline{q}Hd$\tabularnewline
	$\OOp{eW}{}$  & $(\overline{\ell}\sigma^{\mu\nu}e)\sigma^{I}HW_{\mu\nu}^{I}$  & $\ROp{B\ell}{\prime}$  & $\frac{1}{2}(\overline{\ell}\gamma^{\mu}\ii\overleftrightarrow{D}^{\nu}\ell)B_{\mu\nu}$  & $\OOp{eH}{}$  & $(H^{\dagger}H)\overline{\ell}He$\tabularnewline
	$\OOp{eB}{}$  & $(\overline{\ell}\sigma^{\mu\nu}e)HB_{\mu\nu}$  & $\ROp{\widetilde{B}\ell}{\prime}$  & $\frac{1}{2}(\overline{\ell}\gamma^{\mu}\ii\overleftrightarrow{D}^{\nu}\ell)\widetilde{B}_{\mu\nu}$  &  & \tabularnewline
	&  & $\ROp{Be}{}$  & $(\overline{e}\gamma^{\mu}e)\partial^{\nu}B_{\mu\nu}$  &  & \tabularnewline
 &  & $\ROp{Be}{\prime}$  & $\frac{1}{2}(\overline{e}\gamma^{\mu}\ii\overleftrightarrow{D}^{\nu}e)B_{\mu\nu}$  &  & \tabularnewline
	&  & $\ROp{\widetilde{B}e}{\prime}$  & $\frac{1}{2}(\overline{e}\gamma^{\mu}\ii\overleftrightarrow{D}^{\nu}e)\widetilde{B}_{\mu\nu}$  &  & \tabularnewline
\cbottomrule 
\end{tabular}
\par\end{centering}
\caption{\label{tab:or2fermions} Physical and redundant operators with two fermions.}
\end{table}

\begin{table}[h]
\begin{centering}
\begin{tabular}{cccccc}
\ctoprule 
\multicolumn{2}{c}{\cellcolor[gray]{0.90}Four-quark} & \multicolumn{2}{c}{\cellcolor[gray]{0.90}Four-lepton} & \multicolumn{2}{c}{\cellcolor[gray]{0.90}Semileptonic}\tabularnewline
	\cmrule{1-6}
	$\OOp{qq}{(1)}$  & $(\overline{q}\gamma^{\mu}q)(\overline{q}\gamma_{\mu}q)$  & $\OOp{\ell\ell}{}$  & $(\overline{\ell}\gamma^{\mu}\ell)(\overline{\ell}\gamma_{\mu}\ell)$  & $\OOp{\ell q}{(1)}$  & $(\overline{\ell}\gamma^{\mu}\ell)(\overline{q}\gamma_{\mu}q)$\tabularnewline
	$\OOp{qq}{(3)}$  & $(\overline{q}\gamma^{\mu}\sigma^{I}q)(\overline{q}\gamma_{\mu}\sigma^{I}q)$  &$\OOp{ee}{}$  &$(\overline{e}\gamma^{\mu}e)(\overline{e}\gamma_{\mu}e)$  & $\OOp{\ell q}{(3)}$  &$(\overline{\ell}\gamma^{\mu}\sigma^{I}\ell)(\overline{q}\gamma_{\mu}\sigma^{I}q)$\tabularnewline
	$\OOp{uu}{}$  &$(\overline{u}\gamma^{\mu}u)(\overline{u}\gamma_{\mu}u)$  & $\OOp{\ell e}{}$  & $(\overline{\ell}\gamma^{\mu}\ell)(\overline{e}\gamma_{\mu}e)$  & $\OOp{eu}{}$  & $(\overline{e}\gamma^{\mu}e)(\overline{u}\gamma_{\mu}u)$\tabularnewline
	$\OOp{dd}{}$  & $(\overline{d}\gamma^{\mu}d)(\overline{d}\gamma_{\mu}d)$  &  &  & $\OOp{ed}{}$  &$(\overline{e}\gamma^{\mu}e)(\overline{d}\gamma_{\mu}d)$\tabularnewline
	$\OOp{ud}{(1)}$  & $(\overline{u}\gamma^{\mu}u)(\overline{d}\gamma_{\mu}d)$  &  &  & $\OOp{qe}{}$  & $(\overline{q}\gamma^{\mu}q)(\overline{e}\gamma_{\mu}e)$\tabularnewline
	$\OOp{ud}{(8)}$  &$(\overline{u}\gamma^{\mu}T^{A}u)(\overline{d}\gamma_{\mu}T^{A}d)$  &  &  & $\OOp{\ell u}{}$  & $(\overline{\ell}\gamma^{\mu}\ell)(\overline{u}\gamma_{\mu}u)$\tabularnewline
	$\OOp{qu}{(1)}$  & $(\overline{q}\gamma^{\mu}q)(\overline{u}\gamma_{\mu}u)$  &  &  & $\OOp{\ell d}{}$  &$(\overline{\ell}\gamma^{\mu}\ell)(\overline{d}\gamma_{\mu}d)$\tabularnewline
	$\OOp{qu}{(8)}$  &$(\overline{q}\gamma^{\mu}T^{A}q)(\overline{u}\gamma_{\mu}T^{A}u)$  &  &  &$\OOp{\ell edq}{}$  & $(\overline{\ell}e)(\overline{d}q)$\tabularnewline
$\OOp{qd}{(1)}$  &$(\overline{q}\gamma^{\mu}q)(\overline{d}\gamma_{\mu}d)$  &  &  & $\OOp{\ell equ}{(1)}$  & $(\overline{\ell}_{r}e)\epsilon_{rs}(\overline{q}_{s}u)$\tabularnewline
	$\OOp{qd}{(8)}$  &$(\overline{q}\gamma^{\mu}T^{A}q)(\overline{d}\gamma_{\mu}T^{A}d)$  &  &  &$\OOp{\ell equ}{(3)}$  &$(\overline{\ell}_{r}\sigma^{\mu\nu}e)\epsilon_{rs}(\overline{q}_{s}\sigma_{\mu\nu}u) $\tabularnewline
$\OOp{quqd}{(1)}$  &$(\overline{q}_{r}u)\epsilon_{rs}(\overline{q}_{s}d)$  &  &  &  & \tabularnewline
$\OOp{quqd}{(8)}$  &$(\overline{q}_{r}T^{A}u)\epsilon_{rs}(\overline{q}_{s}T^{A}d)$  &  &  &  & \tabularnewline
\cbottomrule
\end{tabular}
\par\end{centering}
\caption{\label{tab:or4fermions} Baryon and lepton number conserving operators with four fermions.}
\end{table}

\begin{table}[h]
	\begin{adjustwidth}{-.5in}{-.5in}
        \begin{center}
\begin{tabular}{cccccc}
\ctoprule
\multicolumn{2}{c}{\cellcolor[gray]{0.90}$\boldsymbol{\Psi^{2} X H}+\textbf{h.c.}$} &
\multicolumn{4}{c}{\cellcolor[gray]{0.90}$\boldsymbol{\Psi^{2}XD}$} \tabularnewline
\cmrule{1-6}
	$\EOp{uG}{}$ & $\bar{q} T^A \sigma^{\mu\nu} u \widetilde{H} \widetilde{G}^A_{\mu\nu}$ &
	$\EOp{Gq}{}$ & $\bar{q}T^A (\sigma^{\mu\nu} \gamma^\rho + \gamma^\rho \sigma^{\mu\nu}) q D_\rho \widetilde{G}^A_{\mu\nu}$ &
	$\EOp{Gd}{}$ & $\bar{d}T^A (\sigma^{\mu\nu} \gamma^\rho + \gamma^\rho \sigma^{\mu\nu}) d D_\rho \widetilde{G}^A_{\mu\nu}$  
\tabularnewline 
	$\EOp{uW}{}$ & $\bar{q} \sigma^I \sigma^{\mu\nu} u \widetilde{H} \widetilde{W}^I_{\mu\nu}$ &
	$\EOp{Gq}{\prime}$ & $\ii \bar{q}(T^A \sigma^{\mu\nu} \slashed{D} - \lvec{\slashed{D}} \sigma^{\mu\nu} T^A) q G^A_{\mu\nu}$&
	$\EOp{Gd}{\prime}$ & $\ii \bar{d}(T^A \sigma^{\mu\nu} \slashed{D} - \lvec{\slashed{D}} \sigma^{\mu\nu} T^A) d G^A_{\mu\nu}$
\tabularnewline 
	$\EOp{uB}{}$ & $\bar{q} \sigma^{\mu\nu} u \widetilde{H} \widetilde{B}_{\mu\nu}$ &
	$\EOp{\widetilde{G}q}{\prime}$ & $\ii \bar{q}(T^A \sigma^{\mu\nu} \slashed{D} - \lvec{\slashed{D}} \sigma^{\mu\nu} T^A) q \widetilde{G}^A_{\mu\nu}$&
	$\EOp{\widetilde{G}d}{\prime}$ & $\ii \bar{d}(T^A \sigma^{\mu\nu} \slashed{D} - \lvec{\slashed{D}} \sigma^{\mu\nu} T^A) d \widetilde{G}^A_{\mu\nu}$ 
\tabularnewline 
	$\EOp{dG}{}$ & $\bar{q} T^A \sigma^{\mu\nu} d H \widetilde{G}^A_{\mu\nu}$ &
	$\EOp{Wq}{}$ & $\bar{q}\sigma^I (\sigma^{\mu\nu} \gamma^\rho + \gamma^\rho \sigma^{\mu\nu}) q D_\rho \widetilde{W}^I_{\mu\nu}$ &
	$\EOp{Bd}{}$ & $\bar{d}(\sigma^{\mu\nu} \gamma^\rho + \gamma^\rho \sigma^{\mu\nu}) d \partial_\rho \widetilde{B}_{\mu\nu}$  
\tabularnewline 
	$\EOp{dW}{}$ & $\bar{q} \sigma^I \sigma^{\mu\nu} d H \widetilde{W}^I_{\mu\nu}$ &
	$\EOp{Wq}{\prime}$ & $\ii \bar{q}(\sigma^I \sigma^{\mu\nu} \slashed{D} - \lvec{\slashed{D}} \sigma^{\mu\nu} \sigma^I) q W^I_{\mu\nu}$&
	$\EOp{Bd}{\prime}$ & $\ii \bar{d}(\sigma^{\mu\nu} \slashed{D} - \lvec{\slashed{D}} \sigma^{\mu\nu}) d B^A_{\mu\nu}$
\tabularnewline 
	$\EOp{dB}{}$ & $\bar{q} \sigma^{\mu\nu} d H \widetilde{B}_{\mu\nu}$ &
	$\EOp{\widetilde{W}q}{\prime}$ & $\ii \bar{q}(\sigma^I \sigma^{\mu\nu} \slashed{D} - \lvec{\slashed{D}} \sigma^{\mu\nu} \sigma^I) q \widetilde{W}^I_{\mu\nu}$&
	$\EOp{\widetilde{B}d}{\prime}$ & $\ii \bar{d}(\sigma^{\mu\nu} \slashed{D} - \lvec{\slashed{D}} \sigma^{\mu\nu}) d \widetilde{B}_{\mu\nu}$
\tabularnewline 
	$\EOp{eW}{}$ & $\bar{\ell} \sigma^I \sigma^{\mu\nu} e H \widetilde{W}^I_{\mu\nu}$ &
	$\EOp{Bq}{}$ & $\bar{q}(\sigma^{\mu\nu} \gamma^\rho + \gamma^\rho \sigma^{\mu\nu}) q \partial_\rho \widetilde{B}_{\mu\nu}$ &
	$\EOp{W\ell}{}$ & $\bar{\ell}\sigma^I (\sigma^{\mu\nu} \gamma^\rho + \gamma^\rho \sigma^{\mu\nu}) \ell D_\rho \widetilde{W}^I_{\mu\nu}$ 
\tabularnewline 
	$\EOp{eB}{}$ & $\bar{\ell} \sigma^{\mu\nu} e H \widetilde{B}_{\mu\nu}$ &
	$\EOp{Bq}{\prime}$ & $\ii \bar{q}(\sigma^{\mu\nu} \slashed{D} - \lvec{\slashed{D}} \sigma^{\mu\nu}) q B_{\mu\nu}$&
	$\EOp{W\ell}{\prime}$ & $\ii \bar{\ell}(\sigma^I \sigma^{\mu\nu} \slashed{D} - \lvec{\slashed{D}} \sigma^{\mu\nu} \sigma^I) \ell W^I_{\mu\nu}$
\tabularnewline 
\cmrule{1-2} 
\multicolumn{2}{c}{\cellcolor[gray]{0.90}$\boldsymbol{\psi^{2} H D^{2}}+\textbf{h.c.}$} &
	$\EOp{\widetilde{B}q}{\prime}$ & $\ii \bar{q}(\sigma^{\mu\nu} \slashed{D} - \lvec{\slashed{D}} \sigma^{\mu\nu}) q \widetilde{B}_{\mu\nu}$&
	$\EOp{\widetilde{W}\ell}{\prime}$ & $\ii \bar{\ell}(\sigma^I \sigma^{\mu\nu} \slashed{D} - \lvec{\slashed{D}} \sigma^{\mu\nu} \sigma^I) \ell \widetilde{W}^I_{\mu\nu}$
	\tabularnewline \cmrule{1-2}
	$\EOp{uH}{}$ & $\bar{q} \sigma^{\mu\nu} D^{\rho} u D^\sigma \widetilde{H} \epsilon_{\mu\nu\rho\sigma}$ &
	$\EOp{Gu}{}$ & $\bar{u}T^A (\sigma^{\mu\nu} \gamma^\rho + \gamma^\rho \sigma^{\mu\nu}) u D_\rho \widetilde{G}^A_{\mu\nu}$ &
	$\EOp{B\ell}{}$ & $\bar{\ell}(\sigma^{\mu\nu} \gamma^\rho + \gamma^\rho \sigma^{\mu\nu}) \ell \partial_\rho \widetilde{B}_{\mu\nu}$ 
\tabularnewline 
	$\EOp{dH}{}$ & $\bar{q} \sigma^{\mu\nu} D^{\rho} d D^\sigma H \epsilon_{\mu\nu\rho\sigma}$ &
	$\EOp{Gu}{\prime}$ & $\ii \bar{u}(T^A \sigma^{\mu\nu} \slashed{D} - \lvec{\slashed{D}} \sigma^{\mu\nu} T^A) u G^A_{\mu\nu}$&
	$\EOp{B\ell}{\prime}$ & $\ii \bar{\ell}(\sigma^{\mu\nu} \slashed{D} - \lvec{\slashed{D}} \sigma^{\mu\nu}) \ell B_{\mu\nu}$
\tabularnewline 
	$\EOp{eH}{}$ & $\bar{\ell} \sigma^{\mu\nu} D^{\rho} e D^\sigma H \epsilon_{\mu\nu\rho\sigma}$ &
	$\EOp{\widetilde{G}u}{\prime}$ & $\ii \bar{u}(T^A \sigma^{\mu\nu} \slashed{D} - \lvec{\slashed{D}} \sigma^{\mu\nu} T^A) u \widetilde{G}^A_{\mu\nu}$&
	$\EOp{\widetilde{B}\ell}{\prime}$ & $\ii \bar{\ell}(\sigma^{\mu\nu} \slashed{D} - \lvec{\slashed{D}} \sigma^{\mu\nu}) \ell \widetilde{B}_{\mu\nu}$
\tabularnewline 
&&
	$\EOp{Bu}{}$ & $\bar{u}(\sigma^{\mu\nu} \gamma^\rho + \gamma^\rho \sigma^{\mu\nu}) u \partial_\rho \widetilde{B}_{\mu\nu}$ &
	$\EOp{Be}{}$ & $\bar{e}(\sigma^{\mu\nu} \gamma^\rho + \gamma^\rho \sigma^{\mu\nu}) e \partial_\rho \widetilde{B}_{\mu\nu}$
\tabularnewline 
&&
	$\EOp{Bu}{\prime}$ & $\ii \bar{u}(\sigma^{\mu\nu} \slashed{D} - \lvec{\slashed{D}} \sigma^{\mu\nu}) u B_{\mu\nu}$&
	$\EOp{Be}{\prime}$ & $\ii \bar{e}(\sigma^{\mu\nu} \slashed{D} - \lvec{\slashed{D}} \sigma^{\mu\nu}) e B_{\mu\nu}$
\tabularnewline 
&&
$\EOp{\widetilde{B}u}{\prime}$ & $\ii \bar{u}(\sigma^{\mu\nu} \slashed{D} - \lvec{\slashed{D}} \sigma^{\mu\nu}) u \widetilde{B}_{\mu\nu}$&
$\EOp{\widetilde{B}e}{\prime}$ & $\ii \bar{e}(\sigma^{\mu\nu} \slashed{D} - \lvec{\slashed{D}} \sigma^{\mu\nu}) e \widetilde{B}_{\mu\nu}$
\tabularnewline \cbottomrule
\end{tabular}
\caption{\label{tab:e2fermions} Evanescent operators with two fermions.}
\end{center}
\end{adjustwidth}
\end{table}
\begin{table}[h]
	\begin{adjustwidth}{-.5in}{-.5in}
\begin{centering}
\begin{tabular}{cccccc}
\ctoprule
\multicolumn{2}{c}{\cellcolor[gray]{0.90}$\boldsymbol{\bar{L}R \bar{R}L}$} &
\multicolumn{2}{c}{\cellcolor[gray]{0.90}$\boldsymbol{\bar{R}R \bar{R}R}$} &
\multicolumn{2}{c}{\cellcolor[gray]{0.90}$\boldsymbol{\bar{L}L \bar{R}R}$} 
	\tabularnewline \cmrule{1-6}
	$\EOp{qu}{}$ & $(\bar{q} u) (\bar{u}q)$&
	$\EOp{uu}{(8)}$ & $(\bar{u} \gamma^\mu T^A u) (\bar{u} \gamma_\mu T^A u)$&
	$\EOp{qu}{[3]}$ & $(\bar{q} \gamma^{\mu\nu\rho} q) (\bar{u} \gamma_{\mu\nu\rho}  u)$
\tabularnewline 
	$\EOp{qu}{(8)}$ & $(\bar{q} T^A u) (\bar{u} T^A q)$&
	$\EOp{uu}{[3]}$ & $(\bar{u} \gamma^{\mu\nu\rho}  u) (\bar{u} \gamma_{\mu\nu\rho}  u)$&
	$\EOp{qu}{[3](8)}$ & $(\bar{q} \gamma^{\mu\nu\rho} T^A q) (\bar{u} \gamma_{\mu\nu\rho} T^A u)$
\tabularnewline 
	$\EOp{qd}{}$ & $(\bar{q} d) (\bar{d}q)$&
	$\EOp{uu}{[3](8)}$ & $(\bar{u} \gamma^{\mu\nu\rho} T^A u) (\bar{u} \gamma_{\mu\nu\rho} T^A u)$&
	$\EOp{qd}{[3]}$ & $(\bar{q} \gamma^{\mu\nu\rho} q) (\bar{d} \gamma_{\mu\nu\rho}  d)$
\tabularnewline 
	$\EOp{qd}{(8)}$ & $(\bar{q} T^A d) (\bar{d} T^A q)$&
	$\EOp{dd}{(8)}$ & $(\bar{d} \gamma^\mu T^A d) (\bar{d} \gamma_\mu T^A d)$&
	$\EOp{qd}{[3](8)}$ & $(\bar{q} \gamma^{\mu\nu\rho} T^A q) (\bar{d} \gamma_{\mu\nu\rho} T^A d)$
	\tabularnewline \cmrule{5-6}
	$\EOp{qu}{[2]}$ & $(\bar{q}\gamma^{\mu\nu} u) (\bar{u} \gamma_{\mu\nu}q)$&
	$\EOp{dd}{[3]}$ & $(\bar{d} \gamma^{\mu\nu\rho}  d) (\bar{d} \gamma_{\mu\nu\rho}  d)$&
\multicolumn{2}{c}{\cellcolor[gray]{0.90}$\boldsymbol{\bar{L}L \bar{L}L}$} 
	\tabularnewline \cmrule{5-6}
	$\EOp{qu}{[2](8)}$ & $(\bar{q}\gamma^{\mu\nu} T^A u) (\bar{u} \gamma_{\mu\nu} T^A q)$&
	$\EOp{dd}{[3](8)}$ & $(\bar{d} \gamma^{\mu\nu\rho} T^A d) (\bar{d} \gamma_{\mu\nu\rho} T^A d)$&
	$\EOp{qq}{(8)}$ & $(\bar{q} \gamma^\mu T^A q) (\bar{q} \gamma_\mu T^A q)$
\tabularnewline 
	$\EOp{qd}{[2]}$ & $(\bar{q}\gamma^{\mu\nu} d) (\bar{d} \gamma_{\mu\nu}q)$&
	$\EOp{ud}{}$ & $(\bar{u} \gamma^\mu d) (\bar{d} \gamma_\mu u)$&
	$\EOp{qq}{(3,8)}$ & $(\bar{q} \gamma^\mu \sigma^I T^A q) (\bar{q} \gamma_\mu \sigma^I T^A q)$
\tabularnewline 
	$\EOp{qd}{[2](8)}$ & $(\bar{q}\gamma^{\mu\nu} T^A d) (\bar{d} \gamma_{\mu\nu} T^A q)$&
	$\EOp{ud}{(8)}$ & $(\bar{u} \gamma^\mu T^A d) (\bar{d} \gamma_\mu T^A u)$&
	$\EOp{qq}{[3](1)}$ & $(\bar{q} \gamma^{\mu\nu\rho} q) (\bar{q} \gamma_{\mu\nu\rho} q)$
	\tabularnewline \cmrule{1-2}
\multicolumn{2}{c}{\cellcolor[gray]{0.90}$\boldsymbol{\bar{L}R \bar{L}R}$} &
	$\EOp{ud}{[3]}$ & $(\bar{u} \gamma^{\mu\nu\rho}  d) (\bar{d} \gamma_{\mu\nu\rho}  u)$&
	$\EOp{qq}{[3](3)}$ & $(\bar{q} \gamma^{\mu\nu\rho} \sigma^I q) (\bar{q} \gamma_{\mu\nu\rho} \sigma^I q)$
	\tabularnewline \cmrule{1-2}
	$\EOp{quqd}{[2]}$ & $(\bar{q}_r \gamma^{\mu\nu} u) \epsilon_{rs} (\bar{q}_s \gamma_{\mu\nu} d)$&
	$\EOp{ud}{[3](8)}$ & $(\bar{u} \gamma^{\mu\nu\rho} T^A  d) (\bar{d} \gamma_{\mu\nu\rho} T^A u)$&
	$\EOp{qq}{[3](8)}$ & $(\bar{q} \gamma^{\mu\nu\rho} T^A q) (\bar{q} \gamma_{\mu\nu\rho} T^A q)$
\tabularnewline 
	$\EOp{quqd}{[2](8)}$ & $(\bar{q}_r \gamma^{\mu\nu} T^A u) \epsilon_{rs} (\bar{q}_s \gamma_{\mu\nu} T^A d)$&
	$\EOp{ud}{\prime\,[3]}$ & $(\bar{u} \gamma^{\mu\nu\rho}  u) (\bar{d} \gamma_{\mu\nu\rho} d)$&
	$\EOp{qq}{[3](3,8)}$ & $(\bar{q} \gamma^{\mu\nu\rho} \sigma^I T^A q) (\bar{q} \gamma_{\mu\nu\rho} \sigma^I T^A q)$
\tabularnewline 
&&
	$\EOp{ud}{\prime\,[3](8)}$ & $(\bar{u} \gamma^{\mu\nu\rho} T^A  u) (\bar{d} \gamma_{\mu\nu\rho} T^A d)$&
	&
\tabularnewline \cbottomrule
\end{tabular}
\par\end{centering}
\caption{\label{tab:e4quarks} Evanescent operators with four fermions involving only quarks. }
	\end{adjustwidth}
\end{table}

\begin{table}[h]
\begin{centering}
\begin{tabular}{cccccc}
\ctoprule
\multicolumn{2}{c}{\cellcolor[gray]{0.90}$\boldsymbol{\bar{L}R \bar{R}L}$} &
\multicolumn{2}{c}{\cellcolor[gray]{0.90}$\boldsymbol{\bar{R}R \bar{R}R}$} &
\multicolumn{2}{c}{\cellcolor[gray]{0.90}$\boldsymbol{\bar{L}L \bar{R}R}$} 
	\tabularnewline \cmrule{1-6}
		$\EOp{\ell u}{}$ & $(\bar{\ell} u)(\bar{u} \ell)$&
	$\EOp{eu}{}$ & $(\bar{e} \gamma^\mu  u) (\bar{u} \gamma_\mu e)$&
	$\EOp{\ell qde}{}$ & $(\bar{\ell} \gamma^{\mu}  q) (\bar{d} \gamma_{\mu}  e)$
\tabularnewline
	$\EOp{\ell d}{}$ & $(\bar{\ell} d)(\bar{d} \ell)$&
	$\EOp{ed}{}$ & $(\bar{e} \gamma^\mu  d) (\bar{d} \gamma_\mu e)$&
	$\EOp{\ell u}{[3]}$ & $(\bar{\ell} \gamma^{\mu\nu\rho}\ell) (\bar{u} \gamma_{\mu\nu\rho}u)$
	\tabularnewline
	$\EOp{qe}{}$ & $(\bar{q} e)(\bar{e} q)$&
	$\EOp{eu}{[3]}$ & $(\bar{e} \gamma^{\mu\nu\rho}  u) (\bar{u} \gamma_{\mu\nu\rho} e)$&
	$\EOp{\ell d}{[3]}$ & $(\bar{\ell} \gamma^{\mu\nu\rho}\ell) (\bar{d} \gamma_{\mu\nu\rho}d)$
	\tabularnewline
	$\EOp{\ell edq}{[2]}$ & $(\bar{\ell} \gamma^{\mu\nu} e)(\bar{d} \gamma_{\mu\nu} q)$&
	$\EOp{ed}{[3]}$ & $(\bar{e} \gamma^{\mu\nu\rho}  d) (\bar{d} \gamma_{\mu\nu\rho} e)$&
$\EOp{qe}{[3]}$ & $(\bar{q} \gamma^{\mu\nu\rho}q) (\bar{e} \gamma_{\mu\nu\rho}e)$
	\tabularnewline
	$\EOp{\ell u}{[2]}$ & $(\bar{\ell} \gamma^{\mu\nu} u)(\bar{u} \gamma_{\mu\nu} \ell)$&
	$\EOp{eu}{\prime\,[3]}$ & $(\bar{e} \gamma^{\mu\nu\rho}  e) (\bar{u} \gamma_{\mu\nu\rho} u)$&
	$\EOp{\ell qde}{[3]}$ & $(\bar{\ell} \gamma^{\mu\nu\rho}  q) (\bar{d} \gamma_{\mu\nu\rho}  e)$
	\tabularnewline\cmrule{5-6}
	$\EOp{\ell d}{[2]}$ & $(\bar{\ell} \gamma^{\mu\nu} d)(\bar{d} \gamma_{\mu\nu} \ell)$&
	$\EOp{ed}{\prime\,[3]}$ & $(\bar{e} \gamma^{\mu\nu\rho}  e) (\bar{d} \gamma_{\mu\nu\rho} d)$&
\multicolumn{2}{c}{\cellcolor[gray]{0.90}$\boldsymbol{\bar{L}L \bar{L}L}$} 
	\tabularnewline\cmrule{5-6}
	$\EOp{qe}{[2]}$ & $(\bar{q} \gamma^{\mu\nu} e)(\bar{e} \gamma_{\mu\nu} q)$&
	&&
	$\EOp{\ell q}{}$ & $(\bar{\ell} \gamma^{\mu}  q) (\bar{q} \gamma_{\mu}  \ell)$
	\tabularnewline \cmrule{1-2}
	\multicolumn{2}{c}{\cellcolor[gray]{0.90}$\boldsymbol{\bar{L}R \bar{L}R}$} &
	&&
	$\EOp{\ell q}{(3)}$ & $(\bar{\ell} \gamma^{\mu}\sigma^I q) (\bar{q} \gamma_{\mu}\sigma^I  \ell)$
	\tabularnewline \cmrule{1-2}
	$\EOp{\ell equ}{[2]}$ & $(\bar{\ell}_r \gamma^{\mu\nu}  e)\epsilon_{rs} (\bar{q}_s \gamma_{\mu\nu}  u)$&
	&&
	$\EOp{\ell q}{[3]}$ & $(\bar{\ell} \gamma^{\mu\nu\rho}q) (\bar{q} \gamma_{\mu\nu\rho}\ell)$
	\tabularnewline
	$\EOp{\ell u qe}{}$ & $(\bar{\ell}_r u)\epsilon_{rs} (\bar{q}_s e)$&
	&&
	$\EOp{\ell q}{[3](3)}$ & $(\bar{\ell} \gamma^{\mu\nu\rho} \sigma^I q) (\bar{q} \gamma_{\mu\nu\rho} \sigma^I \ell)$
	\tabularnewline
	$\EOp{\ell uqe}{[2]}$ & $(\bar{\ell}_r \gamma^{\mu\nu}u)\epsilon_{rs} (\bar{q}_s \gamma_{\mu\nu}e)$&
	&&
	$\EOp{\ell q}{\prime\,[3]}$ & $(\bar{\ell} \gamma^{\mu\nu\rho}\ell) (\bar{q} \gamma_{\mu\nu\rho}q)$
	\tabularnewline
	&&
	&&
	$\EOp{\ell q}{\prime\,[3](3)}$ & $(\bar{\ell} \gamma^{\mu\nu\rho} \sigma^I \ell) (\bar{q} \gamma_{\mu\nu\rho} \sigma^I q)$
	\tabularnewline
\cbottomrule
\end{tabular}
\par\end{centering}
\caption{\label{tab:e4semilep} Semileptonic four-fermion evanescent operators. We use the shorthand notation $\gamma^{\mu_1 \ldots \mu_n}  \equiv \gamma^{\mu_1} \ldots \gamma^{\mu_n}$ with no (anti)symmetrization.}
\end{table}

\begin{table}[h]
\begin{centering}
\begin{tabular}{cccccc}
\ctoprule
\multicolumn{2}{c}{\cellcolor[gray]{0.90}$\boldsymbol{\bar{R}R \bar{R}R}$} &
\multicolumn{2}{c}{\cellcolor[gray]{0.90}$\boldsymbol{\bar{L}L \bar{L}L}$} &
\multicolumn{2}{c}{\cellcolor[gray]{0.90}$\boldsymbol{\bar{L}L \bar{R}R}$} 
	\tabularnewline \cmrule{1-6}
	$\EOp{ee}{[3]}$ & $(\bar{e} \gamma^{\mu\nu\rho}  e) (\bar{e} \gamma_{\mu\nu\rho}  e)$&
	$\EOp{\ell \ell}{(3)}$ & $(\bar{\ell} \gamma^{\mu} \sigma^I \ell) (\bar{\ell} \gamma_{\mu} \sigma^I  \ell)$&
	$\EOp{\ell e}{[3]}$ & $(\bar{\ell} \gamma^{\mu\nu\rho} l) (\bar{e} \gamma_{\mu\nu\rho}  e)$
	\tabularnewline \cmrule{1-2}
\multicolumn{2}{c}{\cellcolor[gray]{0.90}$\boldsymbol{\bar{L}R \bar{R}L}$} &
	$\EOp{\ell \ell}{[3]}$ & $(\bar{\ell} \gamma^{\mu\nu\rho} l) (\bar{\ell} \gamma_{\mu\nu\rho}  \ell)$&
	&
	\tabularnewline \cmrule{1-2}
	$\EOp{\ell \ell}{(3)}$ & $(\bar{\ell} \gamma^{\mu} \sigma^I \ell) (\bar{\ell} \gamma_{\mu} \sigma^I  \ell)$&
	$\EOp{\ell\ell}{[3](3)}$ & $(\bar{\ell} \gamma^{\mu\nu\rho}\sigma^I l) (\bar{\ell} \gamma_{\mu\nu\rho}\sigma^I  \ell)$&
	&
	\tabularnewline
	$\EOp{\ell \ell}{[3]}$ & $(\bar{\ell} \gamma^{\mu\nu\rho} l) (\bar{\ell} \gamma_{\mu\nu\rho}  \ell)$&
	&&
	&
	\tabularnewline
	$\EOp{\ell\ell}{[3](3)}$ & $(\bar{\ell} \gamma^{\mu\nu\rho}\sigma^I l) (\bar{\ell} \gamma_{\mu\nu\rho}\sigma^I  \ell)$&
	&&
	&
	\tabularnewline
\cbottomrule
\end{tabular}
\par\end{centering}
\caption{\label{tab:e4lep} Leptonic four-fermion evanescent operators. We use the shorthand notation $\gamma^{\mu_1 \ldots \mu_n}  \equiv \gamma^{\mu_1} \ldots \gamma^{\mu_n}$ with no (anti)symmetrization.}
\end{table}

\begin{table}[h]
\begin{centering}
\begin{tabular}{cccccc}
\ctoprule
\multicolumn{2}{c}{\cellcolor[gray]{0.90}$\boldsymbol{\bar{L}^c L \bar{L}L^c}$} &
\multicolumn{2}{c}{\cellcolor[gray]{0.90}$\boldsymbol{\bar{R}^c R \bar{R}R^c}$} &
\multicolumn{2}{c}{\cellcolor[gray]{0.90}$\boldsymbol{\bar{R}^c R \bar{L}L^c}$} 
	\tabularnewline \cmrule{1-6}
	$\EOp{qq}{c}$ & $(\overline{q^{c}}_{ar} q_{b s} ) (\bar{q}_{b s} q^{c}_{a r})$&
	$\EOp{uu}{c}$ & $(\overline{u^c}_{a} u_{b}) (\bar{u}_{b} u^c_{a})$&
	$\EOp{udqq}{c}$ & $(\overline{u^c}_{a} d_{b}) (\bar{q}_{b r}\epsilon_{rs} q^c_{as})$
	\tabularnewline 
	$\EOp{qq}{c\,\prime}$ & $(\overline{q^c}_{a r} q_{b s} ) (\bar{q}_{a s} q^{c}_{b r})$&
	$\EOp{dd}{c}$ & $(\overline{d^c}_{a} d_{b}) (\bar{d}_{b} d^c_{a})$&
	$\EOp{udqq}{c\,[2]}$ & $(\overline{u^c}_{a} \gamma^{\mu\nu} d_{b}) (\bar{q}_{b r}\epsilon_{rs} \gamma_{\mu\nu}q^c_{as})$
	\tabularnewline \cmrule{5-6}
	$\EOp{qq}{c\,[2]}$ & $(\overline{q^c}_{a r} \gamma^{\mu\nu} q_{b s} ) (\bar{q}_{b s}\gamma_{\mu\nu} q^c_{a r})$&
	$\EOp{ud}{c}$ & $(\overline{u^c}_{a} d_{b}) (\bar{d}_{b} u^c_{a})$&
\multicolumn{2}{c}{\cellcolor[gray]{0.90}$\boldsymbol{\bar{L}^c R \bar{R}L^c}$} 
	\tabularnewline\cmrule{5-6}
	$\EOp{qq}{c\,\prime [2]}$ & $(\overline{q^c}_{a r}\gamma^{\mu\nu} q_{b s} ) (\bar{q}_{a s}\gamma_{\mu\nu} q^c_{b r})$&
	$\EOp{ud}{c\, \prime}$ & $(\overline{u^c}_{a} d_{b}) (\bar{d}_{a} u^c_{b})$&
	$\EOp{qu}{c}$ & $(\overline{q^c}_{a }\gamma^{\mu} u_{b} ) (\bar{u}_{b }\gamma_{\mu} q^c_{a})$
	\tabularnewline
	&&
	$\EOp{uu}{c\,[2]}$ & $(\overline{u^c}_{a} \gamma^{\mu\nu} u_{b}) (\bar{u}_{b}\gamma_{\mu\nu} u^c_{a})$&
	$\EOp{qd}{c}$ & $(\overline{q^c}_{a }\gamma^{\mu} d_{b} ) (\bar{d}_{b }\gamma_{\mu} q^c_{a})$
	\tabularnewline
	&&
	$\EOp{dd}{c\,[2]}$ & $(\overline{d^c}_{a}\gamma^{\mu\nu} d_{b}) (\bar{d}_{b}\gamma_{\mu\nu} d^c_{a})$&
	$\EOp{qu}{c\,\prime}$ & $(\overline{q^c}_{a }\gamma^{\mu} u_{b} ) (\bar{u}_{a}\gamma_{\mu} q^c_{b})$
	\tabularnewline
	&&
	$\EOp{ud}{c\,[2]}$ & $(\overline{u^c}_{a}\gamma^{\mu\nu} d_{b}) (\bar{d}_{b}\gamma_{\mu\nu} u^c_{a})$&
	$\EOp{qd}{c\,\prime}$ & $(\overline{q^c}_{a }\gamma^{\mu} d_{b} ) (\bar{d}_{a}\gamma_{\mu} q^c_{b})$
	\tabularnewline
	&&
	$\EOp{ud}{c\, \prime [2]}$ & $(\overline{u^c}_{a} \gamma^{\mu\nu} d_{b}) (\bar{d}_{a}\gamma_{\mu\nu} u^c_{b})$&
	&
	\tabularnewline
\cbottomrule
\end{tabular}
\par\end{centering}
\caption{ \label{tab:e4quarksC} Evanescent operators with four fermions involving only quarks and featuring charge conjugation. We use the shorthand notation $\gamma^{\mu_1 \ldots \mu_n}  \equiv \gamma^{\mu_1} \ldots \gamma^{\mu_n}$ with no (anti)symmetrization.}
\end{table}
\begin{table}[h]
\begin{centering}
\begin{tabular}{cccccc}
\ctoprule
\multicolumn{2}{c}{\cellcolor[gray]{0.90}$\boldsymbol{\bar{L}^c L \bar{L}L^c}$} &
\multicolumn{2}{c}{\cellcolor[gray]{0.90}$\boldsymbol{\bar{R}^c R \bar{R}R^c}$} &
\multicolumn{2}{c}{\cellcolor[gray]{0.90}$\boldsymbol{\bar{R}^c R \bar{L}L^c}$} 
	\tabularnewline \cmrule{1-6}
	$\EOp{\ell \ell}{c}$ & $(\overline{\ell^{c}}_{r} \ell_{s} ) (\bar{\ell}_{s} \ell^{c}_{ r})$&
	$\EOp{ee}{c}$ & $(\overline{e^c} e) (\bar{e} e^c)$&
	$\EOp{ue\ell q}{c}$ & $(\overline{u^c} e) (\bar{\ell}_{r}\epsilon_{rs} q^c_{s})$
	\tabularnewline 
	$\EOp{q\ell}{c}$ & $(\overline{q^c}_{r} \ell_{s} ) (\bar{\ell}_{s} q^{c}_{r})$&
	$\EOp{eu}{c}$ & $(\overline{e^c}u) (\bar{u} e^c)$&
	$\EOp{ue\ell q}{c\,[2]}$ & $(\overline{u^c}\gamma^{\mu\nu} e) (\bar{\ell}_{r}\gamma_{\mu\nu}\epsilon_{rs} q^c_{s})$
	\tabularnewline \cmrule{5-6}
	$\EOp{q\ell}{c\,\prime}$ & $(\overline{q^c}_{r}  \ell_{s} ) (\bar{q}_{r}\ell^c_{s})$&
	$\EOp{ed}{c}$ & $(\overline{e^c} d) (\bar{d} e^c)$&
\multicolumn{2}{c}{\cellcolor[gray]{0.90}$\boldsymbol{\bar{L}^c R \bar{R}L^c}$} 
	\tabularnewline\cmrule{5-6}
	$\EOp{\ell\ell}{c\, [2]}$ & $(\overline{\ell^c}_{r}\gamma^{\mu\nu} \ell_{s} ) (\bar{\ell}_{s}\gamma_{\mu\nu} q^c_{r})$&
	$\EOp{ee}{c\, [2]}$ & $(\overline{e^c} \gamma^{\mu\nu}e) (\bar{e} \gamma_{\mu\nu}e^c)$&
	$\EOp{\ell e}{c}$ & $(\overline{\ell^c}\gamma^{\mu} e ) (\bar{e}\gamma_{\mu} \ell^c)$
	\tabularnewline
	$\EOp{q\ell}{c\, [2]}$ & $(\overline{q^c}_{r}\gamma^{\mu\nu} \ell_{s} ) (\bar{q}_{s}\gamma_{\mu\nu} q^c_{r})$&
	$\EOp{eu}{c\, [2]}$ & $(\overline{e^c}\gamma^{\mu\nu}u) (\bar{u}\gamma_{\mu\nu} e^c)$&
	$\EOp{qe}{c}$ & $(\overline{q^c}\gamma^{\mu} e ) (\bar{e}\gamma_{\mu} q^c)$
	\tabularnewline
	$\EOp{q\ell}{c\,\prime [2]}$ & $(\overline{q^c}_{r} \gamma^{\mu\nu} \ell_{s} ) (\bar{q}_{r}\gamma_{\mu\nu}\ell^c_{s})$&
	$\EOp{ed}{c\, [2]}$ & $(\overline{e^c}\gamma^{\mu\nu} d) (\bar{d}\gamma_{\mu\nu} e^c)$&
	$\EOp{\ell u}{c}$ & $(\overline{\ell^c}\gamma^{\mu} u ) (\bar{u}\gamma_{\mu} \ell^c)$
	\tabularnewline
	&&
	&&
	$\EOp{\ell d}{c}$ & $(\overline{\ell^c}\gamma^{\mu} d ) (\bar{d}\gamma_{\mu} \ell^c)$
	\tabularnewline
	&&
	&&
	$\EOp{q e d\ell }{c}$ & $(\overline{q^c}\gamma^{\mu} e ) (\bar{d}\gamma_{\mu} \ell^c)$
	\tabularnewline
\cbottomrule
\end{tabular}
\par\end{centering}
\caption{\label{tab:e4fermC} Semileptonic and leptonic evanescent operators with four fermions  featuring charge conjugation. We use the shorthand notation $\gamma^{\mu_1 \ldots \mu_n}  \equiv \gamma^{\mu_1} \ldots \gamma^{\mu_n}$ with no (anti)symmetrization.}
\end{table}

\clearpage

\bibliographystyle{style} 
\bibliography{refs}

\end{document}